%% file: main.tex
\documentclass[]{ucbthesis}
\pdfoutput=1
\usepackage[sorting=none,citestyle=numeric-comp,bibstyle=ieee,backref=true,isbn=false,url=false]{biblatex}
\usepackage{amsmath}
\usepackage{graphicx}
\usepackage[hang,flushmargin]{footmisc} 
\usepackage[bookmarks=true,bookmarksnumbered=true,colorlinks=true,linkcolor=blue,citecolor=blue]{hyperref}
\usepackage{bookmark}
\usepackage{algorithm,algorithmicx,algpseudocode}
\graphicspath{{AlgorithmFigures/}{ApproachingSQFigures/}{NextGenFigures/}{LevelSetFigures/}{VoltageFigures/}{CloakingFigures/}{SolarOptFigures/}{TimeDepFigures/}{IntroFigures/}}

\bibliography{Solar_Cells,Inverse_Design,Foundational_Books,Level_Set_Computations,Record_Solar_Cells,Fluctuation-Dissipation,Optical_Cloaking,Software,Other,Optoelectronics}

\begin{document}


\title{Photonic Design: From Fundamental Solar Cell Physics to Computational Inverse Design}
\author{Owen Dennis Miller}
\degreesemester{Spring}
\degreeyear{2012}
\degree{Doctor of Philosophy}
\chair{Professor Eli Yablonovitch}
\othermembers{Professor Ming Wu \\
  Professor Tarek Zohdi}
\numberofmembers{3}
\prevdegrees{B.S. (University of Virginia) 2007 \\
  M.S. (University of California, Berkeley) 2009}
\field{Electrical Engineering}
\campus{Berkeley}




\maketitle
\copyrightpage

\include{abstract}

\begin{frontmatter}

\begin{dedication}
\null\vfil
\begin{center}
To my newborn daughter Nicole (and your future siblings).\\I can only hope that you will find something you enjoy as much as I have enjoyed this work.
\end{center}
\vfil\null
\end{dedication}

\tableofcontents
\clearpage
\listoffigures
\clearpage
\listoftables
\clearpage
\listofalgorithms

\begin{acknowledgements}
First, I would like to thank Eli for the opportunity to work in his group for the past five years.  His insights and deep understanding of seemingly all of science will never cease to amaze me.  His knowledge is surpassed only by his enthusiasm for discovery, which is infectious.  He is a scientist of the first order, whom I will be grateful to consider a mentor throughout my career.  He has made a real difference in who I am as a scientist, which I think is rare and significant, and for which I will always be thankful.

I would also like to thank the professors on my qualifying exam committee: Ming Wu, Tarek Zohdi, and Xiang Zhang.  Leaders in their respective fields, their probing questions and insightful suggestions throughout the entire process certainly influenced the direction of this work.

More generally, I have greatly enjoyed the research environment at Berkeley.  My colleagues and friends have been amazingly smart and ambitious people, providing a constant source of stimulating conversation and ideas.  I would first like to thank the students with whom I have worked on a variety of projects.  Samarth Bhargava and Vidya Ganapati have suffered more over-the-shoulder coding than anyone rightly deserves, and I thank them for their patience.  They significantly contributed to the inverse design work presented, and I hope we have a chance to collaborate further in the future.  I would like to thank Avi Niv, Chris Gladden, Majid Gharghi, and Ze'ev Abrams for introducing me to new thermodynamics concepts, and for many interesting solar cell discussion.  I would like to thank many others I have not directly worked with, but with whom I have shared many interesting scientific (and non-scientific) conversations: Sapan Agarwal, Matteo Staffaroni, Amit Lakhani, Roger Chen, Jeff Chou, Nikhil Kumar, Justin Valley, Chris Chase, Arash Jamshidi, Mike Eggleston, James Ferrara, and certainly many others.  I would also like to take a moment to thank Vadim Karagodsky, a superior thinker and generous friend, who passed away a few months ago.  The scientific world lost a bright light far too soon.

Finally, I would like to thank my family for their love and support over the years.  My parents have always been amazingly supportive and encouraging, and in many ways shaped who I am today.  Even though I am the oldest, my siblings have undoubtedly taught me more about life than I have taught them.  And I would like to thank my wife, Betty, who is proof that you should marry someone smarter than yourself.  Without her never-ending love, support, and wisdom, this work certainly never would have been possible, and for which I owe the biggest 'thank-you' of all.

\end{acknowledgements}

\end{frontmatter}

\pagestyle{headings}

\include{Introduction}

\part{The Physics of High-Efficiency Solar Cells}

\include{OnVoltage}

\include{ApproachingSQ}

\include{NextGenCells}

\part{Photonic Inverse Design}

\include{Algorithm}
\include{Cloaking}
\include{SolarOpt}

\bookmarksetup{startatroot}

\printbibliography

\appendix
\include{MaxwellSymm}
\include{TimeDependentShapeCalculus}

\end{document}

%% file: Introduction.tex
\chapter{Introduction}
\epigraph{To myself I am only a child playing on the beach, while vast oceans of truth lie undiscovered before me.}{Isaac Newton}

\noindent
Connecting \emph{structure} to \emph{function} has a long history of driving scientific progress.  Einstein's recognition that a four-dimensional spacetime could explain gravity led to general relativity, revolutionizing modern physics.  Heinrich Hertz's spark-gap radiative structures provided insights into electromagnetic wave propagation, and are in many ways the foundation of radio technology.  Periodic crystals exhibit the semi-insulating, semi-conducting behavior enabling the invention of the transistor, the workhorse of computing.  Understanding the relationship between a structure and its functionality can provide deep scientific understanding, generate new conceptual avenues, and enable breakthrough technologies.

This thesis explores the connections between structure and function in photonic design.  Two approaches are taken.  First, within the context of solar cells, we examine the fundamental physics underlying device operation.  The efficiency limits for rather general photovoltaic technologies have been known for fifty years, and yet the best prototypes have fallen far short of their ideal performance.  This is partly due to practical limitations such as material quality, but shortcomings in design practices have also played a significant role, as will be discussed extensively in Part~I.  Counter-intuitively, solar cells, which convert incident photons into extracted electricity, should be designed to be ideal light \emph{emitters}.  This has profound implications on both material selection and device design, and more generally provides a new lens to examine the future prospects of a variety of next-generation solar technologies.

Part II approaches the more general problem of improving photonic design methods.  Until now, the primary method for designing an electromagnetic structure has been heuristic intuition.  For a given application, a researcher familiar with back-of-the-envelope calculations and a variety of simple textbook examples intuits a structure.  He may then test the structure through simulation, but fundamentally the design space is restricted by the engineer's imagination.  We introduce a new framework for \emph{computational inverse design}: instead of computing the response of a structure, we instead try to find the structure that best provides a desired response.  Through efficient shape calculus techniques, non-intuitive, superior structures can be computed quickly and efficiently.

\section{The Solar Energy Landscape}
Solar power is the world's greatest energy resource.  A continuous energy current of approximately $1.7\times 10^{17}\mathrm{W}$ of sunlight is incident upon the earth \cite{Wurfel2005}.  A year's worth of sunlight thus contains $1.5 \times 10^{18} \mathrm{kWh}$ of energy.  By comparison, the known reserves of oil, coal, and gas are $1.75 \times 10^{15} \mathrm{kWh}$, $1.4 \times 10^{15} \mathrm{kWh}$, and $5.5 \times 10^{15} \mathrm{kWh}$, respectively.  A year of sunlight provides more than a hundred times the energy of the world's entire known fossil fuel reserves.  Harnessing solar power would represent a never-ending energy supply.\footnote{I am not suggesting it would be free!}

The difficulty has always been converting solar energy in an efficient and cost-effective way.  Photovoltaic cells are the most promising avenue, directly converting the photons to electricity.  Yet the solar cell modules of the largest publicly traded company, First Solar, convert solar energy at only about $10\%$ efficiency \cite{Green2011d,Green2012}.  Even the best crystalline silicon solar cell modules have efficiencies around $22\%$, wasting almost $80\%$ of the incident power.  New technologies are needed for solar conversion to compete at cost-parity with fossil fuels.

Efficiency is a primary driver of cost for solar cells.  A more efficient module by definition yields more power per unit area.  A significant fraction of a solar cell's cost scales proportional to the installation area, including the cost of the glass, inverter costs (actually directly proportional to the power), and installation costs, among others \cite{Wang2011a}.  Such costs are fixed relative to the module technology, thereby providing a lower bound on the total costs for a given efficiency.  For example, even if the module cost \emph{is zero}, a $10\%$ efficient module cannot produce electricity at cheaper than about $ \$$$0.06 \mathrm{/kWh}$.    By comparison, a $25\%$ efficient module can cost about $300\$$$/\mathrm{m^2}$ and yet produce electricity for the same cost.  The path to cost-parity is through high-efficiency cells.

Given the focus on high efficiency, it is natural to ask: what is the ultimate limit to a solar cell's energy conversion efficiency?  Fifty years ago Shockley and Queisser provided a formulation to answer this question \cite{Shockley1961}.  For a given material and a few basic assumptions,\footnote{Their analysis can also be generalized to analyze technologies for which their assumptions are violated, as in e.g. multi-carrier generation cells.} they recognized the fundamental losses that occur.  First, for all energies smaller than the material bandgap the incident photons cannot be absorbed.  One does not want arbitrarily small bandgaps, however, as carriers generated by absorption thermalize to the bandgap energy, providing a second loss mechanism. And, finally, there is a required rate of emission \emph{from} the solar cell, set by thermodynamic detailed balancing.\footnote{Note that there are also other small contributors to imperfect conversion efficiency, such as the Carnot factor.}  For a single-junction solar cell under one-sun concentration, these loss mechanisms lead to a limiting efficiency of $33.5\%$.  For a variety of other configurations, such as concentrator or multi-junction solar cells, a similar detailed balancing process yields a different limiting efficiency, usually in the $30\%$ to $50\%$ range.

Theoretical efficiency limits are useful primarily because they provide a means for selecting which technologies to pursue, and they are a driving force for further progress.  Yet implicit in such a process is the assumption that the upper limit provides a realistic estimate of potential performance.  Real systems will never be perfect, but small deviations from perfect should yield only small deviations from ideal efficiencies.

A central theme of Part I of this thesis is that the Shockley-Queisser efficiencies are not robust to small deviations.  Although they provide a simple calculational tool, they sweep important internal dynamics ``under the rug.''  We examine these dynamics, resulting in a surprising conclusion: instead of considering external emission as a loss mechanism, it should actually be designed for.  \emph{Maximizing} external emission results in maximal voltages and efficiencies.  Viewed through this principle, the sensitivity of the ideal efficiencies to small imperfections is understandable and predictable.  Additionally, it provides a pathway to approaching the ideal limits.  For example: a rear surface mirror reflectivity of $98\%$ provides double the external emission as a reflectivity of $96\%$.  Similarly small gains in material quality or geometric configuration can also have substantial impacts.  Solar cells are an exception to the rule of diminishing returns; conversely, incremental enhancements can generate significantly improved performance.

It turns out that understanding the voltage output of the solar cell explains the insights discussed above.  Chap.~\ref{chap:OnVoltage} derives the voltage formula and discusses its determinants.  Although the chapter is primarily occupied with the ray optics regime, Sec.~\ref{sec:NearFieldVolt} introduces recent work demonstrating voltage calculations in the sub-wavelength and near-field regimes.  Chap.~\ref{chap:SQLimit} then builds up a formalism for understanding detailed balance efficiency limits with imperfections.  The path to approaching the Shockley-Queisser efficiency limits, through maximizing luminescent yield, is presented.  Finally, Chap.~\ref{chap:NextGenCells} applies the formalism to a variety of next-generation solar cell technologies, examining the robustness of each and providing a new prism through which to consider what technology to pursue.

\section{Inverse Design}
Scientific design generally progresses through three stages.  When there has been little mathematical apparatus built up, scientists make progress through what could be called the \emph{Edisonian method}: hypothesizing new structures or designs, then experimentally testing them.  Once the foundational mathematics is understood, transition to a second phase can occur, in which new designs are intuited and perhaps tested computationally, but fundamentally the design space is imposed by the scientist.  Finally, once computational design tools have been created, new designs can arrive from computational problem-solving; at this point, the division of labor allows the scientist to recognize the important problem to solve and the requisite constraints, while the computational tools explore the design space for optimal performance.  

Different fields are at different stages of scientific design.  There are still large swaths of biology for which the Edisonian method is the primary tool.  At the opposite end of the spectrum, circuit designers have created an impressive array of tools for automated circuit design, to the point where very large, complex system architectures can be computationally generated and optimized.

\begin{figure}[]
\centering
\includegraphics[height=2.5in]{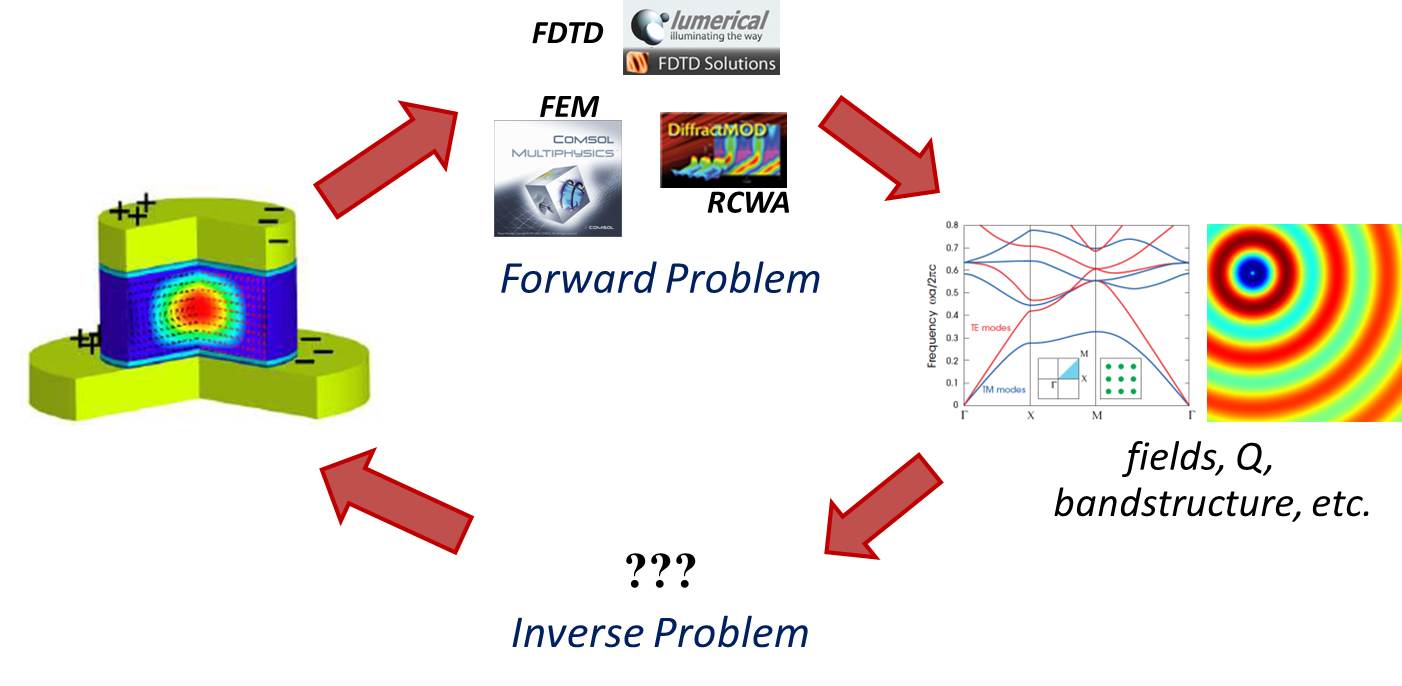}
\caption{``Forward'' and ``inverse'' problems.  There is an abundance of fast, efficient forward solvers, capable of computing the electromagnetic response of a given structure.  New tools for the inverse design problem, such that for a given response, the best structure is found, represent a new paradigm in computational electromagnetics.}
\label{fig:ID_Intro} 
\end{figure}

Photonic design has been in the second phase for a few decades.  Substantial progress has been made in computing the electromagnetic
response of a given structure, such that several commercial
programs provide computational tools for a wide array of problems.  Fig.~\ref{fig:ID_Intro} illustrates some of the computational techniques and tools available for computing the electromagnetic response.  We will call such techniques answers to the ``forward problem,'' where the structure is given and the response is unknown.  But tools for answering the ``inverse problem,'' where one specifies a response and computes a structure, are still very much in their infancy.  Advancements in solving the inverse problem will enable photonic design to reach the final phase of scientific design.

The inverse problem cannot be solved by simply choosing a desired electric field and numerically computing the dielectric structure. It is generally unknown whether such a field can exist, and if so, whether the dielectric structure producing it has a simple physical realization.  Instead, the inverse problem needs to be approached through iteration: start with some initial structure, then iterate until the final structure most closely achieves the desired functionality.

As computational power progressively increases, simulation becomes more accurate and less time-consuming, and computational design takes on more importance in the scientific process.  Part II of this thesis attempts to contribute to this development.

Chap.~\ref{chap:Algorithm} presents a framework for a new inverse design method, formulated through a shape calculus mathematical foundation.  Chapters~\ref{chap:OptCloak}~and~\ref{chap:SolarOpt} apply this framework to two photonic design applications.  Chap.~\ref{chap:OptCloak} demonstrates the utility of the computational approach to optical cloak design, showing its versatility in enabling the designer to decide what constraints and design space to work in.  Chap.~\ref{chap:SolarOpt} applies the optimization method to a new solar cell design: a thin-film solar cell in the sub-wavelength regime, where the ray optical laws are not valid.  A non-intuitive structure is designed, with an angle- and polarization-averaged absorption enhancement of $40$, far larger than enhancements found for other proposed structures in the same regime.  The inverse design framework presented here can be applied to a wide variety of applications, potentially discovering new structures and functionalities.  Whereas the current forefront of electromagnetic computation is the quick solution of the response to a given structure, the inverse problem of computing the structure for a given response may prove much more powerful in the future.

%% file: OnVoltage.tex
\chapter{Luminescent Extraction Determines the Voltage}
\label{chap:OnVoltage}
\epigraph{How wonderful that we have met with a paradox. Now we have some hope of making progress.}{Niels Bohr}

\noindent
The power output of a solar cell is given by its current-voltage product, $P=IV$, equally dependent on both the current and the voltage.  The current is generally straightforward to determine: at what rate are incident photons absorbed, and what percentage of generated carriers are extracted through the contacts?  Determining the voltage, in contrast, requires more subtle understanding.  In this chapter we show that the open-circuit voltage, and therefore also the operating point voltage, is primarily determined by how efficiently the solar cell luminesces.
  
\section{The Relevance of the Open-Circuit Condition}
A key simplification for understanding the voltage derives from using the open-circuit voltage as a proxy for the operating point voltage.  We will justify that simplification here.

Analyzing a solar cell at its optimal power point requires complex dynamics balancing absorption, emission, and charge extraction.  However, instead of directly solving for the operating point current, $I_{OP}$, and voltage, $V_{OP}$, the output power can also be determined by simpler short-circuit and open-circuit conditions.  The power output can be equivalently written
\begin{equation}
P = I_{OP}V_{OP} = I_{SC}V_{OC}FF
\end{equation}
where $I_{SC}$, $V_{OC}$, and $FF$ are the short-circuit current, open-circuit voltage, and fill factor, respectively \cite{Wurfel2005}.  With a simple derivation one can show that the fill factor is itself a function of the open-circuit voltage, reducing the independent parameters to only the short-circuit current and open-circuit voltage.  

Consider a solar cell which can be described by the typical diode equation:
\begin{equation}
\label{eq:JDiode}
J = J_{SC} - J_0 e^{qV/kT}
\end{equation}
The derivation of the $J_{SC}$ and $J_0$ terms will be completed in Sec.~\ref{sec:DetBalance}, but for now they can be left simply in variable form.\footnote{Note that the dark current is typically multiplied by an extra ``$-1$'' factor, which has been omitted for consistency with later sections, and because it is very small relative to $e^{qV/kT}{\approx} e^{40}$.}  First, the open-circuit voltage requires $J=0$, such that $V_{OC}$ is given by
\begin{equation}
\label{eq:VocDiode}
qV_{OC} = kT \ln \left(\frac{J_{SC}}{J_{0}}\right)
\end{equation}
To find the operating point voltage, we must find the voltage for which the power output is maximum.  By setting the derivative $\partial (JV) / \partial V = 0$, the operating point conditions requires:
\begin{equation}
J_{SC} -J_0 e^{qV_{OP}/kT} \left( 1+\frac{qV_{OP}}{kT} \right) = 0
\end{equation}
where $V_{OP}$ is the operating point voltage.  Solving for the voltage and substituting the open-circuit voltage from Eqn.~\ref{eq:VocDiode}
\begin{equation}
\label{eq:VopTrans}
qV_{OP} = qV_{OC} - kT \ln \left( 1 + \frac{qV_{OP}}{kT} \right)
\end{equation}
Eqn.~\ref{eq:VopTrans} is a transcendental equation for $V_{OP}$, and is not separable.  However, one can recursively substitute for $V_{OP}$ on the right-hand side, yielding
\begin{equation}
qV_{OP} = qV_{OC} - kT \ln \left[ 1 + \frac{qV_{OC}}{kT} - \ln \left( 1+ \frac{qV_{OP}}{kT} \right) \right]
\end{equation}
Note that the third term in square brackets will generally be much smaller than the second term, due to the natural log.  Moreover, the difference occurs within a second natural log function, reducing it to a very small correction factor.  Thus to a good approximation:
\begin{equation}
\label{eq:VopVoc}
qV_{OP} = qV_{OC} - kT \ln \left( 1 + \frac{qV_{OC}}{kT} \right)
\end{equation}
Eqn.~\ref{eq:VopVoc} says that the operating point voltage is determined by the open-circuit voltage.  Because the operating point current is directly related to the operating-point voltage, through Eqn.~\ref{eq:JDiode}, this is therefore a proof that the fill factor is determined by the open-circuit voltage.  Note that for real solar cells there will be non-radiative recombination, which contributes non-ideality factors to the exponential voltage dependence of Eqn.~\ref{eq:JDiode}; nevertheless, the open-circuit voltage is still the prime determinant of fill factor and operating point voltage.  Cf. \cite{Green1982} for a variety of more accurate expressions of the fill factor in terms of the open-circuit voltage.  Having shown that the operating point voltage is given by the open-circuit voltage, the remainder of the chapter will explain what determines the open-circuit voltage.

\section{Detailed Balancing}
\label{sec:DetBalance}
Shockley and Queisser were the first to apply the concept of \emph{detailed balancing} to solar cells \cite{Shockley1961}.  Detailed balance dictates that at thermal equilibrium, by definition, every photon absorption event must be countered by a photon emission event, with the balance holding at every frequency and solid angle.  In and of itself, detailed balance is not directly useful for solar cells, which operate far from thermal equilibrium.  However, Shockley and Queisser recognized that the emission spectrum away from thermal equilibrium is different from the emission spectrum at equilibrium only by a scaling factor; this recognition was the key step toward understanding fundamental efficiency limits of solar cells.  

The open-circuit voltage of a solar cell can be derived from the above considerations.  A solar cell at thermal equilibrium with its surrounding environment of temperature $T$ has a constant flux of photons impinging upon it.  The surrounding environment radiates at $T$ according to the tail ($E \gg kT$) of the blackbody formula:
\begin{equation}
b(E) = \frac{2 n_r^2}{h^3 c^2}E^2 \textrm{exp}(-E/kT)
\end{equation}
where $b$ is given in photons per unit area, per unit time, per unit energy, per steradian.  $E$ is the photon energy, $n_r$ is the ambient refractive index, $c$ is the light speed, and $h$ is Planck’s constant.  As Lambertian distributed photons enter the solar cell’s surface at polar angle $\theta$, with energy $E$, the probability they will be absorbed is written as the dimensionless absorbance $a(E,\theta)$.  The flux per unit solid angle of absorbed photons is therefore $a(E,\theta)b(E)$.  In thermal equilibrium there must be an emitted photon for every absorbed photon; the flux of emitted photons per unit solid angle is then also $a(E,\theta)b(E)$.  

When the cell is irradiated by the sun, the system will no longer be in thermal equilibrium.  There will be a chemical potential separation, $\mu$, between electron and hole quasi-Fermi levels.  The emission spectrum, which depends on electrons and holes coming together, will be multiplied by the normalized $np$ product, $(np/n_i^2)$, where $n$, $p$, and $n_i$ are the excited electron and hole concentration, and the intrinsic carrier concentration, respectively.  The Law of Mass Action is $np=n_i^2\exp[\mu/kT]$ for the excited semiconductor in quasi-equilibrium \cite{Kittel1976}.  Then, the total photon emission rate is: 
\begin{equation}
\label{eq:FrontEmRate}
R_{em} = e^{\mu/kT} \iint a(E,\theta)b(E)\textrm{cos}\theta \textrm{d}E \textrm{d}\Omega
\end{equation}
for external solid angle $\Omega$ and polar angle $\theta$.  Eqn.~\ref{eq:FrontEmRate} is normalized to the flat plate area of the solar cell, meaning that the emission rate $R_{em}$ is the emissive flux from only the front surface of the solar cell.  Only non-concentrating solar cells are considered, such that the solid angle integral is taken over the full hemisphere.  There will generally be a much larger photon flux inside the cell, but most of the photons undergo total internal reflection upon reaching the semiconductor-air interface.  If the rear surface is open to the air, i.e. there is no mirror, then the rear surface emission rate will equal the front surface emission rate.  Restricting the luminescent emission to the front surface of the solar cell improves voltage, whereas a faulty rear mirror increases the avoidable losses, significantly reducing the voltage.  Efficient luminescent extraction \emph{through the front surface} yields high voltages.

At open circuit, there is a simple connection between the external photon emission rate, Eqn.~\ref{eq:FrontEmRate}, and the internal carrier recombination rate.  From the open-circuit condition, excited carriers cannot be drawn off as current; instead, they must eventually recombine, either radiatively or non-radiatively.  In the situation that every carrier recombines radiatively and every radiated photon successfully escapes, the net internal recombination rate $R_{recomb}$ equals the external photon emission rate.  However, if the number of external photons produced per excited carrier is reduced to $\eta_{ext}$, which we will call the external fluorescence yield, then we will have
\begin{equation}
R_{recomb} = \frac{1}{\eta_{ext}} R_{em}
\end{equation}
If, for example, only half of the excited carriers recombine and emit photons that make it out of the cell, then the total recombination rate is twice the rate of external emission.

To find the open-circuit voltage we now equate the carrier recombination and generation rates.  Carriers are generated by the incident solar radiation $S(E)$ according to the formula 
\begin{equation}
R_{gen} = \iint a(E,\theta)S(E) \cos \theta \textrm{d}E \textrm{d}\Omega
\end{equation}
Equating the generation and recombination rates, and recognizing that the open-circuit voltage equals the quasi-Fermi level separation ($qV_{OC} = \mu$), the resulting open-circuit voltage is
\begin{equation}
\label{eq:VocRatio}
V_{OC} = \frac{kT}{q} \ln \left( \frac{ \iint a(E,\theta)S(E)\cos \theta \textrm{d}E \textrm{d}\Omega }{ \iint a(E,\theta)b(E) \cos \theta \textrm{d}E \textrm{d}\Omega } \right) + \frac{kT}{q} \ln \left( \eta_{ext} \right)
\end{equation}
Because $\eta_{ext}$ is less than or equal to one, the second term in Eqn.~\ref{eq:VocRatio} represents a loss of voltage due to poor light extraction.   This term was first recognized by Ross \cite{Ross1966,Ross1967,Ross1967a}.  

\section{Entropic Penalties}
\label{sec:EntPen}
With the sun assumed to be a blackbody at $T_S \approx 6000K$ and the absorptivity a step-function equal to one above the bandgap, Eqn.~\ref{eq:VocRatio} can be simplified for more physical intuition.  The ambient blackbody temperature of $300K$, corresponding to $kT \approx 26 meV$, is sufficiently small to approximate the Bose-Einstein distribution denominator of $e^{E/kT}-1$ as $e^{E/kT}$.  In contrast, the solar temperature $kT_S \approx 500 meV$ is too large to approximate the distribution function through its tail, and therefore cannot be similarly approximated.

With the absorptivity assumed to be a step-function independent of angle, the angular integral in the denominator of Eqn.~\ref{eq:VocRatio} simplifies to $\pi$:
\begin{equation}
\label{eq:VocNoSolidAngle}
qV_{OC} = kT \ln \left( \frac{ \Omega_{S} \int S(E) \textrm{d}E }{ \pi \int b(E) \textrm{d}E } \right) + kT \ln \left( \eta_{ext} \right)
\end{equation}
where $\Omega_{S}$ is the solid angle of the sun.  By non-dimensionalizing the numerator of Eqn.~\ref{eq:VocNoSolidAngle} and approximating the denominator as discussed above, Eqn.~\ref{eq:VocNoSolidAngle} becomes:
\begin{equation}
qV_{OC} = kT \ln \left( \frac{ \Omega_{S} (kT_S)^3 \int_{E_g/kT_S}^{\infty} \frac{x^2}{e^x-1} \textrm{d}x }{ \pi (kT)^3 \int_{E_g/kT}^{\infty} x^2 e^{-x} \textrm{d}x } \right) + kT \ln \left( \eta_{ext} \right)
\end{equation}
Finally, integration by parts in the denominator yields
\begin{equation}
\label{eq:VocTerms}
qV_{OC} = E_g - kT \ln \left( \frac{\pi}{\Omega_S} \right) + kT \ln \left( \eta_{ext} \right) + kT \ln \left( \frac{T_S}{T} \right) - 2 kT \ln \left( \frac{E_g}{kT_S} \right) + kT \ln \left( \int_{x_g}^{\infty} \frac{x^2}{e^x-1}\textrm{d}x \right)
\end{equation}
Eqn.~\ref{eq:VocTerms} breaks down the open-circuit voltage into explicit contributions.  Before discussing the contributions, note the similarity of Eqn.~\ref{eq:VocTerms} to the general formula for the Helmholtz free energy:
\begin{equation}
F = U - TS
\end{equation}
where $F$ is the free energy, $U$ is the internal energy, and $S$ is the entropy loss \cite{Kittel1976}.  Although the voltage is often considered an electrical parameter, the process of absorbing photons, generating carriers, and emitting photons (many of which ultimately leave the cell) is also a steady-state process, for which a thermodynamic prescription is appropriate.  The entropy can furthermore be designated as $S = k \ln W$, where $k$ is the Boltzmann factor and $W$ is the configurational phase space of the system.

The system can be examined through either the carriers generated or the incident and exiting photons.  A thermodynamic treatment of the carriers results in the law of mass action ($qV = np/n_i^2$), which is not particularly useful for our analysis, because of the difficulty of tracking the carriers.  Instead, we consider the photon fluxes.

The phase space for optical rays is defined by the optical \'{e}tendue \cite{Winston1978,Shatz2010}.  Optical rays do not occupy the typical six-dimensional phase space of particles.  One of the spatial dimensions is redundant, because the same ray traverses infinitely through one dimension.  By the same token, one momentum component is also redundant.  The momentum spread is equivalent to a specification of the angular spread, such that one can write the differential \'{e}tendue $d\mathcal{E}$ as:
\begin{equation}
\label{eq:Etendue}
d\mathcal{E} = n_r^2 \cos\theta d\mathrm{A} d\Omega
\end{equation}
where $d\mathrm{A}$ is a surface differential, $d\Omega$ is the solid angle subtended, and $\theta$ is the angle between the surface normal of $d\mathrm{A}$ and the ray bundle \cite{Winston1978}.  The refractive index factor $n_r^2$ accounts for the increased density of states in media.  \'{E}tendue can never decrease, leading to theoretical limits for the possible geometric optics concentration factors \cite{Smestad1990}.

In deriving the voltage penalties of Eqn.~\ref{eq:VocTerms}, we will consider entropy generation due to non-idealities.  First, consider that the incident photons occupy only a very small solid angle $\Omega_S \approx 6.8\times 10^{-5} sr$.  Potentially, a solar cell could have an absorptivity of one for all rays within the sun's solid angle, and zero outside the sun's solid angle.  By detailed balance, emission would likewise only occur within $\Omega_S$.  Indeed, this is what would occur in an ideal concentrator.  Recent proposals have also attempted to restrict the emission without concentration \cite{Kosten2011,Peters2011}.  In reality, however, concentrator systems can be impractical due to the significant haze in the sky.  A flat-plate solar cell captures all of the sunlight, and therefore emits back into the entire sky.  Integrating the solid angle with the $\cos\theta$ term in Eqn.~\ref{eq:Etendue}, the emission occurs into a solid angle of $\pi$ steradians.  For the ideal scenario of no haze and no emission outside the solar solid angle, the optical phase space is $W_{\mathrm{ideal}}=A\Omega_{S}$, for a cell surface area $A$.  For the relevant case of a flat-plate solar cell absorbing and emitting from all angles, the phase space is $W_{\mathrm{solid-angle}} = A\pi$, much larger than the ideal case.  The entropy increase is therefore 
\begin{equation}
S_{\mathrm{solid-angle}} = k \ln \left( \frac{W_{\mathrm{ideal}}}{W_{\mathrm{solid-angle}}} \right)  = k \ln \left( \frac{\pi}{\Omega_S} \right)
\end{equation}
which when multiplied by the temperature $T$ is exactly the first penalty term in Eqn.~\ref{eq:VocTerms}.

The second voltage penalty is due to imperfect radiative efficiency.  This term is fairly straightforward.  It can be shown that the optical \'{e}tendue is equivalent to the number of rays in a bundle \cite{Winston1978}.  Given this definition, the relative ratio of $W_{\mathrm{ideal}}$ to the phase space associated with imperfect extraction, $W_{\mathrm{ext}}$, is precisely the inverse of the luminescent extraction efficiency, $1/\eta_{ext}$.  Thus we have
\begin{equation}
S_{\mathrm{ext}} = k \ln \left(\frac{1}{\eta_{ext}}\right) = -k\ln \eta_{ext}
\end{equation}
Therefore the second penalty in Eqn.~\ref{eq:VocTerms} is entropy generation due to a reduction in number of the exiting optical rays.  Importantly, this explains why the \emph{external} yield is the relevant parameter, instead of, e.g., the internal yield.  The external yield is far more sensitive to internal imperfections, placing a higher demand on the solar cell to overcome this entropy penalty.

Finally, the other penalty terms are smaller in magnitude and will not be individually derived here.  The third penalty term is due to ``photon cooling'' \cite{Markvart2008}, while the fourth and fifth are density of states modifications.

The entropic penalty due to mis-match of the solar and emission solid angle is the largest contribution to Eqn.~\ref{eq:VocTerms}, with a value of about $-10.7kT$.  The external yield term is highly variable, but the later terms contribute slightly less than $1kT$ to the voltage.  A first-order approximation to the voltage could then be written
\begin{equation}
qV_{OC} = E_g - 260meV + kT\ln(\eta_{ext})
\end{equation}
where the $260meV$ is almost exact for a $1.4eV$ bandgap and within $\pm kT$ for bandgaps ranging from $1$-$1.8eV$.  More generally, the external luminescence yield is the free parameter that determines how close the open-circuit voltage is to its ideal limit.

\section{The Difficulty of Light Extraction}
\label{sec:DiffLightExtraction}
Sec.~\ref{sec:EntPen} demonstrated that maximum external light extraction is the key to high open-circuit voltage.  In this section we discuss why that is such a difficult task.

\begin{figure}
\centering
\includegraphics[width=5in]{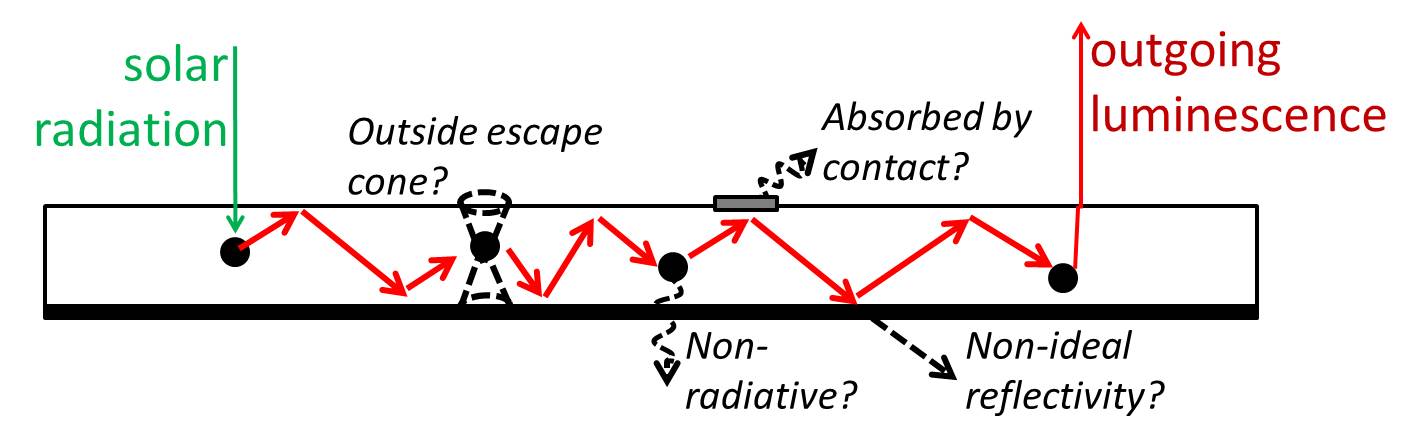}
\caption{The difficulty of light extraction in a solar cell (or light-emitting diode).  Because of the large refractive index of relevant semiconductors, external extraction requires many internal re-absorption and re-emission events, and many reflections from the surfaces.  Small non-idealities result in large external extraction penalties.} 
\label{fig:ExtractionDiff}
\end{figure}

Fig.~\ref{fig:ExtractionDiff} illustrates why $100\%$ light extraction is so difficult.  Consider an incident photon that has been absorbed within the semiconductor.  At the open-circuit condition, no carriers are extracted, and the electron-hole pair must eventually recombine.\footnote{In a semiconductor, the electron-hole pair generated would generally separate, unlikely to form an exciton.  A different electron and hole recombines, although the semantic distinction is unimportant for the current purpose.}  Upon re-emission, however, the newly emitted photon is not guaranteed to leave the cell.  Because of the high refractive indices for relevant semiconductors, there is a significant likelihood of being outside the escape cone, such that the ray undergoes total internal reflection.  The photon must then be re-absorbed before it\footnote{Note that we will speak of a single photon, even though ``it'' undergoes many re-absorption and re-emission events, and consists therefore of many different photons.} can can escape, because in a plane-parallel solar cell a photon emitted outside the escape cone will remain outside the escape cone.  Upon re-absorption, there is no guarantee a photon will be emitted again, as non-radiative processes such as Auger or Shockley-Read-Hall recombination compete with radiative recombination.  Moreover, the photons traversing the cell must avoid imperfects such as non-ideal reflectivity in the mirror or absorbing contacts.  As a consequence, achieving a high external yield requires minimal imperfections of any kind.

The difficulty alluded to above can be made mathematically precise.  There are shortcuts to calculating the external yield through detailed balance at the cell's external surfaces, which will be applied in Chap.~\ref{chap:SQLimit} to handle a variety of geometries.  However, they do not illuminate the photon dynamics particularly well, so an alternate derivation will be provided here.  Assuming a photon has been absorbed, the likelihood of eventual emission will be calculated by averaging over all possible photon paths.

For any geometry, the external luminescence yield, $\eta_{ext}$ can be parameterized by the internal luminescence yield, $\eta_{int}$, and the average probability of an internally emitted photon being re-absorbed, $a_{int}$.  The internal luminescence yield is the probability of a single recombination event being radiative.  This should be contrasted with the external yield, which tracks the probability from initial absorption, through re-absorption and re-emission events, to, finally, possible emission from the cell.

Consider a photon that has been absorbed.\footnote{At open-circuit, as always.}  The probability of internal photon emission is $\eta_{int}$.  Given re-emission, if the photon is inside the escape cone and not re-absorbed before reaching the front surface, the photon will escape.  Otherwise, the photon is re-absorbed, and the process iterates.  The probability of eventual escape, $\eta_{ext}$, is given by the infinite sum
\begin{align}
\eta_{ext} & = \eta_{int} (1-a_{int}) + \eta_{int} a_{int} \eta_{int}  (1-a_{int}) + \ldots \nonumber \\
& = \eta_{int} (1-a_{int}) \sum_{n=0}^{\infty} \left[ \eta_{int}a_{int} \right]^n \nonumber \\
& = \frac{ \eta_{int} (1-a_{int}) }{1-\eta_{int}a_{int}}
\label{eq:etaExtInfSum}
\end{align}

One can calculate the internal absorption probability $a_{int}$ for different geometries.  We will analyze the case of a plane-parallel solar cell with a perfect rear mirror.  $a_{int}$ has two contributions: photons emitted outside the escape cone are absorbed with probability unity;\footnote{There is no possibility to enter the escape cone and no other loss mechanism.} photons within the escape cone can also possibly be absorbed, depending on the optical thickness of the cell.  Instead of directly calculating the probability of absorption, it is easier to calculate the probability of immediate escape, which is $1-a_{int}$.\footnote{Note the distinction between \emph{immediate escape}, which refers to escape before re-absorption, and overall escape, which is the external yield that we are eventually driving towards.}

It can be shown that for the relatively large semiconductor refractive indices in solar cells, $n_r \sim \textrm{$3$--$4$}$, the probability of emission into the escape cone is approximately $1/2n_r^2$ \cite{Yablonovitch1982}.  The probability of immediate escape is thus given by $1/2n_r^2$ times the probability of not being absorbed before reaching the front surface.  Assuming the photons are emitted from carriers uniformly distributed throughout the geometry, one can calculate the average probabilities.

Photons emitted internally are equally likely to be emitted downwards as upwards.  Because of the perfect rear mirror, it is equivalent to treat every photon as emitted upwards, but over a distance $2L$, where $L$ is the cell thickness.  Because of the small escape cone, the light can be approximated as traveling perpendicular to the surface of the cell, resulting in a simple expression for the probability of re-absorption: $1-e^{-\alpha x}$, where $x$ is the distance traveled to the front surface.  The average probability of re-absorption within the escape cone, $a_{lc}$, is then:
\begin{equation}
a_{lc} = \frac{1}{2L}\int_0^{2L}\left(1-e^{-\alpha x} \right) \mathrm{d}x = 1 - \frac{1-e^{-2\alpha L}}{2\alpha L}
\end{equation}
The factor $1-e^{-2\alpha L}$ is identically the probability of absorbing an externally incident photon, which we will call $a_{ext}$ (previously it has simply been referred to as $a$).  Finally, the probability of immediate escape can be written
\begin{equation}
\label{eq:aIntAExtPP}
1-a_{int} = \frac{1}{2n_r^2}\left( \frac{1-e^{-2\alpha L}}{2\alpha L} \right) = \frac{a_{ext}}{4n_r^2\alpha L}
\end{equation}
Inserting Eqn.~\ref{eq:aIntAExtPP} into Eqn.~\ref{eq:etaExtInfSum}, the external luminescence yield can be written
\begin{equation}
\label{eq:etaExtPP}
\eta_{ext}^{PP} = \frac{a_{ext}}{a_{ext}+4n_r^2\alpha L \frac{1-\eta_{int}}{\eta_{int}}}
\end{equation}
Eqn.~\ref{eq:etaExtPP} is the external yield of a plane-parallel solar cell, uniquely determined by the material parameters $\alpha$ and $\eta_{int}$, and the geometrical thickness $L$.\footnote{$a_{ext}$ is a function of $\alpha L$.}

As a sanity check, for the limiting case $\eta_{int} = 1$, the external yield is one.  With no losses, the photons must eventually escape.  However, note the dramatic decrease in $\eta_{ext}$ when the internal yield is slightly less than one.  For a small deviation from ideal, the internal yield can be re-written $\eta_{int} = 1-\gamma$, where $\gamma$ is small.  For the condition $\eta_{int} < 4n_r^2\alpha L / (4n_r^2 \alpha L + a_{ext})$, the second term in the denominator is largest. For full absorption, this crossover occurs at about $99\%$, such that the external yield is approximately
\begin{equation}
\eta_{ext}^{PP} \simeq \frac{a_{ext}}{4n_r^2\alpha L}\frac{1-\gamma}{\gamma} \simeq \frac{a_{ext}}{4n_r^2\alpha L}\frac{1}{\gamma}
\end{equation}
For an internal yield less than approximately $99\%$, the external yield depends on the inverse of the small parameter $\gamma$!  As an example of this extreme dependence, for an internal yield of $99\%$ and an optical thickness $\alpha L = 2.5$, the external yield of the plane-parallel geometry is only about $50\%$.

\section{Random Surface Texturing for Increased Voltage}
\label{sec:RandText}
One of the primary difficulties of light extraction in the plane-parallel geometry was the fact the photons emitted outside the escape cone would not have a chance to escape until a further re-absorption and re-emission event.  One way to improve the extraction, therefore, is to randomly texture the surface of the solar cell.  The narrow escape cone does not increase in size, but a photon emitted outside the escape cone has a chance to be scattered into it by the random roughness.

The external yield of the randomly textured geometry has been derived through steady-state dynamics in \cite{Schnitzer1993a}.  Here, we will recognize that in the weakly absorbing limit, where the randomly textured formulas are derived in any case, the probability of an internally emitted photon being absorbed, $a_{int}$, equals the probability of an externally incident photon being absorbed, $a_{ext}$.  Once the externally incident photon has been scattered upon entering the cell, it is equivalent to having been internally emitted.  Therefore, one can write
\begin{equation}
\label{eq:aIntText}
a_{int} = a_{ext} = \frac{4n_r^2\alpha L}{4n_r^2\alpha L + 1}
\end{equation}
where the external absorptivity is derived in \cite{Yablonovitch1982a}.  Inserting Eqn.~\ref{eq:aIntText} into Eqn.~\ref{eq:etaExtInfSum} yields
\begin{equation}
\label{eq:etaExtText}
\eta_{ext}^{T} = \frac{\eta_{int}}{1+4n_r^2\alpha L \left( 1 - \eta_{int}\right)}
\end{equation}
for the external luminescence yield of a solar cell with a randomly textured front surface, and a perfect rear mirror.  Note that again, for $\eta_{int} = 1-\gamma$, with $\gamma$ small, the external yield is inversely proportional to the small parameter.  Texturing does not provide significant benefit in the almost ideal case, because re-absorption is sufficient to provide randomization.

\begin{figure}
\centering
\includegraphics[width=6.5in]{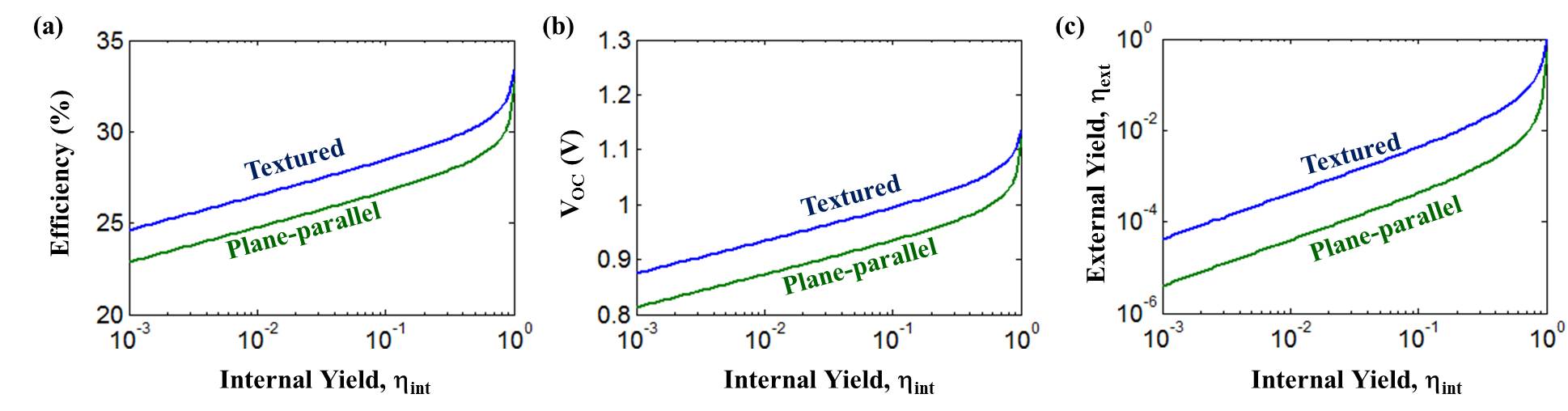}
\caption{The benefits of random surface texturing vs. plane-parallel geometry (without rear reflector).  Assuming a $1.4eV$ bandgap, the optimal plane-parallel and surface textured thicknesses were computed for each value of $\eta_{int}$.  The (a) efficiency, (b) open-circuit voltage, and (c) external yield $\eta_{ext}$ are shown here as a function of internal yield.} 
\label{fig:TextVsPP}
\end{figure}

The regime for which surface texturing provides significant benefits occurs when re-absorption does not provide sufficient randomization.  This occurs if the internal yield is relatively small, such that re-absorption does not lead to re-emission.  A significant boost also comes from the fact that random surface texturing achieves full absorption at much smaller thicknesses.  Making the cell thinner while maintaining full absorption significantly increases the external yield.  If full absorption is assumed for the plane-parallel geometry, the ratio of the two external yields can be written
\begin{equation}
\frac{\eta_{ext}^T}{\eta_{ext}^{PP}} \simeq \frac{4n_r^2\alpha L_{PP} (1-\eta_{int})}{\eta_{int} + 4n_r^2\alpha L_{T} (1-\eta_{int})}
\end{equation}
which for small $\eta_{int}$ can be further approximated by 
\begin{equation}
\frac{\eta_{ext}^T}{\eta_{ext}^{PP}} \simeq \frac{L_{PP}}{L_{T}}
\end{equation}
which is the ratio of the thicknesses of the two cells.  If one were to take, for example, $L_{PP} = 1/2$ (because of the double pass), and $L_{T} = 1/4n^2$, then the relative voltage difference between the two cells $\Delta V$ would be
\begin{equation}
q\Delta V = qV_{OC,\mathrm{textured}} - qV_{OC,\mathrm{plane-parallel}} = kT\ln \left( \frac{\eta_{ext}^T}{\eta_{ext}^{PP}} \right) \simeq kT\ln(2n^2)
\end{equation}
Furthermore, in comparison with a plane-parallel cell \emph{without} a rear mirror, which would require $L_{PP} = 1$, the randomly textured cell would have a voltage boost of
\begin{equation}
q\Delta V \simeq kT\ln (4n^2)
\end{equation}
a difference of approximately $100meV$.

Fig.~\ref{fig:TextVsPP} shows detailed calculations of the benefits from random surface texturing.  For a $1.4eV$ bandgap material, a step-function absorber is assumed.  For each value of internal yield $\eta_{int}$, the optimal optical thicknesses $L_{PP}$ and $L{T}$ are computed.  The efficiencies, open-circuit voltage, and external yield are shown as a function of internal yield.  Note the significant improvement from surface texturing for $\eta_{int}<1$.  The optimal thicknesses are generally $\alpha L_{PP} \approx 5$ (there is not a rear mirror) and $\alpha L_{T} \approx 0.5$.  This translates to a superior external yield for the textured geometry, by the ratio $\eta_{ext}^T/\eta_{ext}^{PP} \approx 10$, resulting in a voltage boost of approximately $kT\ln(10) \approx 60meV$, as seen in Fig.~\ref{fig:TextVsPP}(c).

\section{Photon Extraction Versus Light Trapping}	
There is often confusion about whether enhanced photon extraction is at odds with ``light trapping.''  It would seem that increasing light emission out of the solar cell would reduce the amount of possible light trapping, thereby decreasing the solar cell current.  However, increased absorption and light trapping are actually complementary processes.  This is explained by the asymmetry imposed by the semiconductor's high refractive index relative to air.  
	
The intuitive notion of light trapping as a means of increasing the optical path length will be represented here as an increase in the absorptivity per unit optical thickness.  Mathematically, improved light trapping represents an increase in the quantity $a_{ext}/\alpha L$, where $a_{ext}$ is the probability of absorbing an externally incident photon.  For clarity and simplicity, a step-function absorber is assumed such that the absorption coefficient is $\alpha$ for all energies above the bandgap.	
	
There are cases in which enhancing emission does not affect light trapping.  It is shown that the voltage nevertheless increases, demonstrating why photon extraction, rather than light trapping, is the fundamental determinant of voltage.  We derive here a rather general formula linking external yield and the light-trapping quantity $a_{ext}/\alpha L$.
	
Before deriving the general formula, it is worth noting the link between light trapping and external extraction for the previously considered plane-parallel and randomly textured geometries.  The plane-parallel case effectively has no light trapping; incident light has one pass through the cell (or two with a rear mirror), but then immediately exits.  Similarly, for an internally emitted photon, there is only a small escape cone through which emission can occur.  A photon emitted outside the escape cone cannot be emitted without a re-absorption process.  Essentially, there a number of modes within the semiconductor that do not couple to external plane waves, restricting both light-trapping for absorption and light extraction for emission.  Conversely, a random surface texture provides coupling between all of the semiconductor internal modes and the external plane wave modes.\footnote{In the ray optical limit, the internal modes are also plane wave modes.}  Both light trapping and light extraction are enhanced through this coupling.  These two examples demonstrate the intuition behind the claim that light trapping and light extraction are complementary processes.
	
Having demonstrated that light trapping and emission enhancement are complementary in two extreme cases, the plane-parallel and randomly textured solar cells, we can go further and prove that in general the two quantities are linked.  To do so, we must extend the detailed balance model of Sec.~\ref{sec:DetBalance}.  In Sec.~\ref{sec:DetBalance}, the detailed balance derivation matched absorption and emission through the front surface at thermal equilibrium.  The open-circuit voltage is then known as a function of the absorptivity.  For clarity, we re-label the absorptivity $a(E)$ as $a_{ext}(E)$, emphasizing that it is the probability of an \textit{externally incident} photon being absorbed within the solar cell.  Re-writing the equation at the open-circuit condition with the new notation:
\begin{equation}
\label{eq:VocExtDetBalance}
\int a_{ext}(E)S(E)\mathrm{d}E = \Omega_{ext} e^{qV/kT} \int a_{ext}(E)b(E)\mathrm{d}E + R_{nr}L
\end{equation}
where $\Omega_{ext}$ is the emissive solid angle and $R_{nr}$ is the non-radiative recombination rate per unit volume.

	One can also perform detailed balance within the solar cell, normalizing per unit volume instead of per unit area.  As in Sec.~\ref{sec:DetBalance}, at thermal equilibrium absorption exactly equals emission.  The absorption rate per unit volume within the solar cell is $4\pi n_r^2 \alpha(E) b(E)$ \cite{VanRoosbroeck1954}, where $\alpha$ is the material absorption coefficient, the absorption occurs over $4\pi$ solid angle, and the factor of $n_r^2$ accounts for the increased number of optical modes within the semiconductor.  The non-equilibrium rate again is the equilibrium rate scaled by the Boltzmann $e^{qV/kT}$ term.  As with the external surface derivation, the electon-hole generation and recombination rates are equal at steady-state, and at open-circuit this leads to the equation:
\begin{equation}
\frac{1}{L}\int a_{ext}(E)S(E)\mathrm{d}E = 4\pi n_r^2 e^{qV/kT}\int \alpha(E) b(E) \mathrm{d}E - 4\pi n_r^2 e^{qV/kT}\int a_{\mathrm{int}}(E) \alpha(E) b(E) \mathrm{d}E + R_{nr}
\end{equation}	
The first term on the left-hand side is the volumetric absorption rate.  The first term on the right-hand side is the internal emission rate.  The second term on the right, not necessary in the external surface derivation, is the internal re-absorption rate, equal to the internal emission rate multiplied by the probability of absorbing an internally emitted photon, $a_{int}(E)$.  Re-arranging and simplifying, one finds
\begin{equation}
\label{eq:VocIntDetBalance}
\int a_{ext}(E)S(E)\mathrm{d}E = \pi e^{qV/kT} \int  \left[ 4n_r^2 \alpha(E) L \left( 1-a_{\mathrm{int}}(E) \right) \right] b(E) \mathrm{d}E + R_{nr}L
\end{equation} 
The two detailed balance approaches, through balance at either the external surface or the internal volume, represent the same physical situation.  They must, therefore, be equivalent.  In the case of step-function absorbers, we can set the integrands of Eqn.~\ref{eq:VocIntDetBalance} and Eqn.~\ref{eq:VocExtDetBalance} equal to each other.  This results in the relation:
\begin{equation}
\label{eq:aExtInt}
a_{ext} \Omega_{ext} = 4\pi n_r^2 \alpha L \left(1 - a_{int} \right)
\end{equation}
Note that Eqn.~\ref{eq:aExtInt} already indicates that reducing the internal absorption rate (possibly by increasing emission through the front), will increase the external absorptivity.  The quantity $1-a_{int}$ is the probability of a photon escaping before re-absorption.  However, the photon could escape through either the front of the solar cell, which has solid angle $\Omega_f$, or the rear of the solar cell, with solid angle $\Omega_r$.  Nevertheless, re-arranging Eqn.~\ref{eq:aExtInt}, we can represent the total escape probability as
\begin{equation}
\label{eq:eIntAExt}
e_{int} = \frac{a_{ext}}{4n_r^2\alpha L} \frac{\Omega_{ext}}{\pi}
\end{equation}
Generalizing Eqn.~\ref{eq:etaExtInfSum} for general geometries, the external luminescence yield is:
\begin{align}
\eta_{ext} & = \eta_{int} e_f + \eta_{int} a_{int} \eta_{int} e_f + \ldots \nonumber \\
& = \frac{ \eta_{int} e_f }{1-\eta_{int}a_{int}}
\label{eq:etaExtInfSumGen}
\end{align}
where $e_f$ is the probability of immediate escape through the front.  For evenly distributed carriers and photons throughout the cell, the escape rate through the front will be related to the total escape right by the ratio of solid angles:
\begin{equation}
\label{eq:escapeFront}
e_f = \frac{\Omega_f}{\Omega_f + \Omega_r} e_{int} = \frac{a_{ext}}{4n_r^2\alpha L} \frac{\Omega_f}{\pi}
\end{equation} 
By substituting the equations for $e_f$ (Eqn.~\ref{eq:escapeFront}) and $a_{int}$ (Eqn.~\ref{eq:aExtInt}) into Eqn.~\ref{eq:etaExtInfSumGen}, we can relate the external absorptivity to the external luminescence yield for a general geometry:
\begin{equation}
\label{eq:etaExtAExt}
\eta_{ext} = \frac{a_{ext}}{4n_r^2\alpha L \frac{\pi}{\Omega_f}\frac{1-\eta_{int}}{\eta_{int}} + a_{ext}\frac{\Omega_{ext}}{\Omega_f}}
\end{equation}
Eqn.~\ref{eq:etaExtAExt} is a \emph{monotonically increasing} function of the light-trapping factor $a_{ext}/\alpha L$.\footnote{Technically it is a monotonically non-decreasing function of the light-trapping factor, with the limiting case $\eta_{int}=1$ providing an example for which the external yield does not increase as a function of $a_{ext}/\alpha L$.}  Assuming the emissive solid angles remain constant, any increase in light trapping will also increase the external yield, mathematically confirming the intuition developed previously.  Note that for the plane-parallel and randomly textured geometries with rear mirrors, Eqn.~\ref{eq:etaExtAExt} reduces to Eqns.~\ref{eq:etaExtPP} and \ref{eq:etaExtText}, respectively.  Note that there are methods for improving the external yield without affecting the light-trapping.  For example, consider an optically thick semiconductor without a rear mirror.  If a rear mirror is added, all of the photons previously emitted through the rear surface will now be emitted through the front surface, yielding a large improvement in the external yield.  However, the absorption has not changed, as the second pass through the semiconductor was not needed.  This is one of a variety of examples where external yield is increased without increasing the light trapping factor.

	Therefore light emission and light trapping are not at odds, but are complementary physical phenomena.  For very general geometric structures, the two processes are mathematically related.  Improving the light trapping will automatically increase the external luminescence yield.  However, there are methods for improving the external yield that do not affect the light trapping, and yet increase the open-circuit voltage.  External yield is therefore the relevant parameter for designing and maximizing the voltage.

\section{Voltage Calculation in the Sub-Wavelength Regime}
\label{sec:NearFieldVolt}
In Sec.~\ref{sec:DetBalance}, the conventional Shockley-Queisser detailed balance approach to calculating voltage was presented.  The Shockley-Queisser method, however, is difficult to implement for solar cells in the sub-wavelength regime.  Near-field effects, such as plasmon coupling or radiative quenching, can have significant effects on the emission (and therefore voltage), while being difficult to capture the SQ formalism.  The simplicity of the SQ method derives from the fact that by normalizing to surface area, the internal dynamics of the solar cell are inherently included, without explicit calculation required.  However, consider even the simple case of a planar solar cell with a lossy back mirror.  In the Shockley-Queisser approach, one would need to calculate the absorptivity of blackbody photons radiating from the metal into the cell, from which the emissivity would be derived.  However, how would one normalize the blackbody radiation from the metal?  If the structure were sufficiently complicated as to require a simulation, how could one inject an angled plane wave from the metal and properly calculate the absorptivity?  The failings of the SQ method in such cases requires an alternate method for calculating the voltage.

\begin{figure}
\centering
\includegraphics[width=3in]{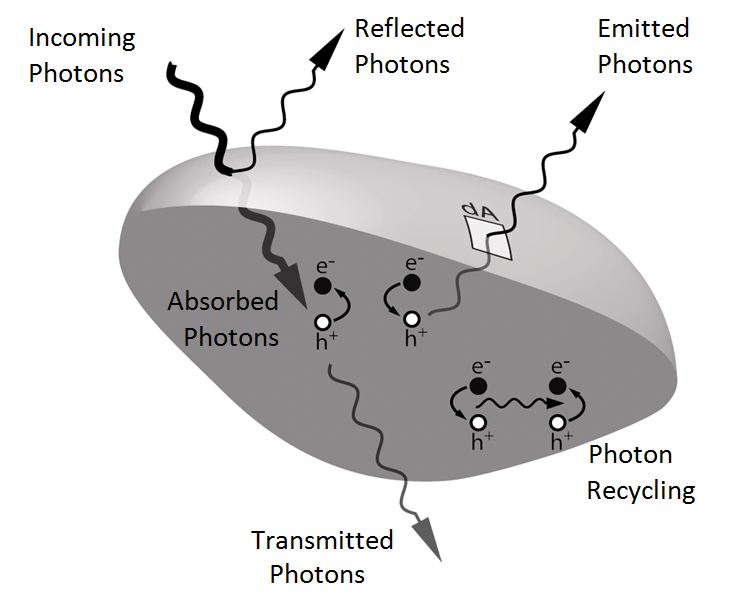}
\caption{Optical processes in a semiconductor. The absorbed fraction of incoming photons excites electron-hole pairs. The emission from recombination of these excited states can be considered as the electromagnetic power leaving the cell surface. Not all radiative recombination contributes to the emission, some is recycled. The population of the excited electrons establishes the (chemical) potential throughout the material volume.} 
\label{fig:SemiProcesses}
\end{figure}

Instead of indirectly calculating the emission through the absorption of external plane waves, one can alternately calculate the emission from local, internal absorption.  There is still a detailed balance relation, but this time connecting the internal blackbody radiation to an incoherent sea of fluctuating dipoles.  If one can calculate the amplitude of the dipoles and their respective emission rates, the cell emissivity as a whole will be known.  This section will present such a method, as well as a computational speed-up for faster calculation.  Fig.~\ref{fig:SemiProcesses} depicts the relevant processes inherent to the method.

The rigorous thermodynamic approach makes use of the fluctuation-dissipation theorem \cite{Kubo1966,Landau1980}.  The fluctuation-dissipation theorem has been used to calculate e.g. radiative heat transfer \cite{Polder1971,Pendry1999,Luo2004,Rodriguez2011} and Casimir forces \cite{Agarwal1975,Genet2000}.  For such an approach to calculate solar cell emission, see \cite{Niv2012}.  Here we will instead present a heuristic derivation that does not get every pre-factor correct, but does present an intuitive picture of the underlying science.

In addition to a detailed balance at the surfaces of the solar cell, there is also an internal detailed balancing at thermal equilibrium.  In thermal equilibrium, the photon absorption rate per unit second must exactly equal the photon emission rate per unit second, at every frequency.  At steady-state there will be a blackbody radiation field of $n_r^2 b(\omega)$ throughout the solar cell, where $b(\omega)$ is the blackbody photon flux per unit solid angle per second per unit frequency, scaled by the semiconductor refractive index $n_r$.  The absorption rate (equivalently, carrier generation rate) is therefore
\begin{equation}
R_{abs}(\omega) = 4\pi n_r^2 \alpha(\omega) b(\omega) 
\end{equation}
At thermal equilibrium there is a detailed balance relation equating the emission and the absorption, such that the emission rate is:
\begin{equation}
\label{eq:IntEmission}
R_{em}(\omega) = 4\pi n_r^2 \alpha(\omega) b(\omega)
\end{equation}
in units of photons per unit volume, per second, and per unit frequency.  The emission arises from the dipole matrix element of the Hamiltonian \cite{Chuang1995}, indicating that in the bulk the emission can be thought of as resulting from inconherent, fluctuating dipoles.  From knowledge of the radiative flux of a single dipole, then, one can derive the effective polarization density at thermal equilibrium.

The power radiated by a single dipole $\mathbf{p}$ is given by \cite{Jackson1999}
\begin{equation}
R_{dipole} = \frac{n_r \omega^4}{12\pi \epsilon_0 c^3}|\mathbf{p}|^2
\end{equation}
In a bulk material at thermal equilibrium, a polarization density $\mathbf{P}$ will arise.  This can be modeled as $N$ dipoles per unit volume per unit frequency, over some volume $V$.  To solve for the polarization density, one can first set the total emission rate from the $N$ dipoles equal to the emission rate of Eqn.~\ref{eq:IntEmission}:
\begin{equation}
\frac{N R_{dipole}}{\hbar \omega} = 4\pi n_r^2\alpha(\omega)b(\omega)
\end{equation}
Canceling equivalent terms and recognizing the absorption coefficient $\alpha(\omega) \approx (\omega/c)\epsilon_I / n_r$, results in a formula for $N$:
\begin{equation}
N=\frac{12}{\pi}\hbar \epsilon_0 \epsilon_I(\mathbf{r},\omega)\frac{1}{e^{\hbar\omega/kT}-1}\frac{1}{|\mathbf{p}|^2}
\end{equation}
where $\epsilon_I$ is the imaginary part of the permittivity.  The total polarization density can be approximated as $|\mathbf{P}(\mathbf{r})|^2 = N|\mathbf{p}|^2$, resulting finally in a formula for the polarization density of
\begin{equation}
\label{eq:P2Approx}
|\mathbf{P}(\mathbf{r})|^2 \approx \frac{12}{\pi} \hbar \left[ \frac{1}{e^{\hbar \omega / kT}-1} \right] \epsilon_0 \epsilon_I(\mathbf{r},\omega) 
\end{equation}
The numerical pre-factors are not exactly correct, but the form of the equation is.  Through a proper thermodynamic analysis, taking into account the statistical nature of the absorption and emission processes, one would find a correlation function for the polarization density of
\begin{equation}
\label{eq:P2Equil}
\left\langle P_k(\mathbf{r},\omega) \overline{P_l(\mathbf{r'},\omega')}\right\rangle_S = \hbar \left[ \frac{1}{2} + \frac{1}{e^{\hbar\omega/kT}-1} \right]\epsilon_0 \epsilon_I(\mathbf{r},\omega) \delta_{kl} \delta(\mathbf{r}-\mathbf{r'})
\end{equation}
where the material has been assumed local and isotropic, the overline denotes complex conjugation, and $<>_S$ denotes the symmetrized correlation function \cite{Callen1951,Agarwal1972,Joulain2005}.

Eqn.~\ref{eq:P2Equil} illustrates the primary concept of this section.  At thermal equilibrium, the blackbody radiator has a modal occupation described by Bose-Einstein statistics.  Moreover, the imaginary part of the permittivity, $\epsilon_I$, dictates absorption, and therefore also the emission rate.  Given just the material permittivity and the temperature, Eqn.~\ref{eq:P2Equil} gives the local polarization density at thermal equilibrium, throughout the absorbing layer.
 
Given the polarization density at equilibrium, one can find the total emission rate.  The emissive flux through all of the surfaces of the solar cell determines the outgoing emission rate:
\begin{equation}
\label{eq:RoutSimp}
R_{out} = \int_A \mathrm{d}\mathbf{r} \int_0^{\infty} \frac{\mathrm{d}\omega}{\hbar\omega}\left\langle \mathbf{S}(\mathbf{r},\omega) \right \rangle \cdot \hat{\mathbf{n}}
\end{equation}
where $\mathbf{S}(\mathbf{r},\omega)$ is the Poynting vector of the fields emanating from the polarization density, and $\hat{n}$ is the outward surface normal.  Ultimately, the field at a surface point $\mathbf{r}$ from the polarization density within the volume $V$ at $\mathbf{r'}$ is given by the electric and magnetic Green dyads $G_{ij}^{EP}(\mathbf{r},\mathbf{r'})$ and $G_{ij}^{HP}(\mathbf{r},\mathbf{r'})$, respectively, using the notation of Sec.~\ref{sec:ShapeCalc}.  Re-writing Eqn.~\ref{eq:RoutSimp} in Einstein notation and introducing the Green's functions leads to
\begin{equation}
\label{eq:RoutGF}
R_{out} = \int_A \mathrm{d}\mathbf{r} \int_0^{\infty} \frac{\mathrm{d}\omega}{2\hbar\omega} \int_V \mathrm{d}\mathbf{r'} \int_V \mathrm{d}\mathbf{r''} \operatorname{Re}\left[\varepsilon_{ijk} \left\langle G_{jl}^{EP}(\mathbf{r},\mathbf{r'}) P_l(\mathbf{r'}) \overline{ G_{km}^{HP}(\mathbf{r},\mathbf{r''}) P_m(\mathbf{r''}) } \right\rangle n_i \right] 
\end{equation}
where $\varepsilon_{ijk}$ is the Levi-Civita symbol.  The polarization sources are known from Eqn.~\ref{eq:P2Equil}.  Inserting them into Eqn.~\ref{eq:RoutGF} yields
\begin{equation}
\label{eq:Rout}
R_{out} = \int_A \mathrm{d}\mathbf{r} \int_0^{\infty} \frac{\mathrm{d}\omega}{2\omega} \frac{1}{e^{\hbar\omega/kT}-1} \int_V \mathrm{d}\mathbf{r'} \epsilon_0 \epsilon_I(\mathbf{r'},\omega) \operatorname{Re}\left[n_i \varepsilon_{ijk} G_{jl}^{EP}(\mathbf{r},\mathbf{r'}) \overline{ G_{kl}^{HP}(\mathbf{r},\mathbf{r'}) } \right] 
\end{equation}
where the zero-point field does not contribute and has not been included.  Eqn.~\ref{eq:Rout} is the outgoing emissive flux at thermal equilibrium for any geometry, where the geometric dependence comes through the Green's functions.  Once the Green's functions are known, the voltage can be determined, as in Sec.~\ref{sec:DetBalance}, through the ratio of the solar generation rate to the thermal equilibrium emission rate:
\begin{equation}
qV_{OC} = kT \ln \left( \frac{R_{abs}}{R_{out}} \right)
\end{equation}
The absorption rate $R_{abs}$ can be found as usual through the absorptivity of the cell with respect to incident plane waves.  Through Eqn.~\ref{eq:Rout}, the emission is now known even when there are near field effects, enabling calculation of the voltage and therefore efficiency of any arbitrary geometry.

\subsection{GaAs Solar Cell Calculation}
\begin{figure}
\centering
\includegraphics[width=4in]{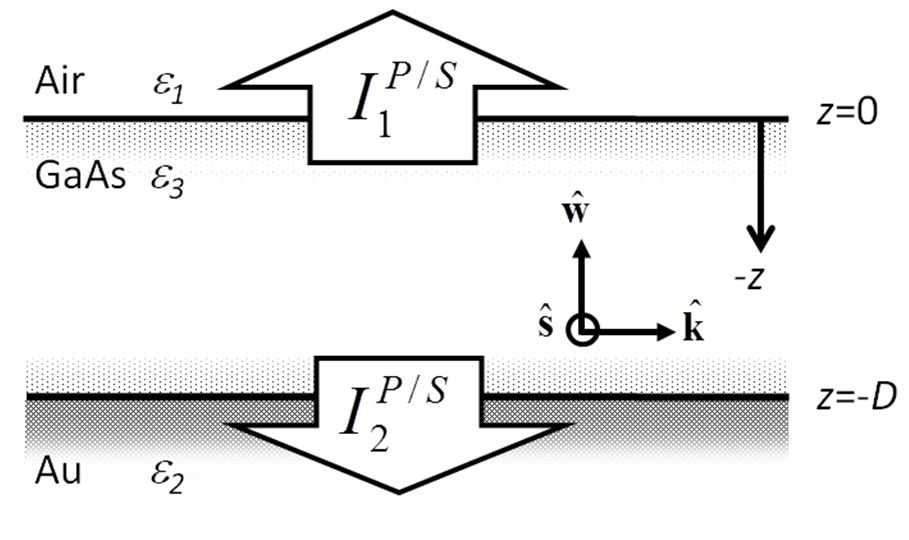}
\caption{Schematic layout of a slab GaAs solar cell with a metallic gold (Au) back reflector. Block arrows represent the four possible emission channels: GaAs to air ($I_1^{P/S}$), and GaAs to Au substrate ($I_2^{P/S}$), for $S$ (TE) and $P$ (TM) polarizations.} 
\label{fig:GaAsArrows}
\end{figure}

We now implement the developed framework in an analytically tractable example.  Given a geometry for which the Green's functions are known, Eqn.~\ref{eq:Rout} gives the emission.  A simple yet relevant geometry for which the Green's functions are known is shown in Fig.~\ref{fig:GaAsArrows}, a multi-layer plane-parallel system.  The system consists of a GaAs slab on gold.  There are two surfaces through which the emission can occur, the front and rear, through which the emission occurs into channels $I_1$ and $I_2$, respectively.  In this system the fields can further be categorized as $S$ (TE) or $P$ (TM), determined by the transverse field component.

The Green's functions for multi-layer systems are well-known \cite{Sipe1987}.  A plane-wave decomposition coupled with matrix analysis permits exact calculation of the fields for arbitrarily many layers.  For imaginary wave vectors, evanescent waves and even surface plasmon modes can emerge.  Switching to CGS-EMU units for consistency with \cite{Niv2012}, the emission rate with the Green's functions is
\begin{equation}
R_{out} = \frac{1}{4\pi^3}\int_0^{\infty} \mathrm{d}\omega \exp \left(\frac{-\hbar\omega}{kT}\right) \int_0^{\infty}k\mathrm{d}k\left(I_1^S + I_1^P + I_2^S + I_2^P\right)
\end{equation}
where $k$ is the magnitude of the wavevector tangent to the interface, and the notation is consistent with Fig.~\ref{fig:GaAsArrows}.  The individual emission channels $I_{1/2}^{P/S}$ contain the Green's function, and work out to
\begin{equation}
I_{1/2}^{S/P} = |\tau_{1/2}^{S/P}|^2\frac{w_{1/2}^{'}}{|w_3|^2} \left[ w_3^{'} \left(1-e^{-2w_3^{''}D}\right) \left( 1+|r_{3,2/1}^{S/P}|^2 e^{-2w_3^{''}D}\right) + 2w_3^{''}g^{S/P}e^{-2w_3^{''}D} \mathrm{Im} \left\lbrace r_{3,2/1}^{S/P} \left( e^{2iw_3^{'}D}-1 \right) \right\rbrace \right]
\end{equation}
where $D$ is the GaAs thickness, $w_i=\sqrt{\epsilon_i (\omega/c)^2-k^2}$ is the magnitude of the wavevector projected onto the z-axis, and the Fresnel reflection and transmission coefficients are $r_{1/2}^{S/P}$ and $t_{1/2}^{S/P}$, respectively.  $g^{S/P}$ is a polarization-dependent term
\begin{displaymath}
g^{S/P} = \left\{
	\begin{array}{cr}
		1 & S\textrm{-}polarized \\
		\frac{k^2-|w_3|^2}{k^2+|w_3|^2} & P\textrm{-}polarized
	\end{array}
	\right.
\end{displaymath}
The Fabry-Perot multiple reflections make up the transmission coefficient $\tau$ through the equation
\begin{equation}
\tau_{1/2}^{S/P} = \frac{t_{3,1/2}^{S/P}}{1-r_{3,1}^{S/P}r_{3/2}^{S/P}\exp(-w_3^{''}D)}
\end{equation}

\begin{figure}
\centering
\includegraphics[width=6.5in]{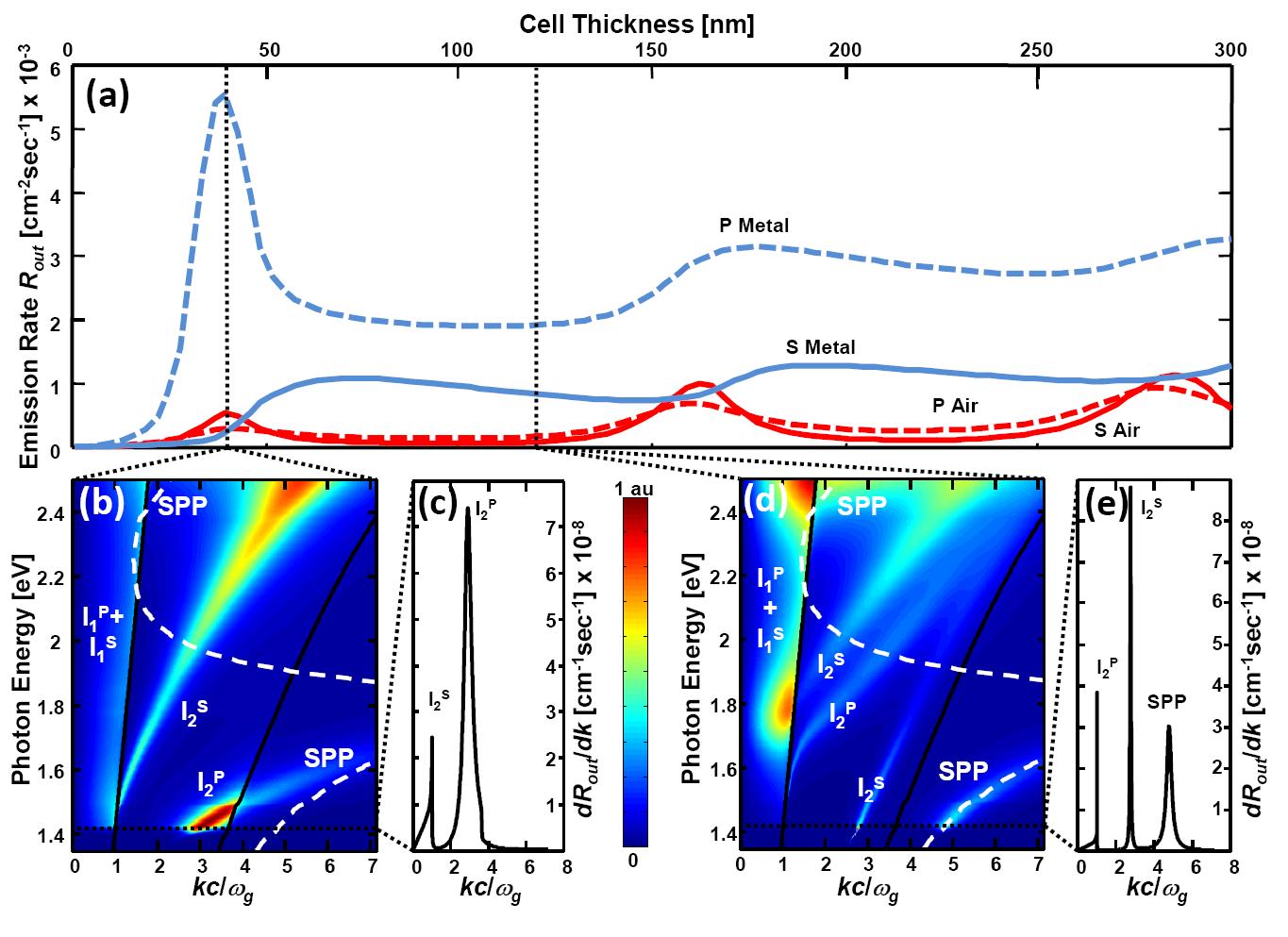}
\caption{(a) Emission rate ($R_{out}$) of the $S$ and $P$ polarizations through GaAs/air and GaAs/Au interfaces as a function of cell thickness; oscillations, the $P$-polarization peak, and other features stem from the unique electromagnetic treatment presented in this work. Panels (b) and (d) show intensity maps of emission channels at thicknesses of $40nm$ and $120nm$, respectively. Individual contributions are labeled. Black lines represent the analytic dispersion of plane waves in air (to the left) and in GaAs (to the right), while white dashed lines depict the analytical SPP dispersion at the GaAs/Au interface. Panels (c) and (e) examine the emission for photons with energies $1.42eV$, near the bandgap of GaAs, for the two thicknesses.} 
\label{fig:EmissionRate}
\end{figure}
Fig.~\ref{fig:EmissionRate}(a) shows the $S$ and $P$-polarization emission rates out of the top and bottom interfaces, i.e. individual contributions of the $I_1^S$, $I_1^P$, $I_2^S$, and $I_2^P$ channels, as a function of GaAs slab thickness. The sum of these four comprises the total emission out of the solar cell. As opposed to a ray optical model, all the electromagnetic aspects of the emission process are captured by the current formalism. These include the cavity-like resonance between the (partially) reflective interfaces responsible for the oscillations in the emission rates, as well as more subtle near field optical effects. One clear signature of such effects is the anomalous peak in the P-polarization emission to the metal at $40nm$ thickness.

We chose two GaAs thicknesses for detailed study: $120nm$ and $40nm$.  Fig.~\ref{fig:EmissionRate}(b) depicts the emission rate as a function of both wavenumber (normalized to $\omega_g/c$) and photon energy, for each of the two thicknesses.  Each map is divided into three distinctive regions.  The leftmost region depicts waves that can propagate in air, the middle section depicts propagating waves in GaAs with evanescent tail in air, and the right section includes wave that are evanescent both in air and GaAs.  As expected, for both thicknesses the emission into air ($I_1^S$ and $I_1^P$) is confined to the leftmost region.

For the $120nm$ thickness, the map divides neatly by modes.  The guided modes (in the middle region) result in emission through the GaAs/Au interface, due to the the loss tangent of the Au.  The map clearly shows coupling into the single-sided surface plasmon polariton (SPP) mode, matching up exactly with the white line representing the SPP dispersion.  There is not particularly strong coupling into any of the modes, and therefore the total emission is relatively small.

At $40nm$ the picture is modified.  There is no longer a single-sided SPP mode; the close proximity of the GaAs/air interface results in the hybrid $I_2^P$-SPP mode that is strongly coupled into from the fluctuating dipoles.  The spectral width over which there is strong coupling results in the enhancement peak visible in Fig.~\ref{fig:EmissionRate}(a).

The signatures of the near field optical effects are clearly observed throughout the emission spectra.  Our formalism inherently captures the local density of states effects, and enables a clear picture to emerge of the underlying physics.

\begin{figure}
\centering
\includegraphics[width=3.5in]{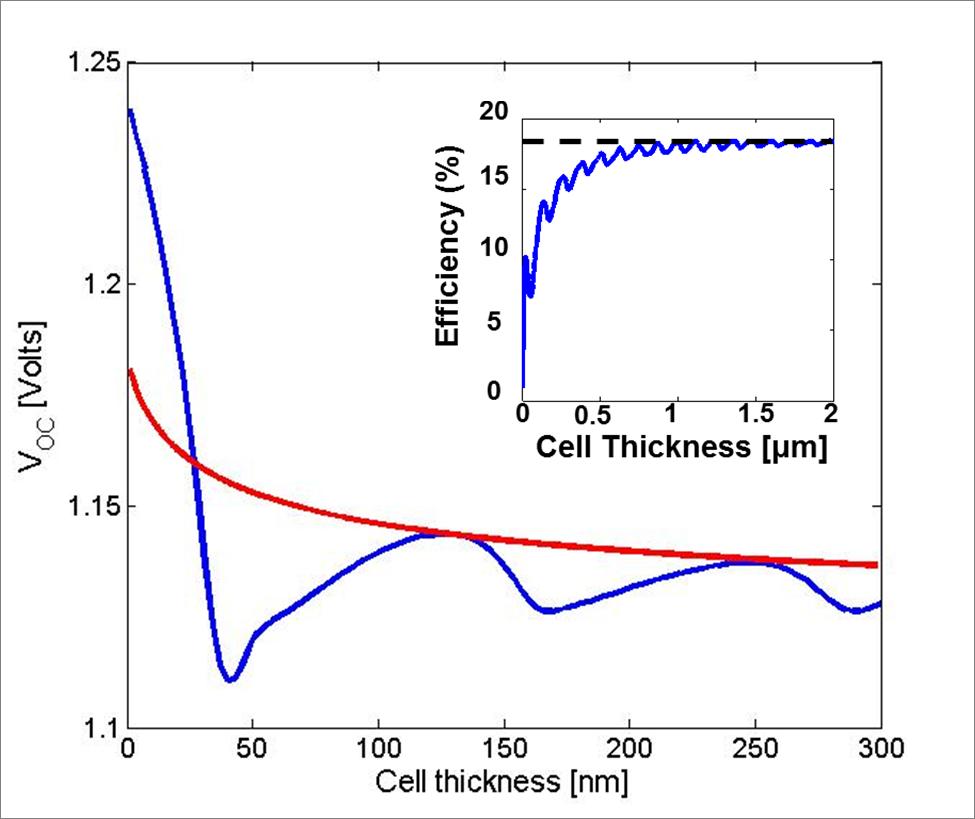}
\caption{Open-circuit voltage vs. cell thickness of FDT formalism compared to a ray based model in blue and red lines, respectively. Oscillations result from both absorption and emission, and tend to relax for thick cells due to GaAs absorption. Inset shows efficiency vs. cell thickness. Our result converges to the expectations of the ray-based models, shown in black dashed line, for very thick cells. Zero efficiency at extremely small thickness despite the rise in voltage is due to the vanishing absorption.}
\label{fig:OpenCircuitVoltage} 
\end{figure}
Fig.~\ref{fig:OpenCircuitVoltage} shows the open-circuit voltage as a function of cell thickness for the GaAs device. For comparison, the red line shows the $V_{OC}$ of the ray-based formalisms \cite{Tiedje1984}. The ray-based model fails to predict the $V_{OC}$ for small GaAs slab thicknesses as it does not account for the electromagnetic nature of the emission process, including near field effects. One such near field effect is the $P$-polarization emission peak of at $40nm$ thickness discussed earlier, which is responsible for the dip in $V_{OC}$ observed at this thickness. The voltage is affected by the electromagnetic phenomena that govern both emission and absorption processes at each thickness. Therefore, the oscillations in $V_{OC}$ are somewhat displaced with respect to those of Fig.~\ref{fig:EmissionRate}(a). They fade away in thicker cells due to GaAs absorption where our prediction and the ray based model are in good agreement. Interestingly, below $30nm$ our approach predicts higher $V_{OC}$ than the ray based model. This observation shows that optic near field effects may suppress the emission out of the cell, a significant fact to consider for the design of future ultra thin devices.

The inset of Fig.~\ref{fig:OpenCircuitVoltage} shows the efficiency of the GaAs slab cell calculated from the I-V relation. The asymptotic efficiency of $18.4\%$ (and $V_{OC}$ of $1.12V$) for a very thick GaAs slab is in agreement with ray optics based models \cite{Tiedje1984}. Therefore, while capturing the electromagnetic nature of the photo voltage and current production in thin solar cells, our analysis converges to the known and expected results in the asymptotic limits of thick devices. We note that in spite of the voltage rise, the efficiency for a very thin slab eventually drops due to diminishing absorption in this device, and thus vanishing photocurrent.

In conclusion, the fluctuation-dissipation theorem connects the thermodynamic and electromagnetic aspects of power generation in solar cells. This yields a rigorous electromagnetic framework for evaluating cell performance under conditions unattainable by previous approaches, especially the optical near-field regime. This analysis accounts for all optical aspects of power generation in solar cells, including modified density of states and dispersive materials. Other non-radiative losses can be incorporated in the usual manner \cite{Ross1967,Tiedje1984}. The analysis is not principally limited to semiconductors and can be applied to any system that can be described with the macroscopic Maxwell equations. The example of an ultra-thin GaAs solar cell demonstrates the power of the method in capturing nano-scale physics. 

\subsection{Computational Implementation}
The formalism presented here lends itself easily to a computational implementation for complex geometries.  The polarization density calculated in Eqn.~\ref{eq:P2Equil} and used in Eqn.~\ref{eq:Rout} can be simulated by properly normalized dipoles radiating incoherently from every point within the absorbing material.  The primary drawback of the approach is the computational cost; for a grid with $N_i$ cells along each dimension $i$, the number of simulations scales as $6N_xN_yN_z$.  However, we will present in this section a means for speeding up the calculation such that the cost is only $4N_xN_y$, effectively reducing the dimensionality by one (the dimension of the relevant interfaces is assumed proportional to $N_xN_y$.

The straightforward approach to computing the emission rate, and therefore the open-circuit voltage, for an arbitrary geometry can be gleaned from re-arranging Eqn.~\ref{eq:Rout}
\begin{equation}
\label{eq:RoutRev}
R_{out} =  \int_V \mathrm{d}\mathbf{r'}  \int_A \mathrm{d}\mathbf{r} \int_0^{\infty} \mathrm{d}\omega f(\mathbf{r'},\omega) \operatorname{Re}\left[n_i \varepsilon_{ijk} G_{jl}^{EP}(\mathbf{r},\mathbf{r'}) \overline{ G_{kl}^{HP}(\mathbf{r},\mathbf{r'}) } \right] 
\end{equation}
where the the many pre-factors have been collected into the function $f(\mathbf{r'},\omega)$, and the order of integration has been changed for clarity.  The fact that the Green's functions needed are for dipoles radiating \emph{from} $\mathbf{r'}$ \emph{to} $\mathbf{r}$ dictates why the order of integration is shown as above.  The incoherent dipoles have to be placed at all possible $\mathbf{r'}$ locations, comprising the entire volume $N_xN_yN_z$ of grid cells.  Each dipole simulation results in the Green's function over the entire surface, collecting the data for all possible $\mathbf{r}$ into a single simulation.  Algorithmically, this procedure is depicted in Alg.~\ref{alg:SlowVoc}.
\begin{algorithm}
\caption{Straightforward computational implementation of Eqn.~\ref{eq:RoutRev}.  This implementation requires $6N_xN_yN_z$ simulations per geometry.}
\label{alg:SlowVoc}
\begin{algorithmic}
\State $R_{out} := 0$
\For{$k = 1:N_x N_y N_z$} 
	\For{$j = x,y,z$}
		\State Simulate $p_j = P_j \Delta j = \Delta j$, 				\State Store $G_{ij}^{EP}(\mathbf{r},\mathbf{r'})$ for all $i,\mathbf{r}$
		\State Simulate $m_j = M_j \Delta j = \Delta j$
		\State Store $G_{ij}^{HP}(\mathbf{r},\mathbf{r'})$ for all $i,\mathbf{r}$
		\State $R_{out} \gets R_{out} + \Delta \mathbf{r}_k\mathbf{'} \int_A \mathrm{d}\mathbf{r} \int_0^{\infty} \mathrm{d}\omega f(\mathbf{r}_k\mathbf{'},\omega) \operatorname{Re}\left[n_i \varepsilon_{ijk} G_{jl}^{EP}(\mathbf{r},\mathbf{r}_k\mathbf{'}) \overline{ G_{kl}^{HP}(\mathbf{r},\mathbf{r}_k\mathbf{'}) }  \right]$
	\EndFor
\EndFor
\end{algorithmic}
\end{algorithm} 
Ideally, instead of looping through the many points within the absorber volume and calculating all of the surface data at once, it would be preferable to loop through the surface area and then calculate the volume data at once.  This is in fact possible, using Lorentz reciprocity as derived in Appendix~\ref{chap:SymmMaxwell}.  The key result is Eqn.~\ref{eq:GreenSymm}, which states that the dipole source and location points can be switched, according to: $G_{ij}^{EP}(\mathbf{r},\mathbf{r'}) = G_{ji}^{EP}(\mathbf{r'},\mathbf{r})$ and $G_{ij}^{HP}(\mathbf{r},\mathbf{r'}) = -G_{ji}^{EM}(\mathbf{r'},\mathbf{r})$.  Inserting their symmetric counterparts into Eqn.~\ref{eq:RoutRev} and changing the order of integration again:
\begin{equation}
\label{eq:RoutFast}
R_{out} =  \int_A \mathrm{d}\mathbf{r} \int_V \mathrm{d}\mathbf{r'} \int_0^{\infty} \mathrm{d}\omega f(\mathbf{r'},\omega) \operatorname{Re}\left[-n_i \varepsilon_{ijk} G_{lj}^{EP}(\mathbf{r'},\mathbf{r}) \overline{ G_{lk}^{EM}(\mathbf{r'},\mathbf{r}) } \right] 
\end{equation}
Now the Green's function represent the fields \emph{from} $\mathbf{r'}$ \emph{to} $\mathbf{r}$, meaning the dipole locations are not limited to the surfaces of the absorber.  Moreover, the polarizations of the dipoles are now indexed by $j$ and $k$, which are connected to the local normal $n_i$ through the Levi-Civita symbol.  For a given local normal $n_i$, there are only two possible combinations of $j$ and $k$ (e.g. for $i=1$ only $(j=2,k=3)$ and $(j=3,k=2)$ are possible) allowed.  This reduces the number of dipoles per iteration to four rather than six, leaving the total computational cost to be $4N_xN_y$, rather than $6N_xN_yN_z$.  Note that the cost could actually be a small multiple greater than $4N_xN_y$, as the surface area of the absorber could be a small multiple of the factor $N_xN_y$.  The faster algorithm is given by Alg.~\ref{alg:FastVoc}.
\begin{algorithm}
\caption{Fast computation of $V_{OC}$ in sub-wavelength solar cells.  Computational implementation of Eqn.~\ref{eq:RoutFast}, taking advantage of Lorentz reciprocity.  This implementation requires only $4N_xN_y$ simulations per geometry.}
\label{alg:FastVoc}
\begin{algorithmic}
\State $R_{out} := 0$
\For{$k = 1:N_x N_y$} 
	\For{$j = k,l \perp i$}
		\State Simulate $p_j = P_j \Delta j = \Delta j$, 				\State Store $G_{ij}^{EP}(\mathbf{r'},\mathbf{r})$ for all $\mathbf{r'}$
		\State Simulate $m_j = M_j \Delta j = \Delta j$
		\State Store $G_{ij}^{EM}(\mathbf{r'},\mathbf{r})$ for all $\mathbf{r'}$
	\EndFor
	\State $R_{out} \gets R_{out} + \Delta \mathbf{r}_k \int_V \mathrm{d}\mathbf{r'} \int_0^{\infty} \mathrm{d}\omega f(\mathbf{r'},\omega) \operatorname{Re}\left[-n_i \varepsilon_{ijk} G_{lj}^{EP}(\mathbf{r'},\mathbf{r}) \overline{ G_{lk}^{EM}(\mathbf{r'},\mathbf{r}) } \right]$
\EndFor
\end{algorithmic}
\end{algorithm} 
Through algorithm~\ref{alg:FastVoc}, the emission rate and therefore open-circuit voltage can be computed for any arbitrary geometry.  It includes all near-field effects, providing an indispensable design tool in the search for next-generation solar technologies.

\section{Conclusions}
This chapter developed the key concepts that determine the output voltage of a solar cell.  Of critical importance is the idea that external luminescence efficiency directly determines the voltage, such that the solar cell should be designed for maximum emission at open-circuit.  Fundamentally, this arises from the thermodynamic link between absorption and emission as dictated by detailed balance.  The difficulty of extracting photons, discussed in Sec.\ref{sec:DiffLightExtraction}, has significant consequences for high-efficiency solar cells.  Chap.~\ref{chap:SQLimit} studies the ramifications as solar cells approach their Shockley-Queisser efficiency limits.

At the sub-wavelength scale, calculating the emission either analytically or computationally becomes more complex.  A vast array of near-field effects must be incorporated, properly accounting for the modified density of states and localized modes.  A new framework, based on the fluctuation-dissipation theorem, has been presented, along with a computational algorithm applicable to arbitrary geometries.  This framework will help characterize and design next-generation solar cells.

%% file: ApproachingSQ.tex
\chapter{Approaching the Shockley-Queisser Efficiency Limit}
\label{chap:SQLimit}
\epigraph{Nothing is more practical than a good theory.}{Kurt Lewin}

\noindent
Absorbed sunlight in a solar cell produces electrons and holes.  But, at the open circuit condition, the carriers have no place to go.  They build up in density and, ideally, they emit external luminescence that exactly balances the incoming sunlight.  Any additional non-radiative recombination impairs the carrier density buildup, limiting the open-circuit voltage.  At open-circuit, efficient external luminescence is an indicator of low internal optical losses.  Thus efficient external luminescence is, counter-intuitively, a necessity for approaching the Shockley-Queisser efficiency limit.  A great Solar Cell also needs to be a great Light Emitting Diode. 
	Owing to the narrow escape cone for light, efficient external emission requires repeated attempts, and demands an internal luminescence efficiency ${\gg}90\%$.

\section{Introduction}

The Shockley-Queisser (SQ) efficiency limit \cite{Shockley1961} for a single junction solar cell is ${\sim}33.5\%$ under the standard AM1.5G flat-plate solar spectrum \cite{SolarSpectrum}.  In fact, detailed calculations in this chapter show that GaAs is capable of achieving this efficiency.  Nonetheless, the record GaAs solar cell had achieved only $26.4\%$ efficiency \cite{Green2010b} in 2010.  Previously, the record had been $26.1\%$ \cite{Bauhuis2009} and prior to that stuck \cite{Green2008} at $25.1\%$, during 1990-2007.  Why then the $7\%$ discrepancy between the theoretical limit $33.5\%$ versus the previously achieved efficiency of $26.4\%$?  

It is usual to blame material quality.  But in the case of GaAs double heterostructures, the material is almost ideal \cite{Schnitzer1993b} with an internal luminescence yield of ${>}99\%$.  This deepens the puzzle as to why the full theoretical SQ efficiency is not achieved?

\section{The Physics Required to Approach the Shockley-Queisser Limit}

Solar cell materials are often evaluated on the basis of two properties: how strongly they absorb light, and whether the created charge carriers reach the electrical contacts, successfully.  Indeed, the short-circuit current in the solar cell is determined entirely by those two factors.  However, the power output of the cell is determined by the product of the current and voltage, and it is therefore imperative to understand what material properties (and solar cell geometries) produce high voltages.  We show here that maximizing the external emission of photons from the front surface of the solar cell proves to be the key to reaching the highest possible voltages \cite{Miller2012}.  In the search for optimal solar cell candidates, then, materials that are good radiators, in addition to being good absorbers, are most likely to reach high efficiencies.

\begin{figure}[]
\centering
\includegraphics[width=3.5in]{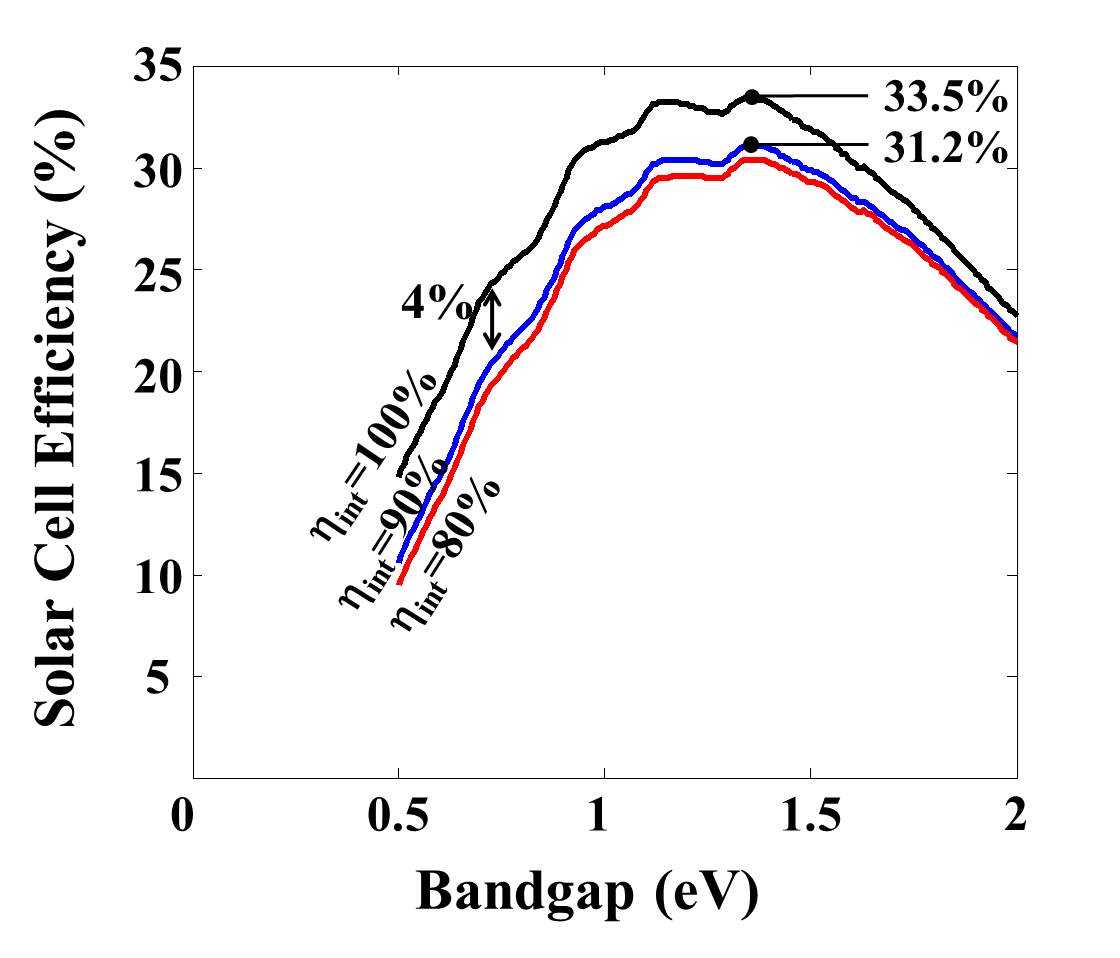}
\caption{The drastic effect of internal luminescence efficiency, $\eta_{int}$, on theoretical solar cell efficiency.  The shortfall is particularly noticeable for smaller bandgaps.  A reduction from $\eta_{int}=100\%$ to $\eta_{int}=90\%$ already causes a large drop in performance, while a reduction from $\eta_{int}=90\%$ to $\eta_{int}=80\%$ causes little additional damage.  Owing to the need for photon recycling, and the multiple attempts required to escape the solar cell, $\eta_{int}$ must be ${\gg}90\%$.} 
\label{fig:EfficiencyBandgap}
\end{figure}

As solar efficiency begins to approach the SQ limit, the internal physics of a solar cell transforms, such that photonic considerations overtake electronic ones.  Shockley and Queisser showed that high solar efficiency is accompanied by a high concentration of carriers, and by strong luminescent emission of photons.  In a good solar cell, the photons that are emitted internally are likely to be trapped, re-absorbed, and re-emitted at open-circuit.  

The SQ limit assumes perfect external luminescence yield at open-circuit.  On the other hand, inefficient external luminescence at open-circuit is an indicator of non-radiative recombination and optical losses.  Owing to the narrow escape cone, efficient external emission requires repeated escape attempts, and demands an internal luminescence efficiency ${\gg}90\%$.  We find that the failure to efficiently extract the recycled internal photons is an indicator of an accumulation of non-radiative losses, which are largely responsible for the failure to achieve the SQ limit in the best solar cells.

In high efficiency solar cells it is important to engineer the photon dynamics.  The SQ limit requires $100\%$ external luminescence to balance the incoming sunlight at open circuit.  Indeed, the external luminescence is a thermodynamic measure \cite{Ross1967,Markvart2008} of the available open-circuit voltage.  Owing to the narrow escape cone for internal photons, they find it hard to escape through the semiconductor surface. Except for the limiting case of a perfect material, external luminescence efficiency is always significantly lower than internal luminescence efficiency.  Then the SQ limit is not achieved.

The extraction and escape of internal photons is now recognized as one of the most pressing problems in light emitting diodes (LED’s) \cite{Zhmakin2011,Crawford2009,Wiesmann2009}.  We assert that luminescence extraction is equally important to solar cells.  \textbf{The Shockley-Queisser limit cannot be achieved unless light extraction physics is designed into high performance solar cells, which requires that non-radiative losses be minimized, just as in LED’s}.

\begin{figure}[]
\centering
\includegraphics[width=6.5in]{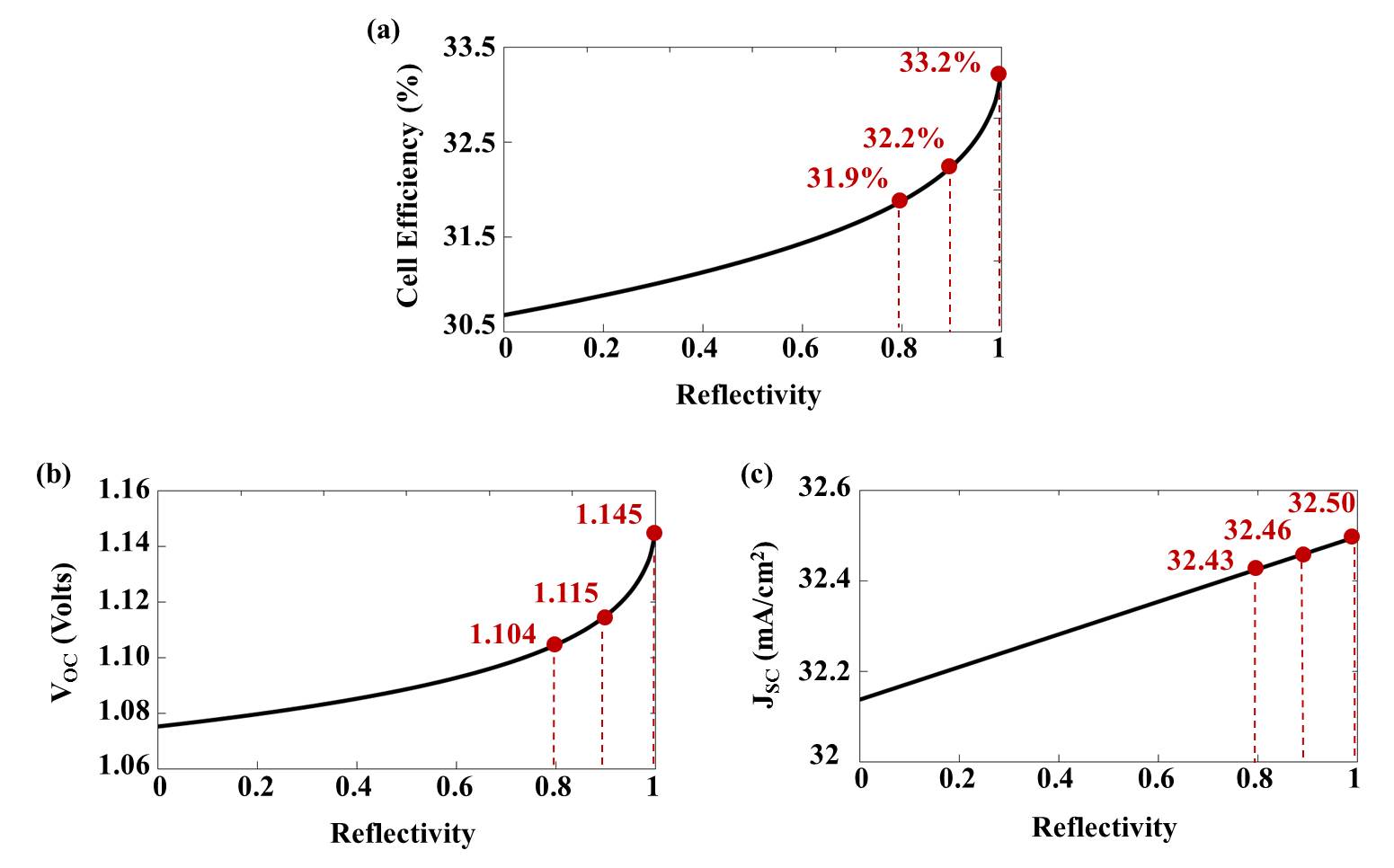}
\caption{The drastic effect of rear mirror reflectivity on cell efficiency and on open-circuit voltage, $V_{OC}$, but not on short-circuit current, $J_{SC}$, for a $3\mu m$ thick GaAs solar cell.  Mirror reflectivity ${\gg}90\%$ makes a big difference, owing to the small escape cone for external emission, and the multiple attempts needed for escape.} 
\label{fig:EfficiencyReflectivity}
\end{figure}

In some way this is counter-intuitive, since an extracted photon cannot contribute to performance.  Paradoxically, $100\%$ external extraction at open-circuit is exactly what is needed to achieve the SQ limit.  The paradox is resolved by recognizing that high extraction efficiency at open-circuit is an indicator, or a gauge, of small optical losses.  Previous record solar cells have typically taken no account of light extraction, resulting in the poor radiative efficiencies calculated in \cite{Green}.  Nonetheless, approaching the $33.5\%$ SQ limit will require light extraction to become part of all future designs.  The present shortfall below the SQ limit can be overcome.
  
A recent paper by Green \cite{Green} reinforces the importance of light extraction.  The record solar cells that have reached the highest efficiencies are also the ones with the highest external luminescence yield.

Although Silicon makes an excellent solar cell \cite{Zhao1998}, Auger recombination fundamentally limits its internal luminescence yield to ${<}20\%$ \cite{Trupke2003}, which prevents Silicon from approaching the SQ limit.  The physical issues presented here pertain to any material that has the possibility of approaching the SQ limit, which requires near unity external luminescence as III-V materials can provide, and that perhaps other material systems can provide as well.

Since light is trapped by total internal reflection, it is likely to be re-absorbed, leading to a further re-emission event.  With each absorption/re-emission event, the solid angle of the escape cone \cite{Yablonovitch1982} allows only $(1/4n^2) {\sim}2\%$ of the internal light to escape.  As a result, $1$ sun incident can produce an internal photon density equivalent to up to $50$ suns.  This puts a very heavy burden on the parasitic losses in the cell.  With only $2\%$ escaping per emission event, even a $90\%$ internal luminescence yield on each cycle would appear inadequate.  Likewise the rear mirror should have ${\gg}90\%$ reflectivity.  This is illustrated in Fig.~\ref{fig:EfficiencyBandgap} and \ref{fig:EfficiencyReflectivity}.

A good solar cell should be designed as a good light emitting diode, with good light extraction.  In a way, this is not surprising.  Most ideal machines work by reciprocity, equally well in reverse.  This has important ramifications.  For ideal materials the burden of high open-circuit voltage, and thereby high efficiency, lies with optical design:  The solar cell must be designed for optimal light extraction under open-circuit conditions.  

The assumption of perfect internal luminescence yield is a seductive one.  The Shockley-Queisser limit gets a significant boost from the perfect photon recycling that occurs in an ideal system.  Unfortunately, for most materials, their relatively low internal luminescence yields mean that the upper bounds on their efficiencies are much lower than the Shockley-Queisser limit.  For the few material systems that are nearly ideal, such as GaAs, there is still a tremendous burden on the optical design of the solar cell.  A very good rear mirror, for example, is of the utmost importance.  In addition, it becomes clear that realistic material radiative efficiencies must be included in a credible assessment of any material's prospects as a solar cell technology.
  
There is a well-known detailed balance equation relating the spontaneous emission rate of a semiconductor to its absorption coefficient \cite{VanRoosbroeck1954}.   Nevertheless, it is not true that all good absorbers are good emitters.  If the non-radiative recombination rate is higher than the radiative rate then the probability of emission will be very low.  Amorphous silicon, for example, has a very large absorption coefficient of about $10^5/cm$, yet the probability of emission at open circuit is approximately $10^{-4}$ \cite{Green}.  The probability of internal emission in high-quality GaAs has been experimentally tested to be $99.7\%$ \cite{Schnitzer1993b}.  GaAs is a unique material in that it both absorbs and radiates well, enabling the high voltages required to reach ${>}30\%$ efficiency.

The idea that increasing light emission improves open-circuit voltage seems paradoxical, as it is tempting to equate light emission with loss.  Basic thermodynamics dictates that materials which absorb sunlight must also emit in proportion to their absorptivity.  Thus electron-hole recombination producing external luminescent emission is a necessity in solar cells.  At open circuit, external photon emission is part of a necessary and unavoidable equilibration process, which does not represent loss at all.  

At open circuit an ideal solar cell would in fact radiate out of the solar cell a photon for every photon that was absorbed.  Any additional non-radiative recombination, or photon loss, would indeed waste photons and electrons.  Thus the external luminescence efficiency is a gauge or an indicator of whether the additional loss mechanisms are present.  In the case of no additional loss mechanisms, we can look forward to $100\%$ external luminescence, and maximum open circuit voltage, $V_{OC}$.  At the power-optimized, solar cell operating bias point \cite{Wurfel2005}, the voltage is slightly reduced, and $98\%$ of the open-circuit photons are drawn out of the cell as real current.  Good external extraction comes at no penalty in current at the operating bias point.

On thermodynamic grounds, it has already been proposed \cite{Ross1967,Rau2007,Kirchartz2008} that the open circuit voltage would be penalized by poor external luminescence efficiency $\eta_{ext}$ as:
\begin{equation}
\label{eq:VoltPenalty}
qV_{OC} = qV_{OC-Ideal} - kT | \ln \eta_{ext} |
\end{equation}
Eqn.~\ref{eq:VoltPenalty} was derived in Chap.~\ref{chap:OnVoltage};  a simpler derivation will be presented here.  Under ideal open-circuit, quasi-equilibrium conditions, the solar pump rate equals the external radiative rate:  $R_{ext} = R_{pump}$.  If the radiative rate is diminished by a poor external luminescence efficiency $\eta_{ext}$, the remaining photons must have been wasted in non-radiative recombination or parasitic optical absorption.  The effective solar pump is then reduced to $R_{pump}\times \eta_{ext}$.  The quasi-equilibrium condition is then $R_{ext}=R_{pump}\times \eta_{ext}$ at open circuit.  Since the radiative rate $R_{ext}$ depends on the carrier density $np$ product, which is proportional to $\exp[qV_{OC}/kT]$, then the poor extraction $\eta_{ext}$ penalizes $V_{OC}$ just as indicated in Eqn.~\ref{eq:VoltPenalty}.

Another way of looking at this is to notice the shorter carrier lifetime in the presence of the additional non-radiative recombination.  We start with a definition
\begin{equation}
\eta_{ext}=\frac{R_{ext}}{R_{ext} + R_{nr}}
\end{equation} 
where $R_{nr}$ is the internal photon and carrier non-radiative loss rate per unit area.  Simple algebraic manipulation shows that the total loss rate $R_{ext} + R_{nr} = R_{ext}/\eta_{ext}$.  Thus a poor $\eta_{ext}<1$ increases the total loss rate in inverse proportion, and the shorter lifetime limits the build-up of carrier density at open circuit.  Then carrier density is connected to $\exp[qV_{OC}/kT]$ as before.

It is important to emphasize that light emission should occur opposite to the direction of the incident photons.  A maximally concentrating solar cell would emit photons only directly back to the sun thus achieving even higher voltages \cite{Araujo1994,Hirst2011}.  However, concentrators miss the substantial fraction of diffuse sunlight, so we focus instead on non-concentrating solar cells.  Such cells absorb both direct and diffuse sunlight, from all incident angles.  The unavoidable balancing emission is that of luminescent photons exiting through the front.  Consequently, light emission \emph{only} from the front surface should be maximized.  Having a good mirror on the rear surface greatly improves the luminescent photon extraction and therefore the voltage.

\section{Theoretical Efficiency Limits of GaAs Solar Cells}

The Shockley-Queisser limit includes a major role for external luminescence from solar cells.  Accordingly, internal luminescence followed by light extraction plays a direct role in determining theoretical efficiency.  To understand these physical effects a specific material system must be analyzed, replacing the hypothetical step function absorber stipulated by SQ.

\begin{figure}[]
\centering
\includegraphics[width=5.5in]{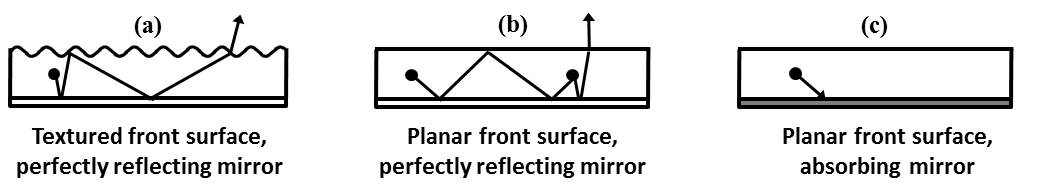}
\caption{Solar performance from three geometries: (a) randomly textured front surface with a perfectly reflecting mirror on the rear surface, (b) planar front surface with a perfectly reflecting mirror on the rear surface, and (c) planar front surface with an absorbing mirror on the rear surface.} 
\label{fig:GeometriesConsidered}
\end{figure}

GaAs is a good material example, where external luminescence extraction plays an important role in determining the fundamental efficiency prospects.  The quasi-equilibrium approach developed by SQ \cite{Shockley1961} is the most rigorous method for calculating such efficiency limits.  Properly adapted, it can account for the precise incoming solar radiation spectrum, the real material absorption spectrum, the internal luminescence efficiency, as well as the external extraction efficiency and light trapping \cite{Tiedje1984}.  Calculations including such effects for Silicon solar cells were completed more than 25 years ago \cite{Tiedje1984}.  Surprisingly, a calculation with the same sophistication has not yet been completed for GaAs solar cells.  

Previous GaAs calculations have approximated the solar spectrum to be a blackbody at $6000 K$, and/or the absorption coefficient to be a step function \cite{Shockley1961,Balenzategui1997,Marti1997}.  The efficiency limits calculated with these assumptions are all less than or equal to $31\%$.

In this chapter the theoretical maximum efficiency of a GaAs solar cell is calculated.  It is shown, using the one-sun AM1.5G \cite{SolarSpectrum} solar spectrum and the proper absorption curve of GaAs, that the theoretical maximum efficiency is in fact $33.5\%$.  Allowing for practical limitations, it should be possible to manufacture flat-plate single-junction GaAs solar cells with efficiencies above $30\%$ in the near future.  As we have already shown, realizing such efficiencies will require optical design such that the solar cell achieves optimal light extraction at open circuit.
  
To explore the physics of light extraction, we consider GaAs solar cells with three possible geometries, as shown in Fig.~\ref{fig:GeometriesConsidered}.  The first geometry, Fig.~\ref{fig:GeometriesConsidered}(a), is the most ideal, with a randomly textured front surface and a perfectly reflecting mirror on the rear surface.  The surface texturing enhances absorption and improves light extraction, while the mirror ensures that the photons exit from the front surface and not the rear.  The second geometry, Fig.~\ref{fig:GeometriesConsidered}(b), uses a planar front surface while retaining the perfectly reflecting mirror.  Finally, the third geometry, Fig.~\ref{fig:GeometriesConsidered}(c), has a planar front surface and an absorbing rear mirror, which captures most of the internally emitted photons before they can exit the front surface.  We will show that this configuration achieves almost the same short-circuit current as the others, but suffers greatly in voltage and, consequently, efficiency.  Thus the optical design affects the voltage more than it does the current.  Note that the geometry of Fig.~\ref{fig:GeometriesConsidered}(c) is equivalent to the common situation in which the active layer is epitaxially grown on top of an electrically passive substrate, which absorbs without re-emission.

\begin{figure}[]
\centering
\includegraphics[width=3in]{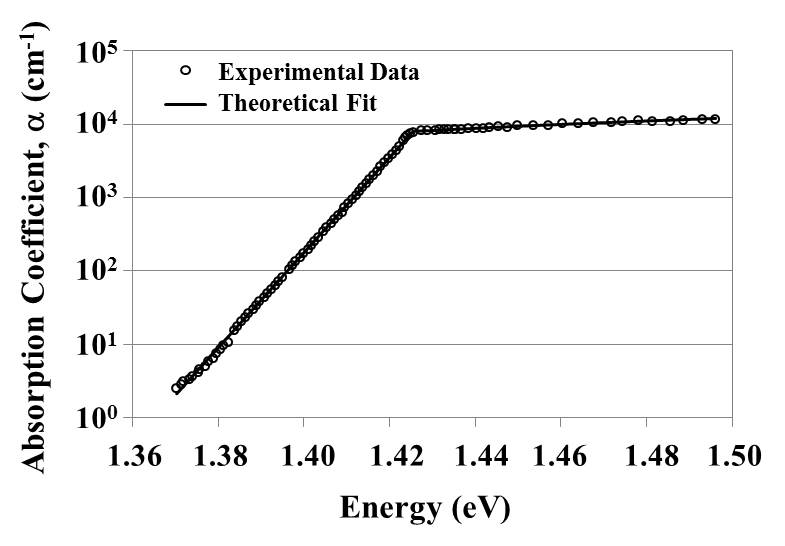}
\caption{GaAs absorption coefficient as a function of energy, with the experimental data from \cite{Sturge1962} and the theoretical fit according to Eqn.~\ref{eq:alphaFit}.  GaAs has a direct bandgap at $Eg=1.42 eV$, with an ``Urbach tail'' that falls off exponentially for lower energies.} 
\label{fig:GaAsAlpha}
\end{figure}

GaAs has a $1.4 eV$ bandgap that is ideally suited for solar cells.  It is a direct-bandgap material, with an absorption coefficient of $8000 cm^{-1}$ near its (direct) band-edge.  By contrast, the absorption coefficient of Si is ${\sim}10^4$ times weaker at its indirect band-edge.  Fig.~\ref{fig:GaAsAlpha} is a semi-log plot of the GaAs absorption coefficient as a function of energy; the circles are experimental data from \cite{Sturge1962} while the solid line is a fit to the data using the piecewise continuous function: 
\begin{equation}
\label{eq:alphaFit}
\alpha = \left\{
	\begin{array}{lr}
		\alpha_0 \exp\left(\frac{E-E_g}{E_0}\right) & E < E_g \\
		\alpha_0 \left( 1+\frac{E-E_g}{E'}\right) & E > E_g
	\end{array}
	\right.
\end{equation}
where $\alpha_0 = 8000/cm$, the Urbach energy is $E_0 = 6.7meV$, and $E' = 140meV$.  The exponential dependence of the absorption coefficient below the bandgap is characteristic of the ``Urbach tail'' \cite{Urbach1953}.

Efficient external emission can be separated into two steps: first, the semiconductor should have a substantially higher probability of recombining radiatively, rather than non-radiatively.  We define the internal luminescence yield, $\eta_{int}$, similarly to the external luminescence yield, as the probability of radiative recombination versus non-radiative recombination:
\begin{equation}
\label{eq:RintDef}
\eta_{int} = \frac{R_{int}}{R_{int}+R_{nr}}
\end{equation}
where $R_{int}$ and $R_{nr}$ are the radiative and non-radiative recombination rates per unit volume, respectively.  The internal luminescence yield is a measure of intrinsic material quality.  The second factor for efficient emission is proper optical design, to ensure that the internally radiated photons eventually make their way out to external surface of the cell.  Maximizing both factors is crucial for high open-circuit and operating point voltages.

We now derive the external luminescence yield for the three different geometries.  At open-circuit, $R_{pump}$ and the recombination rates, $R_{ext} + R_{nr}$, are equal, and this allowed the derivation of Eqn.~\ref{eq:VoltPenalty} for a general open-circuit voltage.  In operation, however, current will be drawn from the solar cell and the two rates will not be equal.  The current will be the difference between pump and recombination terms:
\begin{equation}
\label{eq:sjIV}
J = q \left( R_{pump}-R_{ext}-R_{nr} \right)
 = \int_0^{\infty} a(E)S(E)\textrm{d}E - \frac{1}{\eta_{ext}}q\pi e^{qV/kT} \int_0^{\infty}a(E)b(E)\textrm{d}E
\end{equation}
where the external luminescence from the cell is a Lambertian that integrates to $\pi$ steradians, and the absorptivity $a(E,\theta)$ has been assumed independent of polar angle $\theta$, which is clearly valid for the randomly textured surface.  It is approximately independent of incident angle for a planar front surface because the large refractive index of GaAs refracts the incident light very close to perpendicular inside the solar cell.

\subsection{Randomly Textured Surface}
Randomly texturing the front surface of the solar cell, Fig.~\ref{fig:GeometriesConsidered}(a), represents an ideal method for coupling incident light to the full internal phase space of the solar cell.  The absorptivity of the textured cell has been derived in \cite{Yablonovitch1982a}:
\begin{equation}
\label{eq:AbsRandText}
a(E) = \frac{4n^2\alpha L}{4n^2\alpha L + 1}
\end{equation}
Although only strictly valid in the weakly absorbing limit, the absorptivity is close enough to one for large $\alpha L$ that Eqn.~\ref{eq:AbsRandText} can be used for all energies.

To derive the external luminescence yield, all of the recombination mechanisms must be identified.  We have assumed a perfectly reflecting rear mirror, so the net radiative recombination is the emission from the front surface, given by Eqn.~\ref{eq:FrontEmRate}.  The only fundamental non-radiative loss mechanism in GaAs is Auger recombination.  The Auger recombination rate per unit area is $CLn_i^3\exp\left[3qV/2kT\right]$, where $L$ is the thickness of the cell, $C=7\times 10^{-30} cm^6 s^{-1}$ is the Auger coefficient \cite{Strauss1993}, and intrinsic doping is assumed to minimize Auger recombination. The external luminescence yield can then be written:
\begin{equation}
\label{eq:ExtYieldMirror}
\eta_{ext}\left(V\right) = \frac{\pi e^{qV/kT} \int_0^{\infty} a(E)b(E)\textrm{d}E}{\pi e^{qV/kT} \int_0^{\infty} a(E)b(E)\textrm{d}E + CLn_i^3 e^{3qV/2kT} }
\end{equation}

\subsection{Planar Front Surface with Perfectly Reflecting Mirror}
A second interesting configuration to consider is that of Fig.~\ref{fig:GeometriesConsidered}(b), which has a planar front surface and a perfectly reflecting rear mirror.  Comparison with the first configuration allows for explicit determination of the improvement introduced by random surface texturing.  Not surprisingly, surface texturing only helps for very thin cells.

The absorptivity of the planar cell is well-known:
\begin{equation}
a(E) = 1-e^{-2\alpha L }
\end{equation}
where the optical path length is doubled because of the rear mirror.  Using this absorptivity formula, Eqn.~\ref{eq:FrontEmRate} still represents the external emission rate.  As a consequence, the external luminescence yield follows the same formula, Eqn.~\ref{eq:ExtYieldMirror}, albeit with a different absorptivity, $a(E)$, for the planar front surface versus the textured solar cell.

\subsection{Planar Front Surface with Absorbing Mirror}
We have emphasized the importance of light extraction at open circuit to achieve a high voltage.  To demonstrate the effects of poor optical design on efficiency, we also consider the geometry of Fig.~\ref{fig:GeometriesConsidered}(c).  No extra recombination mechanism has been introduced, but the rear mirror now absorbs light rather than reflecting it internally.  (Or equivalently, it transmits light into a non-radiating, optically lossy, substrate.)

\begin{figure}[]
\centering
\includegraphics[width=3in]{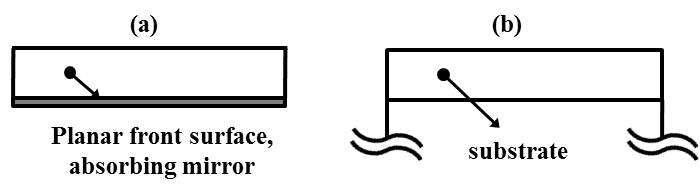}
\caption{Optically equivalent configurations, assuming the substrate does not reflect, nor re-radiate, absorbed photons.  In both cases, almost all of the internally emitted light will be lost out of the rear surface of the solar cell.} 
\label{fig:EquivGeometries}
\end{figure}

One could explicitly calculate the probability of internally emitted light escaping, tracking the photons to calculate the external luminescent yield (as in Sec.~\ref{sec:DiffLightExtraction}).  However, a simpler approach is to realize that the geometry with an absorbing rear-mirror is equivalent to a setup with an absorbing non-luminescent substrate supporting the active material, as depicted in Fig.~\ref{fig:EquivGeometries}.  Viewed either way, the absorptivity is: $a(E) = 1-e^{-\alpha L}$, where the light now has the opportunity for only one pass through the semiconductor to become absorbed.

To calculate the external luminescence yield, one can use the rate balancing method described earlier.  The recombination terms for emission out of the front surface and Auger processes are still present.  Now there is also a term for emission out of the rear surface.  By the same reasoning as for front surface emission, the emission out of the rear surface balances the thermal radiation coming from below: $a'(E,\theta')\times b'(E) \times \exp[qV/kT]$ which includes a further boost by the quasi-equilibrium factor $\exp[qV/kT]$.  At the rear surface, the density of states of the internal blackbody radiation $b'(E)\equiv n_r^2 b(E)$ is increased by $n_r^2$, where $n_r$ is the refractive index of the semiconductor.  The rear absorption $a'(E,\theta)$ is also modified as shown in the following equation for the total number of incident photons absorbed per unit area:
\begin{equation}
2\pi n_r^2 \int_0^{\infty} b(E) \int_0^{\pi/2} \left( 1-e^{-\frac{f(\theta ')\alpha L}{\cos\theta '}} \right) \cos \theta ' \sin \theta ' \textrm{d}\theta '
\end{equation}
where the $2\pi$ prefactor arises from the azimuthal integral, and $f(\theta ')$ equals one (two) for photons inside (outside) the escape cone, accounting for the different path lengths traveled by internal photons at angles greater or less than the critical angle $\theta_c$, defined by the escape cone at the top surface.  The internal path length by oblique rays is increased by the factor $1/\cos\theta '$.  A similar expression for the rear absorption is found in \cite{Marti1997}.  The external luminescence yield is now the ratio of the emission out of the front surface to the sum of the emission out of either surface plus Auger recombination:
\begin{equation}
\label{eq:ExtYieldAbsMirror}
\eta_{ext}\left(V\right) = \frac{\pi e^{qV/kT} \int_0^{\infty}a(E)b(E) \textrm{d}E }{ \pi e^{qV/kT} \int_0^{\infty} b(E) \left[ a(E) + 2n_r^2 \int_0^{\pi/2} \left( 1-e^{-\frac{f(\theta ')\alpha L}{\cos \theta '}} \right) \sin \theta ' \cos \theta ' \textrm{d} \theta ' \right] \textrm{d}E + CLn_i^3 e^{3qV/2kT} }
\end{equation}
which is an explicit function of the quasi-Fermi level separation $qV$.

\subsection{Equivalence of Photon and Carrier Equations}
The importance of photon management in general, and luminescent extraction in particular, have been emphasized.  Yet in Eqn.~\ref{eq:ExtYieldMirror}, for example, simple photonic quantities such as the semiconductor refractive index are not present.  The refractive index certainly plays a major role in photon extraction, through determination of the critical angle for total internal reflection, so how is it possibly not in the external yield formula?  This sub-section demonstrates that the dependence on refractive index is hidden because the equations are written in terms of the carriers; formulating the yield by directly accounting for every photon path yields an equivalent result while also directly illuminating the importance of photonic quantities.

A photon-counting derivation of the external yield of a planar geometry with a perfectly reflecting mirror was already completed in Chap.~\ref{chap:OnVoltage}.  Eqn.~\ref{eq:etaExtPP} is a formula for the yield that \emph{does} directly incorporate photonic quantities such as refractive index.  It is relatively straightforward to show that Eqn.~\ref{eq:etaExtPP} is equivalent to Eqn.~\ref{eq:ExtYieldMirror}.

By determination of the internal yield $\eta_{int}$ and further manipulation, the equivalence can be made definite.  The internal yield as previously defined is the ratio of radiative recombination events per unit volume to total recombination events (c.f. Eqn.~\ref{eq:RintDef}).  The radiative recombination rate per unit volume can be determined by internal detailed balancing, in contrast with the surface-normalized detailed balancing presented in Sec.~\ref{sec:DetBalance}.  The emission is again expressed through the absorption process, except this time it is the proportional to the material absorption coefficient:
\begin{equation}
\label{eq:SVReqn}
R_{int} = 4\pi \alpha n_r^2 e^{qV/kT} \int_{E_g}^{\infty} b(E) \mathrm{d}E
\end{equation}
Eqn.~\ref{eq:SVReqn} is known as the van Rooesbrouck-Shockley relation \cite{VanRoosbroeck1954}.  The non-radiative recombination rate is assumed to be the Auger recombination rate
\begin{equation}
R_{nr} = Cn_i^3 e^{3qV/2kT}
\end{equation}
The quantity $(1-\eta_{int})/\eta_{int}$ of Eqn.~\ref{eq:etaExtPP} can now be written
\begin{equation}
\label{eq:RnrToRint}
\frac{1-\eta_{int}}{\eta_{int}} = \frac{R_{nr}}{R_{int}} = \frac{C n_i^3 e^{3qV/2kT}}{4\pi \alpha n_r^2 e^{qV/kT} \int_{E_g}^{\infty} b(E) \mathrm{d}E }
\end{equation}
Inserting Eqn.~\ref{eq:RnrToRint} into Eqn.~\ref{eq:etaExtPP} and re-arranging
\begin{equation}
\eta_{ext}\left(V\right) = \frac{\pi e^{qV/kT} \int_0^{\infty} a_{ext} b(E)\textrm{d}E}{\pi e^{qV/kT} \int_0^{\infty} a_{ext} b(E)\textrm{d}E + CLn_i^3 e^{3qV/2kT} }
\end{equation}
which is identically Eqn.~\ref{eq:ExtYieldMirror} for the step-function absorber considered here.

The equivalence between the photon and carrier equations is important.  It is often easier to derive important quantities such as current, voltage, or luminescence yield through carrier generation and recombination, resulting in e.g. Eqns.~\ref{eq:ExtYieldMirror},\ref{eq:ExtYieldAbsMirror}.  Nevertheless, underlying the carrier quantities are the photon dynamics discussed in Chap.~\ref{chap:OnVoltage}, which ultimately dictate the voltage and efficiency.

\section{Discussion}

\begin{figure}[]
\centering
\includegraphics[width=3.5in]{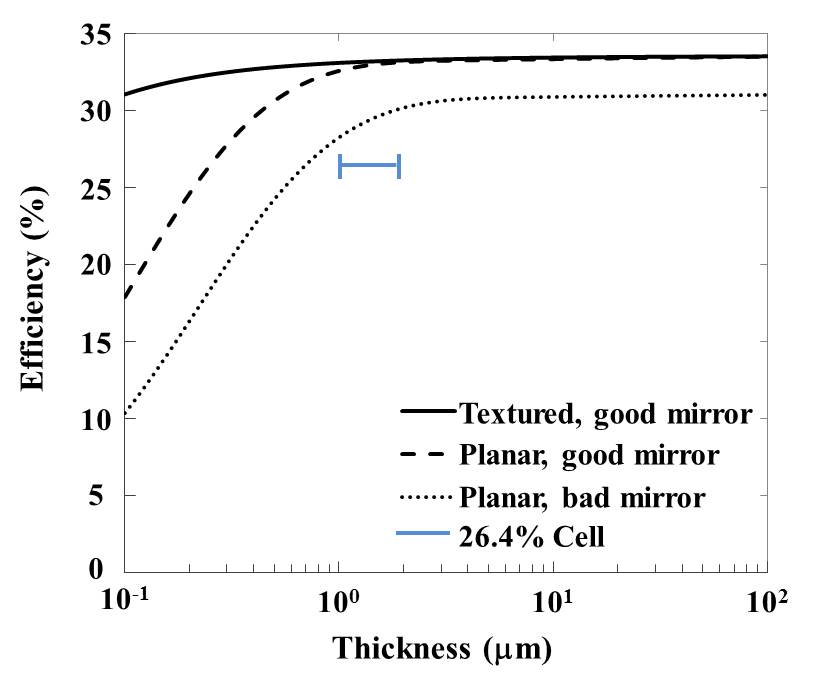}
\caption{GaAs solar cell efficiency as a function of thickness.  Random surface texturing does not increase the limiting efficiency of $33.5\%$, although it enables the high efficiencies even for cell thicknesses less than one micron.  Having an absorbing mirror on the rear surface incurs a voltage penalty and reduces the theoretical limiting efficiency to $31.1\%$.  There is still a sizeable gap between the $26.4\%$ cell and the theoretical limit.  The cell thickness was not specified in \cite{Green2010b}, and has been estimated as $\textrm{$1$--$2$}\mu m$.} 
\label{fig:EfficiencyThickness}
\end{figure}
Given the absorptivity and external luminescence yield of each geometry, calculation of the solar cell’s I-V curve and power conversion efficiency is straightforward using Eqn.~\ref{eq:sjIV}.   The power output of the cell, $P$, is simply the current multiplied by the voltage.  The operating point (i.e. the point of maximum efficiency) is the point at which $dP/dV = 0$.  Substituting the absorption coefficient data and solar spectrum values into Eqn.~\ref{eq:sjIV}, it is simple to numerically evaluate the bias point where the derivative of the output power equals zero.

Fig.~\ref{fig:EfficiencyThickness} is a plot of the solar cell efficiencies as a function of thickness for the three solar cell configurations considered.  Also included is a horizontal line representing the best GaAs solar cell fabricated up to 2010, which had an efficiency of $26.4\%$ \cite{Green2010b}.  The maximum theoretical efficiency is $33.5\%$, more than $7\%$ larger in absolute efficiency.  An efficiency of $33.5\%$ is theoretically achievable for both planar and textured front surfaces, provided there is a mirrored rear surface.  

\begin{figure}[]
\centering
\includegraphics[width=6.5in]{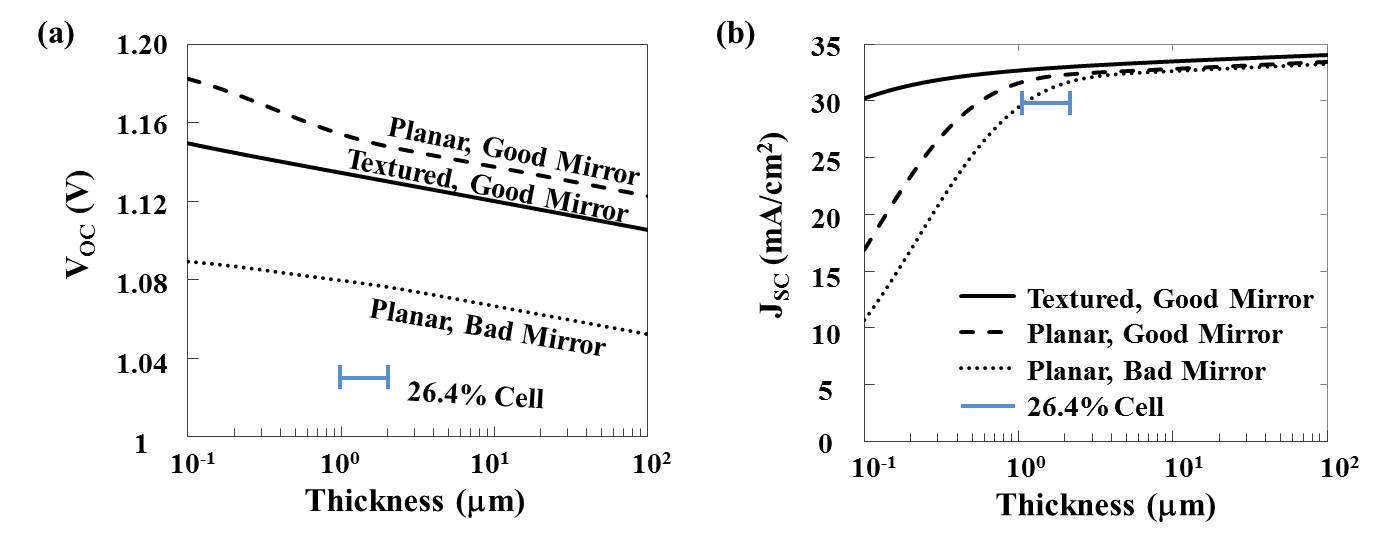}
\caption{ (a) Open-circuit voltage ($V_{OC}$), and (b) short-circuit current ($J_{SC}$), as a function of thickness for each of the solar cell configurations considered.  The planar cell with an absorbing mirror reaches almost the same short-circuit current as the other two configurations, but it suffers a severe voltage penalty due to poor light extraction and therefore lost photon recycling.  There is considerable opportunity to increase $V_{OC}$ over the previous record $26.4\%$ cell.  (The textured cell/good mirror has a lower voltage than a planar cell/good mirror owing to the effective bandgap shift observed in Fig.~\ref{fig:PlanarVsTextured}.  This slight voltage drop is not due to poor $\eta_{ext}$.) } 
\label{fig:VocJscThickness}
\end{figure}

Although surface texturing does not increase the maximum efficiency, it does help maintain an efficiency greater than $30\%$ even for solar cells that are only a few hundred nanometers thick.  The cell with a planar surface and bad mirror on its rear surface reaches an efficiency limit of only $31.1\%$, exhibiting the penalty associated with poor light extraction.  To understand more clearly the differences that arise in each of the three configurations, the short-circuit currents and open-circuit voltages of each are plotted in Fig.~\ref{fig:VocJscThickness}.

Fig.~\ref{fig:VocJscThickness} and Tab.~\ref{tab:EfficiencyVocJsc} display the differences in performance between a planar solar cell with a perfect mirror and one with an absorbing mirror.  Although the short-circuit currents are almost identical for both mirror qualities, at thicknesses greater than $2-3\mu m$, the voltage differences are drastic.  Instead of reflecting photons back into the cell where they can be re-absorbed, the absorbing mirror constantly removes photons from the system.  The photon recycling process attendant to a high external luminescence yield is almost halted when the mirror is highly absorbing.  

\begin{figure}[]
\centering
\includegraphics[width=6.5in]{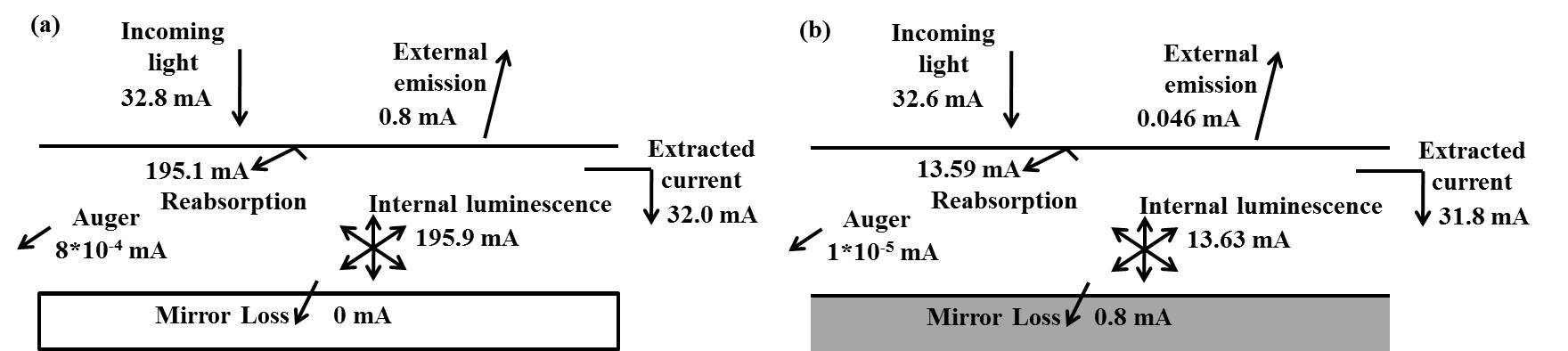}
\caption{ Diagram of currents at the operating point for a planar $10\mu m$ thick, $1cm^2$ solar cell (a) with a perfectly reflecting mirror and (b) with an absorbing mirror.  The two solar cells produce almost the same amount of current, but the cell with a perfectly reflecting mirror achieves a higher voltage.  The internal luminescence and re-absorption demonstrate the impact of front surface emission on carrier densities.  In GaAs, Auger recombination is negligible. } 
\label{fig:PlanarCurrents}
\end{figure}

Fig.~\ref{fig:PlanarCurrents} presents such intuition visually, displaying the internal and external currents of a $10\mu m$ thick GaAs solar cell at its maximum power point, for $0\%$ and $100\%$ reflectivity.  In both cases, the cells absorb very well, and the short-circuit currents are almost identical.  The extracted currents, too, are almost identical.  In each case $0.8mA$ of current is “lost” (i.e. the difference between short-circuit and operating currents); but in the case of the good mirror the current is lost to front surface emission.  There is a strong buildup of photon density when the only loss is emission through the front surface, allowing much higher internal luminescence and carrier density.  A higher operating voltage results.

From Fig.~\ref{fig:EfficiencyThickness}, it is clear that surface texturing is not helpful in GaAs, except to increase current in the very thinnest solar cells.  	In most solar cells, such as Silicon cells, surface texturing provides a mechanism for exploiting the full internal optical phase space.  The incident sunlight is refracted into a very small solid angle within the cell, and without randomizing reflections, photons would never couple to other internal modes.     

\begin{figure}[]
\centering
\includegraphics[width=6.5in]{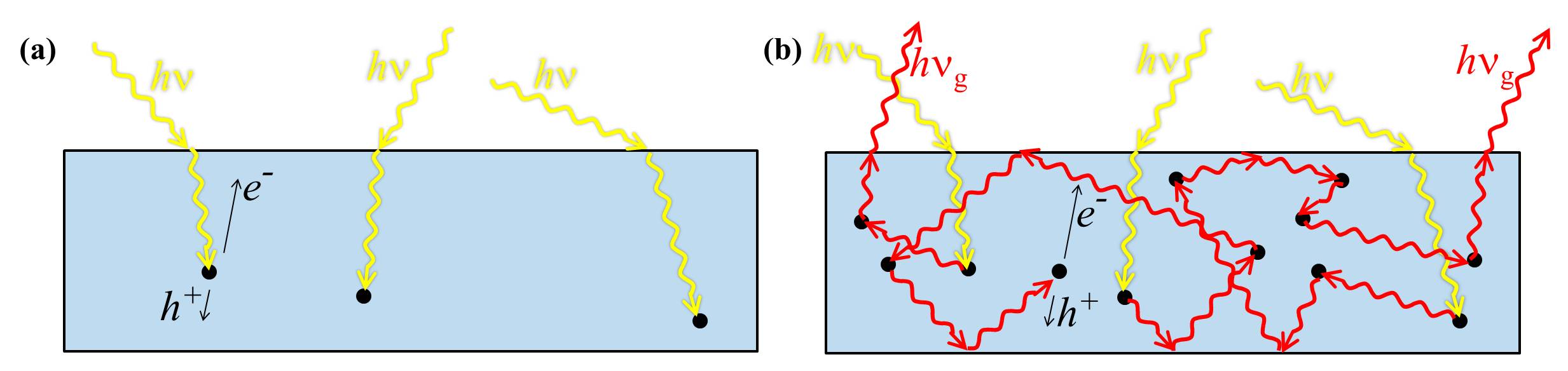}
\caption{ Qualitative illustration of the different photon dynamics in plane-parallel solar cells with (a) low luminescence yield $\eta_{ext}$ and (b) high $\eta_{ext}$, respectively.  In (a), the lack of photon extraction reduces carrier density and thus voltage.  Conversely in (b), the internal photons achieve full angular randomization even without surface texturing, and the high external emission is indicative of high carrier density build-up. } 
\label{fig:PhotonDynamics}
\end{figure}

GaAs is such an efficient radiator that it can provide the angular randomization by photon recycling.  After absorbing a photon, the photon will likely be re-emitted, and the re-emission is equally probable into all internal modes.  Whereas most materials require surface roughness to efficiently extract light, the radiative efficiency of GaAs ensures light extraction based on photon recycling.  Such photon dynamics are illustrated in Fig.~\ref{fig:PhotonDynamics}. 

Marti \cite{Marti1997} and Johnson \cite{Johnson2007} already emphasized the benefits of photon recycling toward efficiency, but we believe that external luminescence yield is the more comprehensive parameter for boosting solar cell efficiency and voltage.

\begin{figure}[]
\centering
\includegraphics[width=3.5in]{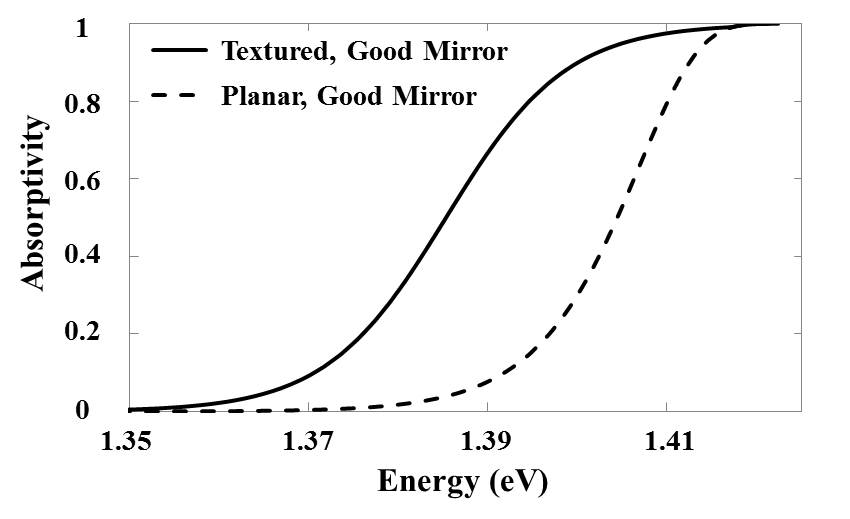}
\caption{ Absorptivity of a $10\mu m$ thick cell as a function of photon energy near the bandgap.  A textured cell absorbs well even at $1.39 eV$, effectively reducing the band-gap.  This explains the lower voltages but higher currents of the textured cell, relative to the plane-parallel cell. } 
\label{fig:PlanarVsTextured}
\end{figure}

It seems surprising that the planar solar cell would have a higher voltage than the textured cell.  This is due to a second-order effect seen in Fig.~\ref{fig:PlanarVsTextured}.  Textured cells experience high absorption even below the bandgap, due to the longer optical path length provided.  Texturing effectively reduces the bandgap slightly, as shown in Fig.~\ref{fig:PlanarVsTextured}, accounting for the lower open-circuit voltage but larger short-circuit current values seen in Fig.~\ref{fig:VocJscThickness}.

GaAs is an example of one of the very few material systems that can reach internal luminescence yields close to $1$; a value of $99.7\%$ has been experimentally confirmed \cite{Schnitzer1993b}.  However, it is the external luminescence yield that determines voltage, and that yield depends on both the quality of material and also the optical design. Absorbing contacts, or a faulty rear mirror, for example, will remove photons from the system that could otherwise be recycled.  Additionally, an optically textured design \cite{Yablonovitch1982} can provide the possibility for extraction of luminescent photons, before they could be lost.

Light trapping is normally thought of as a way to absorb more light and increase the current in a thinner cell.  But the concentration of carriers in a thinner cell also provides a voltage increase, $qV \sim kT\ln(4n^2)$, an effect that was implicitly used when light trapping was first incorporated into the fundamental calculation of Silicon efficiency \cite{Tiedje1984}.  Thus texturing improves the voltage in most solar cells.  Nonetheless, one of the main results of this article is that the voltage boost can come with OR without surface texturing in GaAs.  The reason is that efficient internal photon recycling provides the angular randomization necessary to concentrate the light, even in a plane-parallel GaAs cell.  Thus, short-circuit current in GaAs can benefit from texturing, but GaAs voltage accrues the same benefit with, or even without, texturing.  

The distinction between voltage boost by texturing, and voltage boost by photon recycling was already made by Lush and Lundstrom \cite{Lush1991} who predicted the higher voltages and the record efficiencies that have recently been observed \cite{Green2012} in thin film III-V solar cells.  However, the over-arching viewpoint in this article is that voltage is determined by external luminescence efficiency.  That viewpoint accounts in a single comprehensive manner for the benefits of nano-texturing, photon re-cycling, parasitic optical reflectivity, and imperfect luminescence, while being thermodynamically self-consistent.

\begin{table}[ht]
\centering
\begin{tabular}{cc}
\includegraphics[width=6.5in]{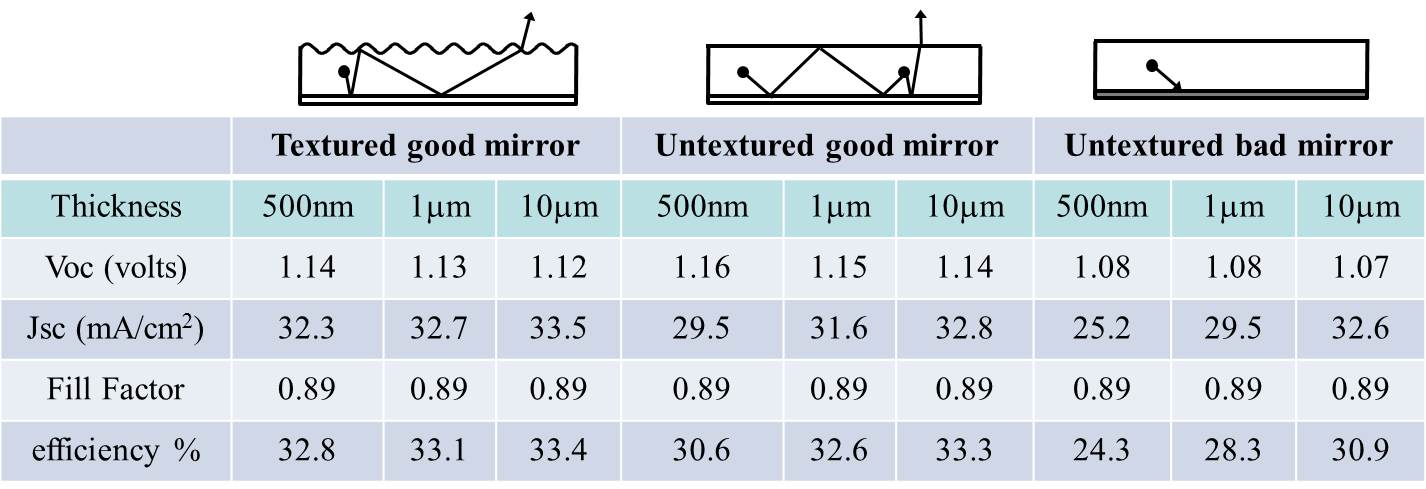}
\end{tabular}
\caption{Table of $V_{OC}$, $J_{SC}$, and efficiency values for three possible geometries and relevant cell thicknesses.  A good rear mirror is crucial to a high open-circuit voltage, and consequently to efficiencies above $30\%$.}
\label{tab:EfficiencyVocJsc}
\end{table}

In the case of perfect photon recycling, there is surprisingly little thickness dependence of $V_{OC}$.  This is to be contrasted with the textured case, where the voltage boost might require light concentration and carrier concentration within a thin cell.  Under perfect photon re cycling, photons are lost only at the surface, and the photon density and carrier density are maintained at the maximum value through the full depth.  The solar cell can be permitted to become thick, with no penalty.  In practice a thick cell would carry a burden, and an optimum thickness would emerge.

\section{A New Single-Junction Efficiency Record}
The prior \cite{Green2010b} one-sun, single-junction efficiency record, $26.4\%$, was set by GaAs cells that had $V_{OC}=1.03V$.  Alta Devices has recently made a big improvement in GaAs efficiency to $28.3\%$ \cite{Kayes,Green2012}.  The improvement was not due to increased short-circuit current; in fact, the Alta Devices cell had $J_{SC} = 29.5 mA/cm^2$, \emph{less} short-circuit current than the $29.8 mA/cm^2$ of the previous record cell.  However, the Alta Devices cell had a measured open-circuit voltage $V_{OC}=1.11V$, an $80mV$ improvement over the $1.03V$ open-circuit voltage of the previous record cell, showing in part the benefit of light extraction.  
\begin{figure}[]
\centering
\includegraphics[width=4in]{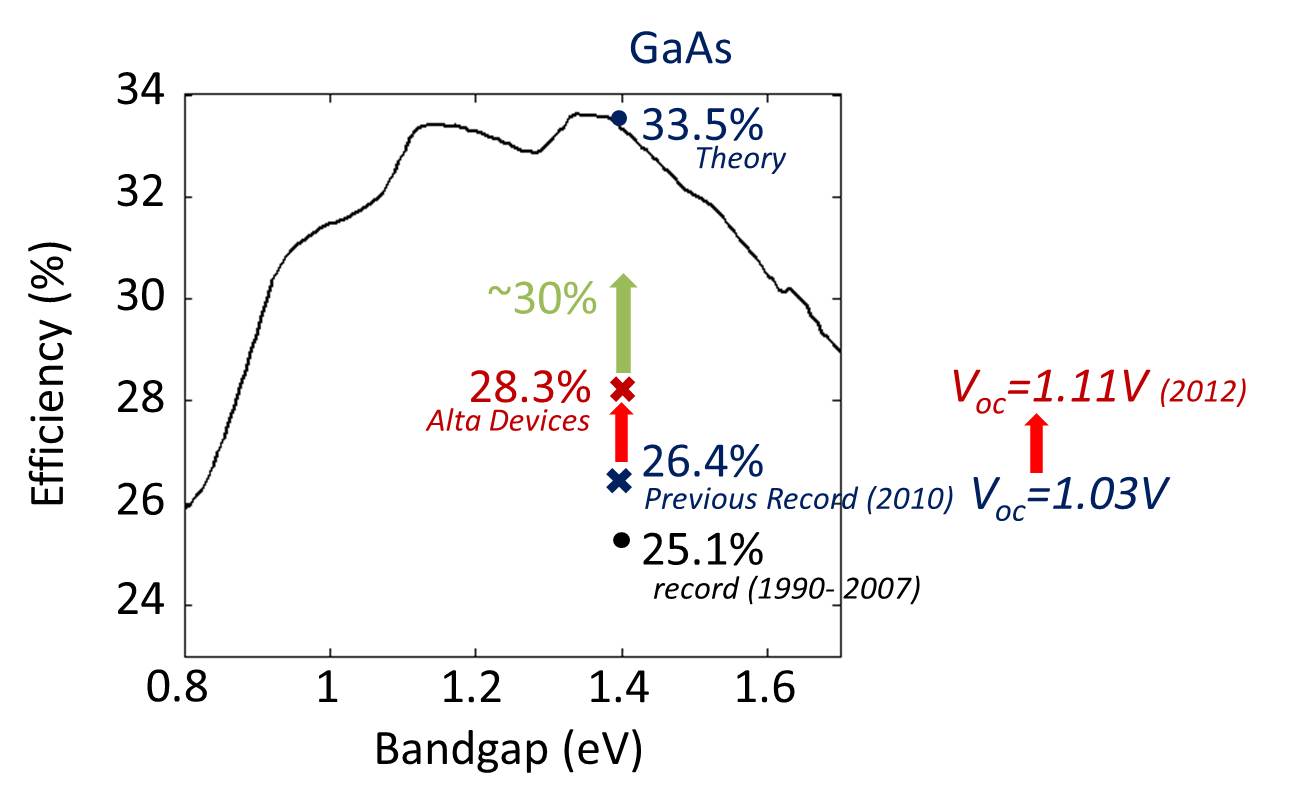}
\caption{ Single-junction, flat-plate solar cell efficiency records over time \cite{Green2008,Green2010b,Green2012}.  Each is a GaAs solar cell.  Alta Devices recently dramatically increased the efficiency record through open-circuit voltage improvement, due to superior photon management.  } 
\label{fig:EfficiencyRecord}
\end{figure}	

\section{Conclusions}
We have shown how to include photon recycling and imperfect radiation properties into the quasi-equilibrium formulation of Shockley and Queisser.  High voltages $V_{OC}$ are achieved by maximizing the external luminescence yield of a system.  Using the standard solar spectrum and the measured absorption curve of GaAs, we have shown that the theoretical efficiency limit of GaAs is $33.5\%$, which is more than $4\%$ higher than that of Silicon \cite{Kerr2003}, and achieves its efficiency in a cell that is 100 times thinner.

Internally trapped radiation is necessary, but not sufficient, for the high external luminescence that allows a cell to reach voltages near the theoretical limits.  The optical design must ensure that the only loss mechanism is photons exiting at the front surface.  A slightly faulty mirror, or equivalently absorbing contacts or some other optical loss mechanism, sharply reduces the efficiency limit that can be achieved.  To realize solar cells with efficiency greater than $30\%$, the optical configuration will need to be very carefully designed.
	
The Shockley-Queisser formulation is still the foundation of solar cell technology.  However, the physics of light extraction and external luminescence yield are clearly relevant for high performance cells and will prove important in the eventual determination of which solar cell technology wins out in the end.  In the push for high-efficiency solar cells, a combination of high-quality GaAs and optimal optical design should enable single-junction flat plate solar cells with greater than $30\%$ efficiency.

%% file: NextGenCells.tex
\chapter{Analysis of next-generation solar cells}
\label{chap:NextGenCells}
\epigraph{For a successful technology, reality must take precedence over public relations, for nature cannot be fooled.}{Richard Feynman}

\noindent
There are three loss mechanisms in single-junction solar cells that result in maximum possible efficiencies of only $33\%$, rather than the Carnot efficiency of almost $95\%$ \cite{Green2006a,Balenzategui1997}.  First, photons with energies smaller than the semiconductor's bandgap energy are not absorbed and therefore not collected.  Second, carriers generated from absorption of high energy photons quickly thermalize to the band-edge, reducing the available free energy.  Finally, the solar cell's emissivity must match it's absorptivity, resulting in significant luminescence.  Because this final condition is a result of thermodynamic detailed balancing, it is not a process to be avoided but rather one to be maximized, as discussed extensively in Chaps.~\ref{chap:OnVoltage},~\ref{chap:SQLimit}.  However, solar cell concepts attempting to circumvent the first two loss mechanisms are abundant.

The Shockley-Queisser method \cite{Shockley1961} is the canonical technique for understanding efficiency limits for each different concept, with each limit depending on the specific technology. The framework developed in Chap.~\ref{chap:SQLimit} can similarly be applied to a variety of technologies, in an attempt to understand deviations from the ideal assumptions and the physics required to approach the efficiency limits.  In this chapter, such analysis is applied to third-generation \cite{Green2006a} solar cells dependent on carrier multiplication, up-conversion, and multi-junction technologies.

A common theme throughout the chapter is the independence of the voltage penalty due to poor extraction, Eqn.~\ref{eq:VoltPenalty}, relative to bandgap.  A luminescence yield of $2\%$, for example, results in a voltage penalty of $-kT\ln(0.02)\approx 100mV$, regardless of the bandgap.  For smaller bandgap solar cells, this is a much greater percentage of the open-circuit voltage, resulting in a larger efficiency penalty.  Large-bandgap cells are relatively robust to imperfect photon management, whereas small-bandgap cells are susceptible to small imperfections.  Consequently, technologies dependent on larger bandgap cells, such as multi-junction tandem cells, have a greater likelihood of approaching their efficiency limits than do technologies dependent on smaller bandgap cells, such as down-converting or up-converting solar cells.

\section{Carrier Multiplication}
\label{sec:CarrMult}
\begin{figure}
\centering
\includegraphics[width=2.5in]{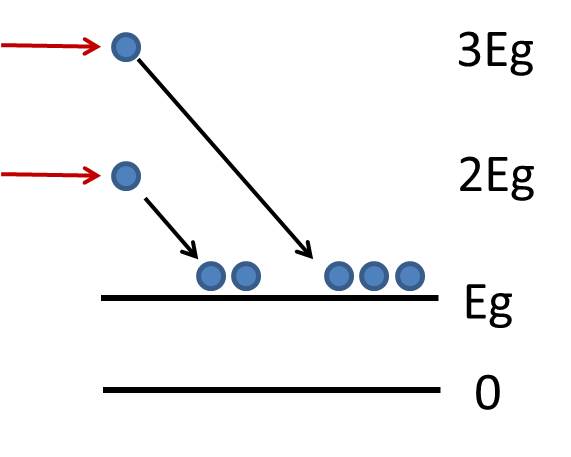}
\caption{Schematic of carrier multiplication techniques for enhancing solar cell efficiency.  Instead of allowing high-energy photons to thermally relax to the band-edge, multiple carriers are generated through impact ionization or a similar process.} 
\label{fig:CarrMultSchematic}
\end{figure}

A prime example of the susceptibility of small bandgaps to imperfect luminescent yield occurs in carrier multiplication schemes.  Multiple-carrier solar cells try to enhance the energy conversion efficiency through generation of more than one carrier per high-energy photon \cite{Hanna2006,Nozik2008,Wurfel1997,Sambur2010,Beard2010}.  A photon with energy $2E_g$ could potentially create two carriers instead of one, if, for example, impact ionization occurred faster than thermalization through phonon scattering.  A photon with energy $3E_g$ could potentially create up to three carriers per photon, and similarly for every multiple of $E_g$ an extra carrier per photon could be generated.

Theoretical calculations of the maximum efficiency possible in a carrier multiplication scheme were completed in \cite{Werner1994,Landsberg2002,Hanna2006}.  We develop here a generalized calculation, accounting for imperfect luminescent yield.  To include the effects of carrier multiplication, the absorptivity is modified.  We can define $a_0$ to be the default absorptivity near the bandgap, such that for a plane-parallel solar cell one would have
\begin{equation}
a_0 = 1-e^{-2\alpha_0 L}
\end{equation}
where for simplicity the absorption coefficient $\alpha$ has been assumed to be a step-function with height $\alpha_0$ near the band-edge.  The absorptivity as a function of energy would then be
\begin{equation}
\begin{array}{ll}
a(E) & = \left\{
\begin{array}{lc}
	0 & E < E_g \\
	a_0 & E_g < E < 2E_g \\
	2a_0 & 2E_g < E < 3E_g \\
	\vdots & 
\end{array}
\right. \\
& = \displaystyle\sum_{m=1}^{M} a_0 H(E-mE_g)
\end{array}
\end{equation}
where $H(\cdot)$ is the Heaviside step function and $M$ is the maximum carrier multiplication factor allowed.  

\begin{figure}
\centering
\includegraphics[width=2.5in]{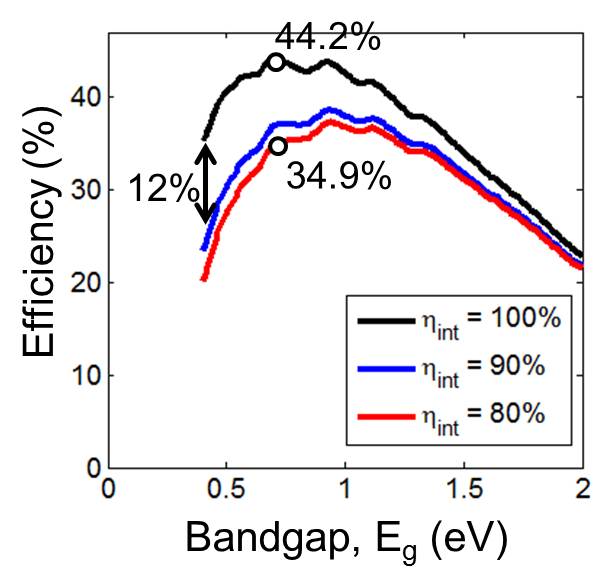}
\caption{Efficiency of solar cells with a carrier multiplication factor of six for internal luminescence yields $\eta_{int}=100\%$ (black), $90\%$ (blue), and $80\%$ (red).  There is a significant deterioration of the efficiency for imperfect photon management, especially at the smaller bandgaps that would otherwise have the potential for more carrier multiplication.} 
\label{fig:CarrMultEfficiencies}
\end{figure}
Just as the absorption is modified by carrier multiplication, careful consideration of the emission must also be taken into account.  Once it has been assumed that there is an effective process for converting a high energy photon into two carriers, the reverse process of two carriers combining into one and emitting a high-energy photon must also be accounted for.  This can be done with a thermodynamic analysis \cite{Luque1997} or a more physical description \cite{Green2006a}.  The physical intuition is that just as emission near the band-edge occurs through annihilation of a single electron-hole pair, with the $np$ product proportional to $e^{qV/kT}$, at a higher multiple $m$ of the bandgap, the emission will be dominated by processes involving $m$ electron-hole pairs, scaling therefore as $e^{mqV/kT}$.  The current-voltage relation for a multi-carrier generation solar cell, analogous to Eqn.~\ref{eq:sjIV} for a single-carrier generation cell, is then
\begin{equation}
J = a_0 \displaystyle\sum_{m=1}^{M}\int_{0}^{\infty} H(E-mE_g) S(E)\textrm{d}E - \frac{1}{\eta_{ext}}q\pi a_0 \displaystyle\sum_{m=1}^{M} e^{mqV/kT} \int_0^{\infty} H(E-mE_g) b(E)\textrm{d}E
\end{equation}

The analysis of the external luminescence yield mimics that of Chap.~\ref{chap:SQLimit}.  Even though there are now many quasi-Fermi levels at each multiple of the bandgap, the emission still primarily occurs at the bandgap, scaling as $e^{qV/kT}$.  The internal yield can thus be approximated as independent of the voltage, to first order.  To model the external yield, some thickness $\alpha L$ must be chosen, determining both $a_0$ and $\eta_{ext}$.  In the calculations below, for each bandgap and each internal yield the optical thickness $\alpha L$ was optimized for maximal efficiency.  

The efficiency itself was calculated by finding the operating point, where $dP/dV = d(JV)/dV = 0$.  The AM1.5G spectrum \cite{SolarSpectrum} was used as the solar source.  In \cite{Hanna2006}, it was shown that the ideal carrier multiplication system converts one photon into up to six carriers at a time.  Beyond six, there are not enough high-energy photons in the solar spectrum to make a difference.  
\begin{figure}
\centering
\includegraphics[width=5.5in]{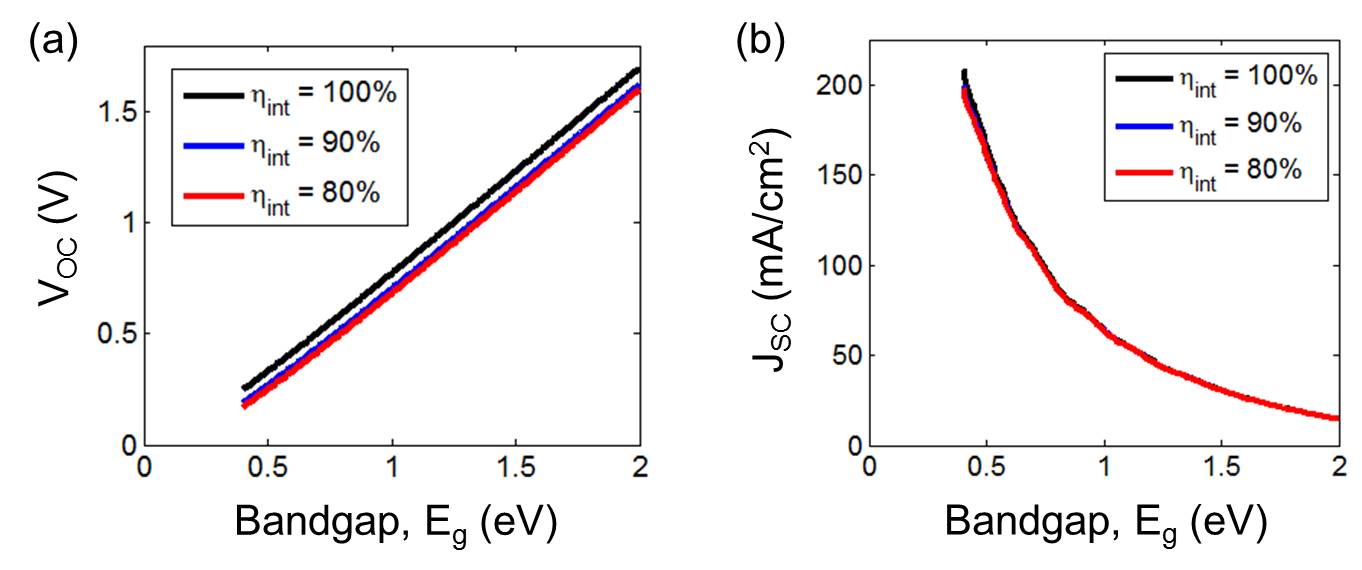}
\caption{(a) $V_{OC}$ and (b) $J_{SC}$ for solar cells with a carrier multiplication factor of six and internal luminescence yields $\eta_{int}=100\%$ (black), $90\%$ (blue), and $80\%$ (red).  There is approximately a $70mV$ penalty in $V_{OC}$ for $90\%$ rather than $100\%$ internal yield, whereas the penalty in $J_{SC}$ is negligible.} 
\label{fig:CarrMultVocJsc}
\end{figure}

The efficiency as a function of bandgap, for internal luminescent yields of $100\%$, $90\%$, and $80\%$ is given in Fig.~\ref{fig:CarrMultEfficiencies}.  The open-circuit voltage and short-circuit current as a function of bandgap for the same systems are shown in Fig.~\ref{fig:CarrMultVocJsc}.  Even though the ideal carrier multiplication solar cell can achieve greater than $44\%$ efficiency at its radiative limit, it is clearly not a robust maximum.  For even a $90\%$ internal yield, the solar cell efficiency drops more than $7\%$ at the optimal bandgap and about $12\%$ at the smallest bandgap.  At $80\%$ internal yield, the efficiency at $0.7eV$ bandgap is only $34.9\%$, scarcely higher than the SQ limit for a conventional solar cell.

The $V_{OC}$ and $J_{SC}$ plots of Fig.~\ref{fig:CarrMultVocJsc} reveal the culprit: a significant loss in $V_{OC}$.  In exactly the same way that single-junction cells were penalized by poor external yield, multi-carrier generation solar cells are also penalized by the difficulty of light extraction.  The $V_{OC}$ penalty results in a greater efficiency penalty in the multi-carrier generation circumstance, however.  For an internal yield of $90\%$ (corresponding to an external yield of approximately $6\%$ in the plane-parallel geometry), the open-circuit voltage is penalized by $kT\ln0.06 \approx 70mV$, as seen in Fig.~\ref{fig:CarrMultVocJsc}(a).  Note that a $90\%$ internal yield would be extraordinarily high, considering that Auger recombination will likely be a prominent factor due to the relative frequency of impact ionization.  

The open-circuit voltage penalty, due to entropy production, is \emph{independent of bandgap}.  Carrier multiplication schemes, because they use photons at higher multiples of the bandgaps to create extra carriers, thrive in the ideal limit at smaller bandgaps.  A bandgap-independent voltage penalty it therefore be a much greater \emph{relative} penalty at small bandgaps, leading to the significant degradation of efficiency seen in Fig.~\ref{fig:CarrMultEfficiencies}.

\begin{figure}
\centering
\includegraphics[width=2.5in]{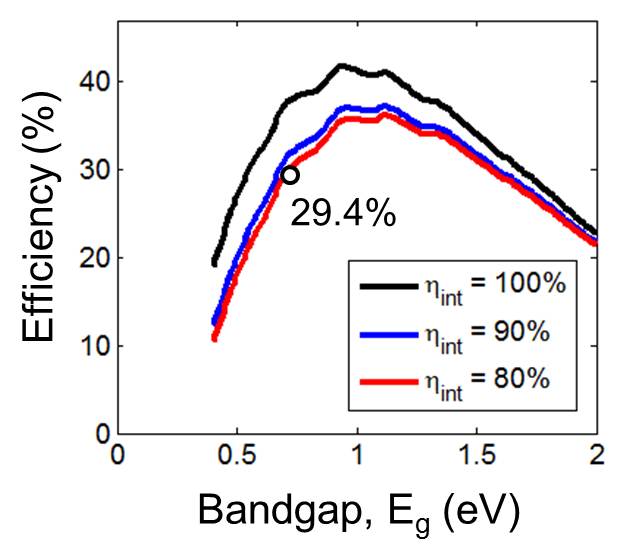}
\caption{Efficiency of solar cells with a carrier multiplication factor of two for internal luminescence yields $\eta_{int}=100\%$ (black), $90\%$ (blue), and $80\%$ (red).} 
\label{fig:CarrMultEfficiencies2x}
\end{figure}

The story is similar in the more realistic case of a carrier multiplication factor of two, as seen in Fig.~\ref{fig:CarrMultEfficiencies2x}.  If one were to choose a $0.7eV$ bandgap, based on the optimum of the $6x$ case, and achieve $80\%$ internal yield, the limiting efficiency would still only be $29.4\%$.  Relying on the theoretical efficiencies calculated in the radiative limit to guide design decisions, such as which bandgap to choose, can be dangerous without a proper understanding of the internal physics required to achieve that limit.  In the case of multi-carrier generation solar cells, the small bandgaps needed to justify such a scheme are not robust to small deviations from ideal conditions.

\section{Up-conversion}
\begin{figure}
\centering
\includegraphics[width=5.5in]{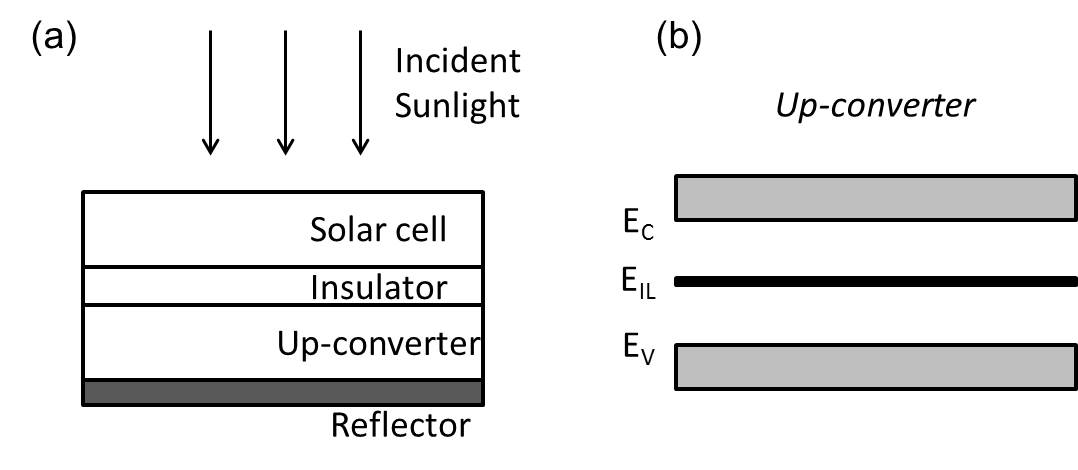}
\caption{(a) Schematic of up-converting solar cells. The up-converter absorbs photons with energies less than the bandgap of the solar cell, through an intermediate level as shown in the band diagram in (b).  See \cite{Trupke2002} for the exact assumptions.} 
\label{fig:UpConvSchematic}
\end{figure}

Up-converting solar cells attempt to increase the photovoltaic efficiency by capturing sub-bandgap photons without decreasing the bandgap \cite{Trupke2002,Baluschev2006,Trupke2006,Shalav2007}.  Although there are a variety of potential up-converting structures, a common one is shown in Figure~\ref{fig:UpConvSchematic}.  An up-converting layer, electrically insulated from a single-junction solar cell, sits below the junction.  A mirror below the up-converting layer prevents any photons from escaping through the rear surface.

The converter is assumed to have a band diagram as in Fig.~\ref{fig:UpConvSchematic}(b).  There is an impurity level that provides absorption of sub-bandgap photons that pass through the solar cell.  Consecutive absorption of sub-bandgap photons, creating a carrier first in the impurity level and then in the conduction band of the converter, leads to the eventual emission of a photon with an energy nearly the converter bandgap energy.  For maximum efficiency, the converter bandgap and the solar cell bandgap should be equal.  To avoid statistical averaging, photon selectivity is assumed, requiring converter valence and conduction bands of finite widths.  The widths are set such that sub-bandgap can be absorbed either in the transition from the valence band to the impurity level \emph{or} in the transition from the impurity level to the conduction band, but not both.  The bandgap of the solar cell is $E_g$, and the energy differences from $E_V$ to $E_{IL}$ and from $E_{IL}$ to $E_C$ will be denoted $E_1$ and $E_2$, respectively.

The up-converting system can be analyzed through the Shockley-Queisser limit by again employing detailed balance at thermal equilibrium, and scaling up the luminescence at the non-equilibrium condition.  Completed in detail in \cite{Trupke2002}, the basic steps are developed here to demonstrate how to include imperfect luminescence efficiency as a generalization of the Shockley-Queisser method.

Consider the up-converting system at steady-state.  By definition, the number of carriers at every level in the system (i.e. the conduction/valence bands of the solar cell and the conduction/valence bands and impurity level of the up-converter) must be constant.  Therefore, the generation and recombination rates for each level must be equal.  With this knowledge a system of equations can be derived containing all of the unknown system quantities.

\begin{figure}
\centering
\includegraphics[width=3.5in]{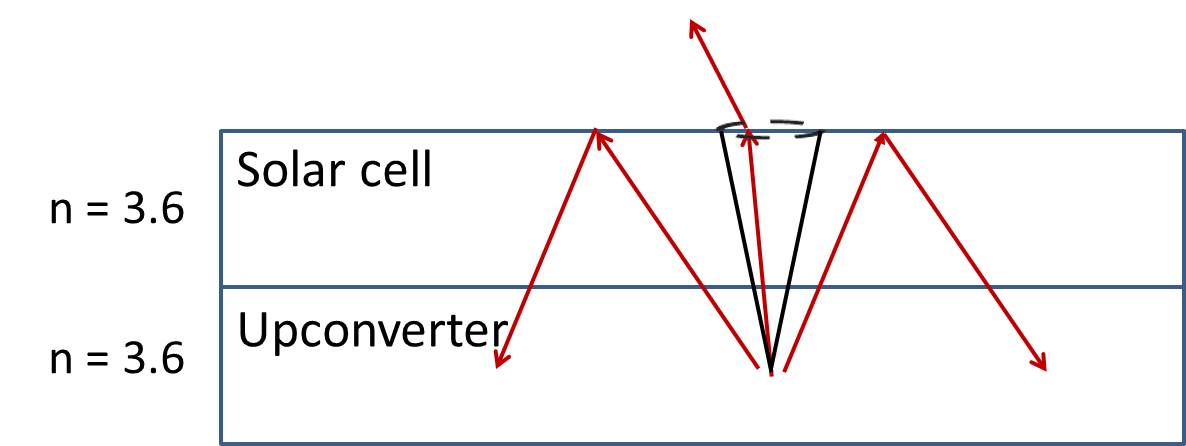}
\caption{Escape cone for sub-bandgap photons emitted from the up-converter.  The escape cone is determined by the refractive index of the air, regardless of the refractive index of the solar cell.  Photon extraction is therefore very difficult and susceptible to small imperfections.} 
\label{fig:UpConvLumEff}
\end{figure}

There are a number of non-idealities that could be included.  Any of the bands within either the solar cell or the up-converter could have additional loss mechanisms associated with them.  Only the simplest loss mechanism is considered here, in which sub-bandgap photons emitted from the up-converter have a less than unity escape efficiency.  Fig.~\ref{fig:UpConvLumEff} demonstrates the difficulty of extraction from the up-converter due to its large refractive index.  Imperfect extraction could occur due to imperfect rear mirror reflectance, non-ideal up-converter materials, absorption of photons in the contacts, etc.  All other processes in the solar cell are assumed to be ideal, as in \cite{Trupke2002}. 

First, the steady-state condition within the up-converter can be determined.  There is no current being drawn from the up-converter; it simply acts as an LED driven by absorption of sub-bandgap photons.  There is a constant exchange of photons between the up-converter and the solar cell, with the luminescence dictated by detailed balancing as in Sec.~\ref{sec:DetBalance}.  In the up-converter, steady-state carrier populations can be determined by quasi-Fermi levels within each of the three bands.  The chemical potential difference between the conduction and valence band will be denoted $\mu^{CV}$, while the difference between the conduction band and impurity level will be $\mu^{CI}$.  Given these definitions, steady-state conditions at the conduction band and impurity level, respectively, require:
\begin{subequations}
\label{eq:UpConvSystem}
\begin{align}
\begin{array}{rl}
\Omega_{sun} & \int_{E_2}^{E_g} a_{UC}(E) S(E) \mathrm{d}E +  \int_{E_g}^{E_g+E_2}  a_{UC}(E) \Omega_{int} a_{SC}(E) b(E) e^{\mu/kT} \mathrm{d}E \\
& = \Omega_{em} e^{\mu^{CI}/kT} \int_{E_2}^{E_g} a_{UC}(E) b(E)  \mathrm{d}E + \Omega_{int} e^{\mu^{CV}/kT} \int_{E_g}^{E_g+E_2}  a_{UC}(E) b(E) \mathrm{d}E
\end{array} \\
\begin{array}{rl}
\Omega_{sun} & \int_{E_1}^{E_2} a_{UC}(E)S(E)\mathrm{d}E + \Omega_{ext} e^{\mu^{CV}/kT} \int_{E_2}^{E_g} a_{UC}(E)b(E)\mathrm{d}E \\
= & \Omega_{sun} \int_{E_2}^{E_g} a_{UC}(E)S(E)\mathrm{d}E + \frac{1}{\eta_{ext}}\Omega_{ext} e^{(\mu^{CV}-\mu^{CI})/kT} \int_{E_1}^{E_2} a_{UC}(E)b(E)\mathrm{d}E
\end{array}
\end{align}
\end{subequations}
where $\Omega_{sun}$, $\Omega_{int}$, and $\Omega_{em}$ are the solid angles subtended by the sun, by internal radiative transfer between the up-converter and solar solar, and for emissive radiation from the solar cell, respectively; $\mu$ is the chemical potential of the solar cell.  In each equation the terms on the left-hand side represent carrier generation processes (through either absorption or emission) at the relevant level or band, while the right-hand side represents carrier removal processes.  Given the solar cell chemical potential, Eqns.~\ref{eq:UpConvSystem} can be solved for the quantities $\mu^{CV}$ and $\mu^{CI}$, resulting in the emission from the up-converter.  The solar cell can then be analyzed through typical detailed balance, but with the extra luminescence from the up-converter also contributing.  Consequently, the current-voltage J-V relationship is
\begin{equation}
\label{eq:UpConvJV}
\begin{array}{ll}
J/q = & \Omega_{sun} \int_{E_g}^{\infty} a_{SC}(E) S(E) \mathrm{d}E + \Omega_{int} e^{\mu^{CV}/kT}\int_{E_g}^{E_g+E_2}a_{SC}(E)a_{UC}(E)b(E)\mathrm{d}E \\
& - \Omega_{em} e^{\mu/kT} \int_{E_g}^{\infty}a_{SC} b(E) \mathrm{d}E - \Omega_{int} e^{\mu/kT} \int_{E_g}^{E_g+E_2} a_{SC} b(E) \mathrm{d}E
\end{array}
\end{equation}

The solid angles of the system are dependent on the desired circumstance.  For a non-concentrated system without any emission restriction, one would have $\Omega_{sun}=6.8\times 10^{-5}sr$ and $\Omega_{em}=\pi$.  For a concentrator system with concentration $C$, there are effectively $C$ suns in the system, resulting in $\Omega_{sun}=C\times 6.8\times 10^{-5}sr$.  Note that this sets an upper limit on the concentration factor of $C_{max}=\pi/6.8\times 10^{-5} \approx 46,200$.  For a non-concentrated system with emission restricted to only the solar solid angle, one would have $\Omega_{sun}=\Omega_{em}=6.8\times 10^{-5}sr$.  In all cases, the internal radiative transfer between the up-converter and the solar cell implies a solid angle of $\Omega_{int} = \pi n_r^2$, where the extra factor including the solar cell refractive index $n_r$ accounts for the increased blackbody density of states at thermal equilibrium.

Although the solar cell and up-converter were treated as perfect absorbers in \cite{Trupke2002}, understanding the effects of imperfect luminescence efficiency require a model of the internal dynamics.  Consequently, a step-function absorber will be assumed, with both plane-parallel and textured geometries considered.  The external luminescence yield is given by Eqn.~\ref{eq:etaExtPP}.

Eqns.~\ref{eq:UpConvSystem},~\ref{eq:UpConvJV} are solved simultaneously to generate the $J$-$V$ curve for the up-converting system.  Three system configurations are considered: the maximum concentration case ($\Omega_{sun}=\pi$), the minimum emission case ($\Omega_{em} = 8.5\times 10^{-5}$), and a non-concentrating case in which a small impurity \emph{band} is allowed instead of a single level.  For each configuration, efficiencies as a function of bandgap are calculated for three different material/structure combinations: perfect $100\%$ internal luminescence yield, $90\%$ internal luminescence yield (in the up-converter) in a plane-parallel geometry, and $90\%$ in a randomly textured geometry.  

\begin{figure}
\centering
\includegraphics[width=6in]{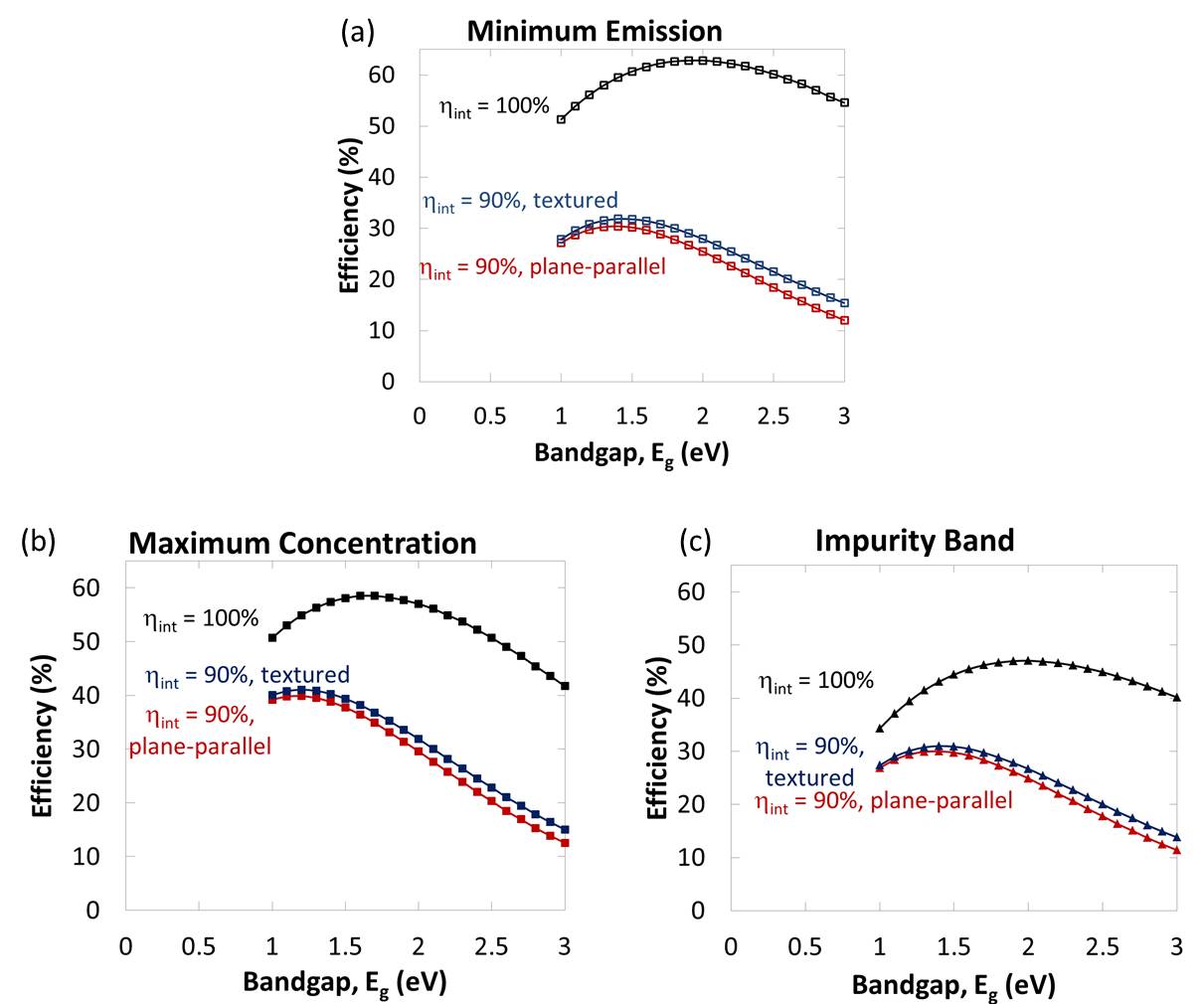}
\caption{Calculated efficiencies for three up-converting system configurations: (a) minimum emission ($\Omega_{em}=6.8\times 10^{-5}$), (b) maximum concentration ($\Omega_{sun} = \pi$), (c), non-concentrating, but with an impurity band over which carriers can thermalize, instead of a single impurity level.} 
\label{fig:UpConvEff}
\end{figure}

The calculated efficiency limits are shown in Fig.~\ref{fig:UpConvEff}.  To achieve full absorption, the absorption coefficient was defined such that $\alpha L = 2.5$ in the plane-parallel case, and $\alpha L = 1$ in the randomly textured case, where $L$ is the thickness of the up-converting layer.  The efficiencies for $\eta_{int} = 100\%$ corroborate the values in \cite{Trupke2002}.  The severe penalty for small deviations from $\eta_{int}=100\%$ are immediately apparent, for both plane-parallel and randomly textured configurations.  The penalty from imperfect luminescence yield within the solar cell is not even included, as only non-ideal up-converter yield was modeled.  Whereas the theoretical efficiency limits for the maximum concentration and minimum emission cases at $\eta_{int}=100\%$ are $58.5\%$ and $62.8\%$, their maximum efficiencies fall to less than $40\%$ and $30\%$, respectively, for an internal luminescence yield of $\eta_{int}=90\%$.  A similar decrease is seen for a system in which an impurity band exists in the up-converter, allowing for carriers to thermalize and loosening the restrictions on $E_1$ and $E_2$ to achieve photon selectivity.

\begin{figure}
\centering
\includegraphics[width=5in]{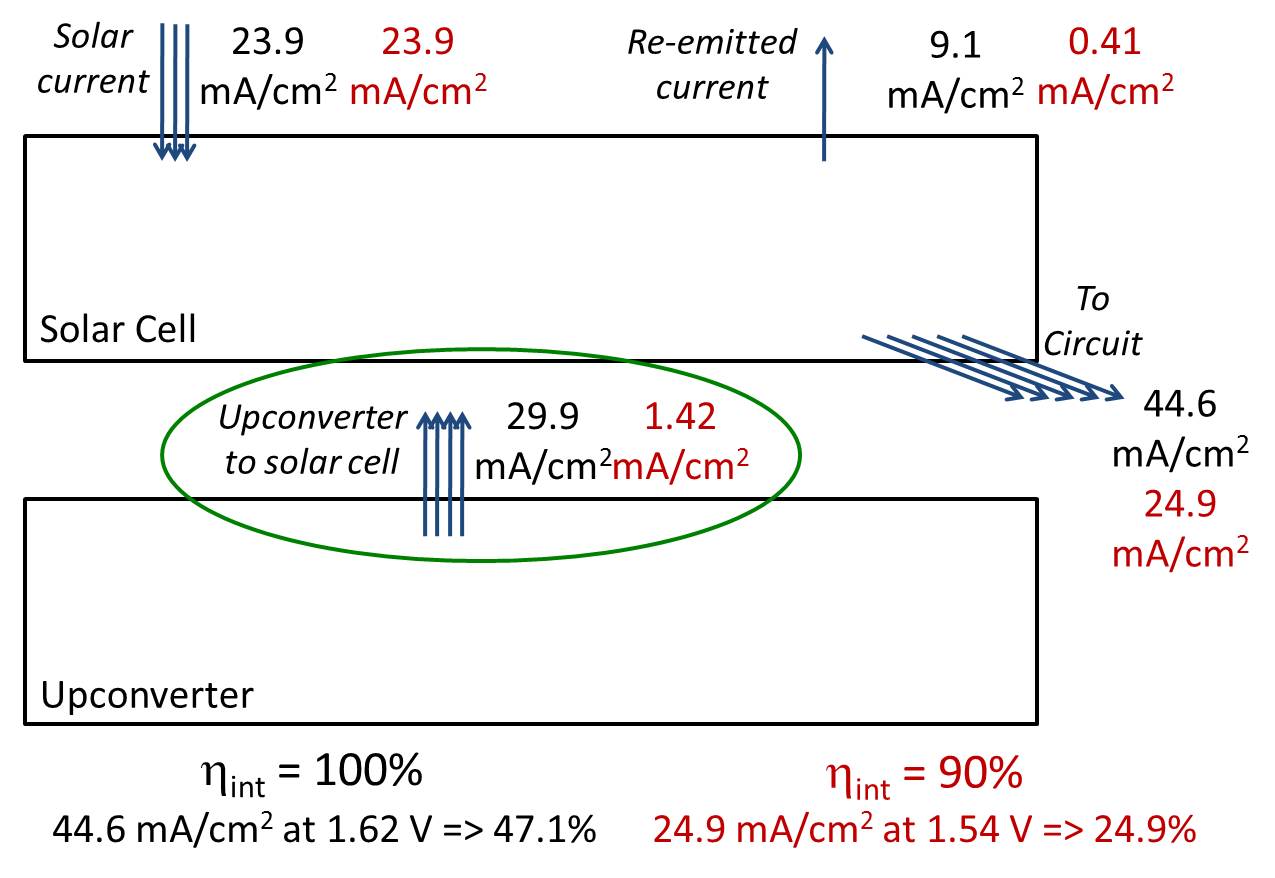}
\caption{Operating electronic and photonic currents for a non-concentrating up-converting system, for a solar cell with bandgap $E_g=2eV$, and an impurity level in the up-converter.} 
\label{fig:UpConvCurrents}
\end{figure}

The extraordinary loss of efficiency fundamentally results from the use of the up-converter as a light-emitting diode, driven by the sun.  Fig.~\ref{fig:UpConvCurrents} shows the relevant currents at operating point for the non-concentrating, impurity band up-converter system, at both $\eta_{int} = 100\%$ and $\eta_{int}=90\%$.  The bandgap is taken to be $2eV$, the optimal bandgap under ideal conditions.  Under ideal conditions, the up-converter provides significant benefit to both the current and the voltage, through luminescent exchange with the solar cell.  Whereas a stand-alone $2eV$ solar cell would typically miss the substantial current available from photons below the bandgap, the up-converter successfully sends almost $30mA/cm^2$ back to the solar cell, while also emitting sub-bandgap photons as thermodynamically required.  However, this luminescent emission is extremely susceptible to loss.

As emphasized in Chap.~\ref{chap:SQLimit}, there is a voltage penalty of $kT\ln \eta_{ext}$ for poor luminescent extraction in conventional single-junction solar cells, resulting in a sharp voltage reduction for relatively small imperfections.  The penalty for the up-converter system is actually even more severe.  The luminescence of the up-converter depends linearly on the extraction efficiency (by definition), such that imperfect light extraction results in a penalty of $\eta_{ext}$ in the emission.  This luminescence contributes directly to both the current and voltage of the solar cell.  Therefore the penalty in efficiency is proportional to $\eta_{ext} kT \ln \eta_{ext}$, quickly eroding the advantage of the up-converter.  In the example of Fig.~\ref{fig:UpConvCurrents}, even at $90\%$ internal luminescent yield, the external luminescence (in the plane-parallel case) is only about $6\%$, implying a decrease in the luminescence of more than a factor of $15$.  For a $6000K$ blackbody, there are more than $80mA/cm^2$ of incident photons with energy less than $2eV$, yet even this small degradation of the system results in a photon current of only $1.42 mA/cm^2$ generated from the up-converter.

Through understanding the luminescence of the up-converter one can also understand the trends in Fig.~\ref{fig:UpConvEff}.  At $\eta_{int}=90\%$, internal re-emission still provides a mechanism for exploring the internal phase space without any random reflections.  For $\alpha L = 1$, the external yield of a textured cell is $\approx 15\%$, compared to the $6\%$ described previously for the plane-parallel geometry.  Texturing helps, but only to a small extent in this scenario.  

The bandgap dependence of the efficiency penalty is also explainable.  In contrast to the single-junction solar cells of Chap.~\ref{chap:SQLimit} or the multiple exciton solar cells of Sec.~\ref{sec:CarrMult}, the penalty associated with imperfect luminescent yield actually increases with bandgap.  This is because the high-bandgap solar cells depend critically on receiving up-converted light, representing the photons missed because they were smaller than the bandgap.  As the external extraction diminishes rapidly to $\simeq 1\%$, the up-converting system reverts essentially to a single-junction solar cell, with little benefit at small bandgaps and the attendant penalty at higher bandgaps.

\section{Multi-junction}

\begin{figure}
\centering
\includegraphics[width=3in]{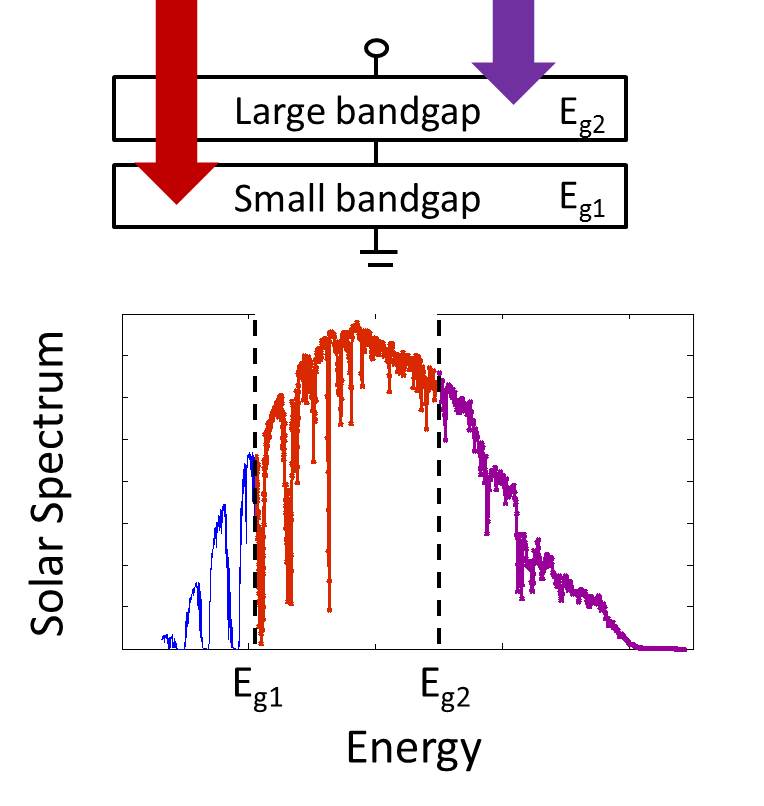}
\caption{Multi-junction solar cells more efficiently convert solar energy, by extracting higher voltage from the high-energy photons with energy $E>E_{g2}$.} 
\label{fig:MultiJuncSchematic}
\end{figure}

Multi-junction solar cells are a third technology that can overcome the $33.5\%$ single-junction efficiency limit.  One of the fundamental loss mechanisms of the single-junction solar cell is the energy that goes to thermalization after a high-energy photon has been absorbed.  Counter-balancing the pressure to increase the bandgap is the loss that occurs to not absorbing photons below the bandgap, leading to an ideal intermediate bandgap in the infrared ($1.1$-$1.4eV$ range) as seen in Fig.~\ref{fig:EfficiencyBandgap}.  

Alternatively, one can imagine splitting the solar spectrum and sending it to multiple separate cells, as in Fig.~\ref{fig:MultiJuncSchematic}.  In Fig.~\ref{fig:MultiJuncSchematic}, photons with energy $E$ less than the small bandgap, $E_{g1}$ still are not absorbed.  However, photons with energy $E_{g1}<E<E_{g2}$ are absorbed by the bottom cell and thermalize to the $E_{g1}$ band-edge, while photons with energy $E>E_{g2}$ are absorbed by the top cell and thermalize to the $E_{g2}$ band-edge.  In this arrangement, less energy is lost to thermalization, due to the two band-edges in the system.  Out of the great variety of solar cell techniques, the highest-efficiency photovoltaic systems are three-junction solar cells with concentration factors of $\approx 400$, reaching efficiencies of $43.5\%$ \cite{Green2012}.

Here we will focus our analysis on two-junction solar cells, again attempting to understand the robustness of detailed balance efficiency calculations.  The two-junction analysis proceeds similar to that of the previous configurations.  There are many possible spectrum splitting configurations, which can yield very different efficiency calculations \cite{Vos1980,Baruch1995,Kurtz2008}.  We will assume the larger bandgap material is sitting directly on the smaller bandgap material, and that each has a refractive index $n_r=3.5$.  The two junctions are in series and current-matched, as the majority of experimental systems are.  A perfect rear mirror completes the system.  The sun is treated as a blackbody at $5780K$.

The currents through the two junctions are the differences between the carrier generation and recombination rates:
\begin{align}
\label{eq:MultiJuncJ1}
	J_1 & = \int_{E_{g1}}^{E_{g2}} a_1(E) S(E) \mathrm{d}E - \frac{\pi}{\eta_{ext}^{(1)}} e^{qV_1/kT} \int_{E_{g1}}^{\infty} a_1(E)b(E) \mathrm{d}E + \pi n_r^2 e^{qV_2/kT} \int_{E_{g2}}^{\infty} a_1(E) a_2(E) b(E) \mathrm{d}E \\
	J_2 & = \int_{E_{g2}}^{\infty} a_2(E) S(E) \mathrm{d}E - \frac{\pi}{\eta_{ext}^{(2)}} e^{qV_2/kT} \int_{E_{g2}}^{\infty} a_2(E)b(E) \mathrm{d}E
\label{eq:MultiJuncJ2}
\end{align}
where $J_1$ is the current through the smaller bandgap junction and $J_2$ is the current through the larger bandgap junction.  In Eqns.~\ref{eq:MultiJuncJ1}, \ref{eq:MultiJuncJ2}, the first term represents absorption of the incident sunlight, while the second term represents total recombination (radiative and non-radiative).  The current $J_1$ also has a third term, representing the radiative transfer from the larger bandgap solar cell to the smaller one.  

This radiative transfer not only leads to extra current for the smaller bandgap solar cell, it also leads to different external yields $\eta_{ext}^{(1)}$ and $\eta_{ext}^{(2)}$.  The internal yield for the smaller bandgap cell, $\eta_{ext}^{(1)}$, is exactly the same as the external yield from a typical plane-parallel solar cell with perfect rear mirror, given by Eqn.~\ref{eq:etaExtPP}.  The external yield of the larger bandgap solar cell is smaller, because photons escape through the rear of the junction and are absorbed in the other cell, instead of being re-directed back into the cell by a mirror.  To calculate the yield, we can use Eqn.~\ref{eq:etaExtAExt}, with $\Omega_{ext}=\pi(1+n_r^2)$ and $\Omega_f = \pi$, giving:
\begin{equation}
\label{eq:etaExt2}
\eta_{ext}^{(2)} = \frac{a_2}{ a_2(1+n_r^2)+4n_r^2\alpha L_2 \frac{1-\eta_{int}}{\eta_{int}}}
\end{equation}
where $\alpha L_2$ is the optical thickness of the cell.  The internal yield $\eta_{int}$ is taken to be the same for each cell.  The refractive index factor in the denominator of Eqn.~\ref{eq:etaExt2} accounts for the extra density of states leading to enhanced emission through the rear rather than the top of the higher bandgap cell.  To achieve full absorption, the optical thicknesses of the two cells are taken to be $\alpha L_1=2.5$ and $\alpha L_2=5$, respectively.

\begin{figure}
\centering
\includegraphics[width=6.5in]{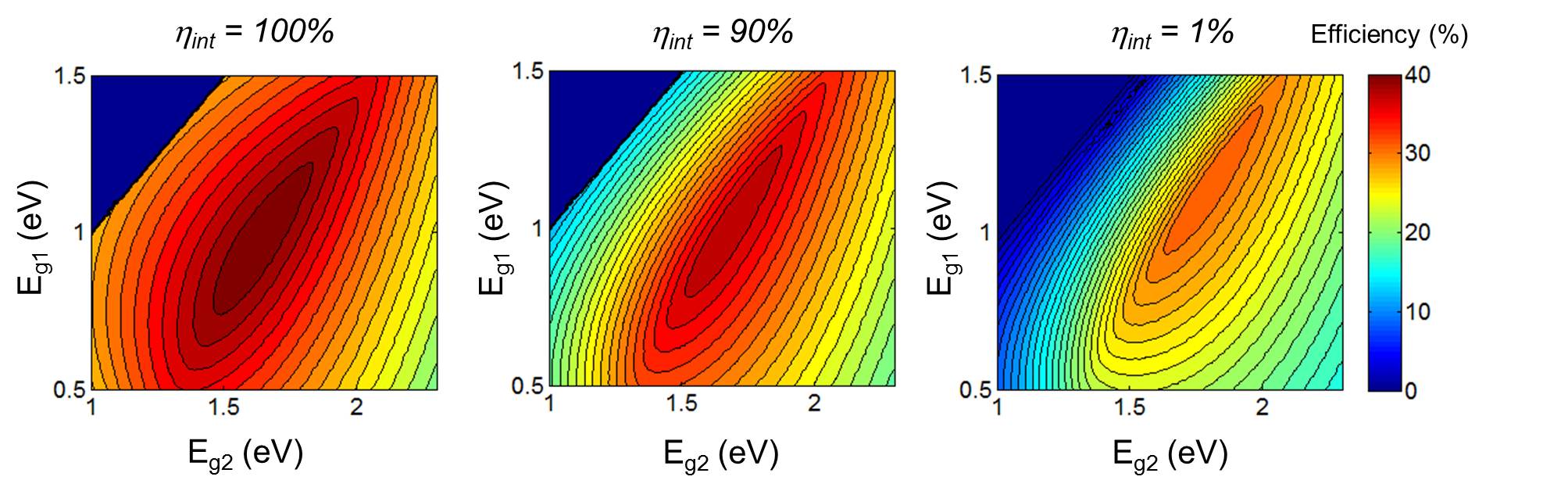}
\caption{Two-junction solar cell efficiencies as a function of the two bandgaps, $E_{g1}$ and $E_{g2}$., for internal yield $\eta_{int} = 100\%, 90\%, 1\%$.  The optimal bandgap increases as the internal yield decreases.  At larger bandgaps, the efficiency is relatively robust to imperfect light extraction.} 
\label{fig:MultiJuncEffBg}
\end{figure}

To calculate the efficiency as a function of the two bandgaps, a system of equations must be solved simultaneously.  First, the currents through the two cells must be equal, for the current-matching requirement.  Second, at the operating point the power through the cell should be maximized.  This can be represented mathematically as an optimization problem:
\begin{equation}
\label{eq:optEqn}
\begin{aligned}
& \underset{V_1, V_2}{\text{maximize}}
& & P(V_1,V_2) = J_2(V_2)(V_1+V_2) \\
& \text{subject to}
& & J_1(V_1,V_2) = J_2(V_2) 
\end{aligned}
\end{equation} 
where the power $P$ to be maximized could have also been written $J_1(V_1+V_2)$, because $J = J_1 = J_2$.  We can convert Eqn.~\ref{eq:optEqn} into an unconstrained optimization through Lagrangian calculus
\begin{equation}
\label{eq:optLagrangian}
\begin{aligned}
& \underset{V_1, V_2, \lambda}{\text{maximize}}
& & F(V_1,V_2,\lambda) = J_2(V_2)(V_1+V_2) + \lambda \left[ J_1(V_1,V_2) - J_2(V_2) \right]
\end{aligned}
\end{equation} 
resulting in three equations by setting $\partial F / \partial V_1$, $\partial F / \partial V_2$, and $\partial F / \partial \lambda = 0$.  Solving for $\lambda$ results in the final two-equation system enabling calculation of the operating point for each possible pair of bandgaps
\begin{subequations}
\begin{align}
J_1(V_1,V_2) & = J_2(V_2) \\
\frac{\partial P(V_1,V_2)}{\partial V_2} \frac{\partial J_1(V_1,V_2)}{\partial V_1} & = \frac{\partial P(V_1,V_2)}{\partial V_1} \left[ \frac{\partial J_1(V_1,V_2)}{\partial V_2} - \frac{\partial J_2(V_2)}{\partial V_2} \right]
\end{align}
\end{subequations}

\begin{table}[ht]
\centering
\begin{tabular}{cc}
\includegraphics[width=3.5in]{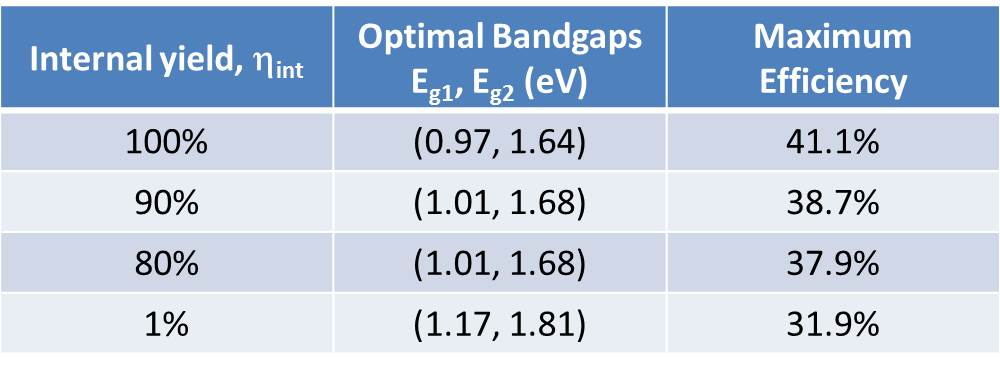}
\end{tabular}
\caption{Table of optimal bandgap and maximum efficiency for two-junction solar cells with varying $\eta_{int}$.  Whereas the ideal bandgaps in the Shockley-Queisser radiative limit are $1.64eV$ and $0.97eV$, at e.g. $\eta_{int} = 1\%$ the ideal bandgaps are $1.81eV$ and $1.17eV$. }
\label{tab:MultiJuncEffTable}
\end{table}

The calculated efficiency as a function of the two junction bandgaps is shown in Fig.~\ref{fig:MultiJuncEffBg}, for three different values of the internal yield: $\eta_{int} = 100\%, 90\%$, and $1\%$.  The efficiency for $\eta_{int} = 100\%$ is relatively large over a broad range of bandgaps for each junction.  Per Table~\ref{tab:MultiJuncEffTable}, the optimal bandgaps at $100\%$ internal yield are $E_{g1}=0.97eV$ and $E_{g2} = 1.64eV$.  Moving to $\eta_{int}=90\%$, the optimal bandgaps shift higher to $1.01eV$ and $1.68eV$, respectively.  At $1\%$ internal yield, the optimal bandgaps move all the way to $1.17eV$ and $1.81eV$, where the solar cell is able to achieve almost $32\%$ efficiency.  

\begin{figure}
\centering
\includegraphics[width=2.8in]{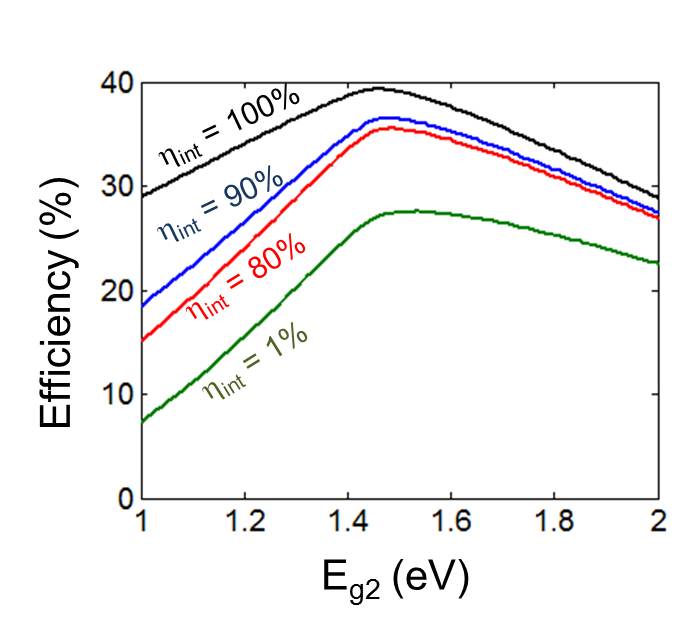}
\caption{Two-junction solar cell efficiencies as a function of the $E_{g2}$, for $E_{g1}$ fixed at $0.7eV$ and for internal yield $\eta_{int} = 100\%, 90\%, 80\%,$ and $1\%$.  The difference in robustness to internal yield for larger bandgaps versus smaller bandgaps is clear.} 
\label{fig:MultiJuncEffEg2}
\end{figure}

As with single-junction solar cells, multi-junction solar cells with larger bandgaps are more robust to imperfect light extraction.  A penalty of $kT\ln\eta_{ext}$ is relatively greater for a small bandgap junction, resulting in the significant penalty as seen in Fig.~\ref{fig:MultiJuncEffEg2}.  For $E_{g1}=0.7eV$, one can see that for $\eta_{int} = 100\%$, $E_{g2}=1eV$ and $E_{g2}=2eV$ produce almost exactly the same efficiency, close to $30\%$.  Hover, at $\eta_{int}=1\%$, $E_{g2}=2eV$ generates almost $15\%$ greater absolute efficiency than $E_{g2}=1eV$, a drastic difference that illustrates the relative robustness of larger bandgaps.

%% file: Algorithm.tex
\chapter{A New Photonic Inverse Design Method}
\label{chap:Algorithm}
\epigraph{God does not care about our mathematical difficulties.  He integrates empirically.}{Albert Einstein}
\newcommand{\epsr}{\overline{\overline{\mathbf{\epsilon}}}}
\newcommand{\mur}{\overline{\overline{\mathbf{\mu}}}}
\newcommand{\Eadj}{E^{\text{A}}}
\newcommand{\Eb}{\mathbf{E}}
\newcommand{\Hb}{\mathbf{H}}

\noindent
Maxwell's equations are the fabric of electromagnetism.  Once formulated, they were the basis for predicting that light is an electromagnetic phenomena, and they have enabled every photonic technology since.  For a scientist or engineer working with photons (of any frequency), proficiency with Maxwell's equations is a requirement.

Consequently, there has been considerable research into fast and efficient computational solutions of Maxwell's equations.  An assortment of methods, from finite-difference time-domain (FDTD) to boundary element method (BEM), have emerged as viable techniques.  They are all meant to solve the \emph{forward problem}: for a given geometry, what are the resulting fields and/or eigen-frequencies?

However, scientists and engineers are typically more interested in the \emph{inverse problem}: for a desired electromagnetic response, what geometry is needed?  Answering this question breaks from the emphasis on the forward problem and forges a new path in computational electromagnetics.

The inverse problem is more difficult than the forward problem.  For the forward problem, with a given geometry and set of sources, physical reality demands that there must be one and only one solution.  However, one cannot specify a set of fields and expect for a geometry to exist or, if it exists, to be unique.  To overcome these issues, it is common to attempt a slightly modified version of the inverse problem: what geometry \emph{most closely} achieves the desired electromagnetic response? 

From this viewpoint, inverse design problems become optimization problems, with the desired functionality subject to the constraint that all fields and frequencies must be solutions of Maxwell's equations.  Mathematically, the general electromagnetic inverse design problem can be written:
\begin{equation}
\label{eq:ConstrPDE}
\begin{aligned}
& \underset{\epsr, \mur}{\text{maximize}}
& & F(\mathbf{E},\mathbf{H},\omega) \\
& \text{subject to}
& & \nabla \cdot \epsr \mathbf{E} = \rho \\
& & & \nabla \cdot \mur \mathbf{H} = 0 \\
& & & \nabla \times \mathbf{E} = -j\omega\mur \mathbf{H} \\
& & & \nabla \times \mathbf{H} = \mathbf{J} + j\omega\epsr \mathbf{E}
\end{aligned}
\end{equation} 
in which the objective is to find the permittivity tensor $\epsr$ and the permeability tensor $\mur$ that maximize the function $F(\mathbf{E},\mathbf{H},\omega)$, subject to the four Maxwell partial differential equations (PDEs).  This approach to inverse design is referred to as \emph{PDE-constrained optimization}.  We will concentrate on time-harmonic problems (with time-dependence $e^{j\omega t}$), but inverse design for time-dependent problems is discussed in Appendix~\ref{chap:TimeDep}.\footnote{An even more general function would also include a dependence on the geometry itself, through $\epsr$ and $\mur$.  This would be  particularly appropriate if a \emph{regularization} term is included, to impose constraints on the geometry.  The derivatives, however, are straightforward and therefore not explicitly included.}  

There are many possible approaches to the photonic inverse design problem, Eqn.~\ref{eq:ConstrPDE}.  The measure of any approach is its efficiency and effectiveness in arriving at a design.  For photonic design problems, the most time-consuming part of the inverse design is the solution of the forward problem, which is generally repeated for many different designs.  The number of simulations required by an approach is therefore a good measure of its efficiency. 

Because the number of simulations is a primary constraint, stochastic algorithms (i.e. genetic algorithms, simulated annealing) are generally unfeasible approaches.  Genetic algorithms have been applied to optimization of two-dimensional problems \cite{Goh2007,Hakansson2006,Poletti2005,Triltsch2006} and computationally small three-dimensional problems \cite{Rinne2008}.  However, the technique usually requires tens of thousands of simulations, rendering it unrealistic for three-dimensional optimizations with a sizeable parameter space.

In contrast, \emph{shape calculus} provides a framework for efficient optimization with many fewer simulations.  Shape calculus is the analogue of the usual differential calculus of finite-dimensional spaces, expanded to the infinite-dimensional space of shape and topology.  It turns out that only two simulations are required to compute the derivative of an objective function with respect to the entire infinite-dimensional space; moreover, typically fewer than a hundred iterations are required to converge to an optimal design.  Shape calculus often enables a 100- or 1000-fold decrease in the number of simulations required, compared with stochastic methods.

This chapter describes our new method for photonic inverse design.  Shape calculus, the mathematical basis of the design methodology, comprises Sec.~\ref{sec:ShapeCalc}.  An intuitive physical description of the method is presented, demonstrating how shape calculus computes derivatives in the most efficient way possible.  Sec.~\ref{sec:InfoLoss} is a brief discussion of how the efficiency is possible, through an informatic lens.  The final two sections discuss practical implementation considerations, with Sec.~\ref{sec:LevelSet} demonstrating the utility of level set geometric representations, and Sec.~\ref{sec:CompImpOpt} discussing the computational algorithm.  Symmetries of the Maxwell's equations turn out to be a crucial aspect of the method, and are derived and discussed in Appendix~\ref{chap:SymmMaxwell}. 

\section{Shape Calculus}
\label{sec:ShapeCalc}

Shape calculus can be derived in abstract form for rather general linear differential operators \cite{Delfour1988,Cea1973,Cea2000,Sokolowski1999,Giles2000,Hackl2007,Herzog2010}.  However, in this thesis we will take the opposite approach.  We will focus explicitly on electromagnetic applications governed by Maxwell's equations, which will enable an intuitive physical understanding of shape calculus to emerge.

We will start with a simple two-dimensional example, removing the vector nature of the problem to clarify the underlying optimization mechanism.  We will then show how this extends to the more general and useful class of three-dimensional inverse design problems, with small but important modifications to the picture.

\subsection{Simple example}
\begin{figure}[]
\centering
\includegraphics[height=2.5in]{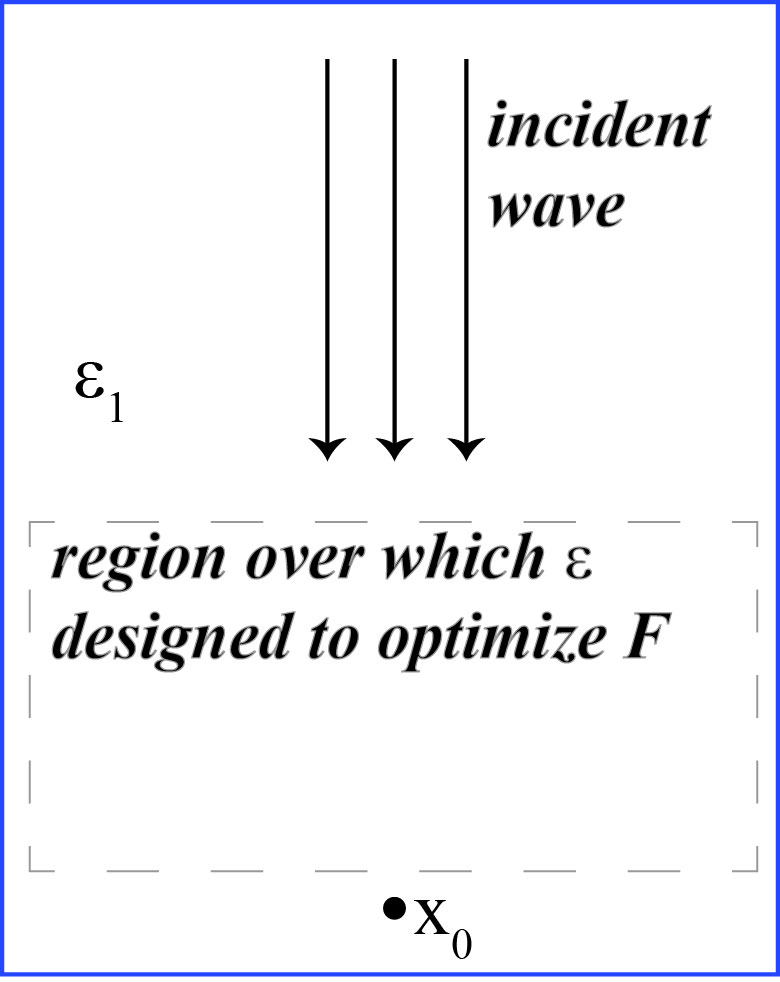}
\caption{Setup for a simple example demonstrating the intuition behind shape calculus.  In this example a wave is assumed incident from the top.  There is a region over which the permittivity is to be designed, to maximize a ``merit function'' $F$.  $F$ is taken to be the field concentration at $x_0$.}
\label{fig:OptSimpleExample} 
\end{figure}

It is useful to consider first a simple example that demonstrates the essence of shape calculus-based optimization.  Fig.~\ref{fig:OptSimpleExample} illustrates the setup.  Translational symmetry is assumed in the out-of-plane dimension, reducing it to a two-dimensional problem.  A wave is incident from the top, with a transverse electric (TE) polarization such that the electric field is a scalar everywhere.  The permittivity and permeability are also assumed scalars.  The objective is to maximize the electric field intensity at some point $\mathbf{x_0}$, so that a ``merit function'' $F = \frac{1}{2}|E(\mathbf{x_0})|^2$ is suitable (the factor of $1/2$ is for normalization purposes).

Consider now the effect of introducing a small dielectric inclusion, with radius $r$ and dielectric constant $\epsilon_2$, at some $\mathbf{x'}$.  Is this a smart placement?  To decide, we must consider its effect on the merit function.  We try to calculate the \emph{change} in the merit function; if the change is positive, the inclusion is a good choice, whereas if it is negative then the inclusion is a poor choice.  The change in the merit function $\delta F$, also called the \emph{variation} in $F$, is simply the difference between the merit function of the new geometry  and the merit function of the old geometry:
\begin{equation}
\label{eq:deltaFscalar}
\delta F = \frac{1}{2} | E^{\text{new}}(\mathbf{x_0})|^2 - \frac{1}{2}|E^{\text{old}}(\mathbf{x_0})|^2
\end{equation}
The new electric field can be written as the sum of the electric electric field and the difference between the two: $E^{\text{new}}(\mathbf{x_0}) = E^{\text{old}}(\mathbf{x_0}) + \delta E(\mathbf{x_0})$.  Because the dielectric inclusion has been assumed small, one can expand Eqn.~\ref{eq:deltaFscalar} and ignore terms that are second order (and potentially higher-order for other $F$) in $\delta E$:
\begin{subequations}
\label{eq:linearization}
\begin{align}
\delta F & = \frac{1}{2} \left[ E^{\text{old}}(\mathbf{x_0}) + \delta E(\mathbf{x_0}) \right]\overline{\left[E^{\text{old}}(\mathbf{x_0}) + \delta E(\mathbf{x_0})\right]} - \frac{1}{2} E^{\text{old}}(\mathbf{x_0})\overline{E^{\text{old}}(\mathbf{x_0})} \\
& = \frac{1}{2}\left[ E^{\text{old}}(\mathbf{x_0})\overline{\delta E(\mathbf{x_0})} + \overline{E^{\text{old}}(\mathbf{x_0})}\delta E(\mathbf{x_0}) + |\delta E(\mathbf{x_0})|^2 \right] \\
& \approx \operatorname{Re}\left[\overline{E^{\text{old}}(\mathbf{x_0})}\delta E(\mathbf{x_0})\right]
\end{align}
\end{subequations}
where the overline denotes complex conjugation.  The step from Eqn.~\ref{eq:linearization}(b) to Eqn.~\ref{eq:linearization}(c) is called \emph{linearization}: the higher-order terms are ignored, which is valid for small geometrical changes.  This process enables optimization even for non-linear electromagnetics problems, by linearizing around the local geometry (and the local field solution) and iteratively making small changes.

\begin{figure}[]
\centering
\includegraphics[width=6in]{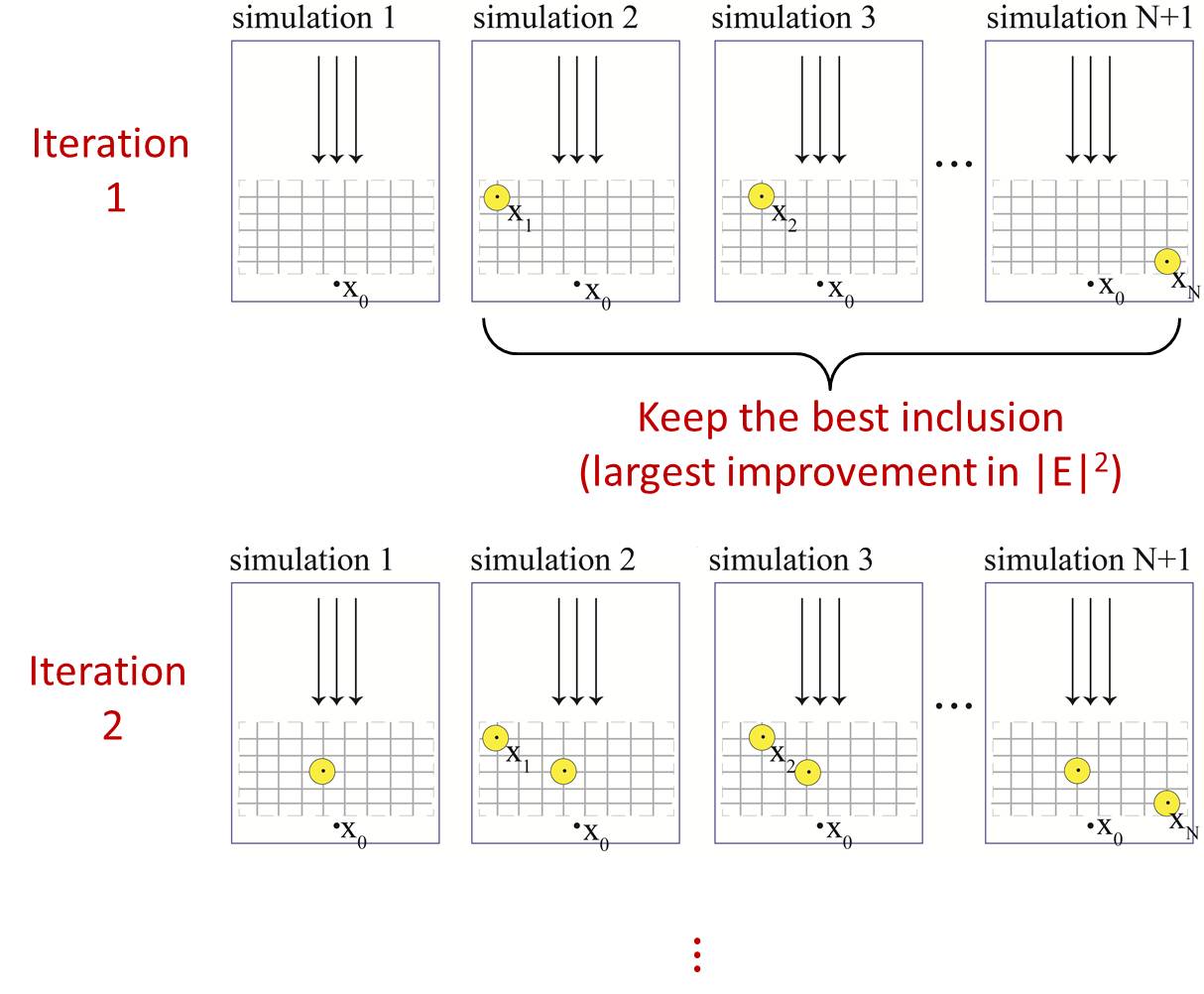}
\caption{Brute force optimization method.  Starting from an area with background permittivity $\epsilon_1$, new inclusions with permittivity $\epsilon_2$ are tested one-by-one.  The best inclusion is kept, and the process iterates many times until a final solution is found.  The shape calculus optimization method yields the same information with only two simulations.} 
\label{fig:BruteForce}
\end{figure}

Although linearization simplifies the form of $\delta F$, there is still not a clear optimization method.  Fig.~\ref{fig:BruteForce} depicts the obvious candidate, a brute force optimization.  A brute force optimization would proceed as follows: from an initial geometry, test a separate inclusion at every possible $\mathbf{x'}$.  Calculate $\delta F$ for each separate inclusion, and update the geometry with the inclusion that had the largest $\delta F$ (i.e. maximally increased the merit function).  Upon updating the geometry, iterate this process, again checking every possible inclusion but with the new geometry.  This somewhat exhaustive search, however, requires far too many simulations to be possible.  Instead, \emph{shape calculus gathers as much information as the brute force method, while requiring only two simulations per iteration.}  The key to the reduction in simulation time is exploitation of symmetry.

The change in the electric field at $\mathbf{x_0}$ between the two geometries, $\delta E(\mathbf{x_0})$, arises from the interaction of the dielectric inclusion inserted at $\mathbf{x'}$.  The original geometry had some field $E^{\text{old}}(\mathbf{x'})$ at $\mathbf{x'}$.  When a cylindrical dielectric inclusion in inserted, the effect is the addition of an induced dipole moment $p^{\text{ind}}$ at $\mathbf{x'}$, with amplitude \cite{Jackson1999}
\begin{equation}
\label{eq:IndDipScalar}
p^{\text{ind}} \simeq \pi r^2 (\epsilon_2-\epsilon_1) E^{\text{old}}(\mathbf{x'})
\end{equation}
To find $\delta E(\mathbf{x_0})$, one must know the field at $\mathbf{x_0}$ from the induced dipole at $\mathbf{x'}$.  The field at $\mathbf{x_0}$ from a dipole of unit amplitude at $\mathbf{x'}$ is given by the Green's function $G(\mathbf{x'},\mathbf{x_0})$.\footnote{Note that this normalization is different from the standard Green's function normalization, for which the source is $p=1/\mu_0 \omega^2$ instead of $p=1$, and is chosen for its notational simplicity in optimization applications.}  The full $\delta E$ is then given by the product of the Green's function with the induced dipole moment, i.e. $\delta E(\mathbf{x_0}) = G(\mathbf{x_0},\mathbf{x'})p^{\text{ind}}(\mathbf{x'})$.  

The formula for the change in merit function from introducing a dielectric inclusion at $\mathbf{x'}$ is now
\begin{equation}
\label{eq:deltaFint1}
\delta F = \operatorname{Re}\left[\overline{E^{\text{old}}(\mathbf{x_0})} G(\mathbf{x_0},\mathbf{x'})p^{\text{ind}}(\mathbf{x'}) \right]
\end{equation}
$E^{\text{old}}(\mathbf{x_0})$ and $p^{\text{ind}}(\mathbf{x'})$ can be calculated in a single simulation.\footnote{Eqn.~\ref{eq:IndDipScalar} dictates that $p^{\text{ind}}$, to first order, only depends on $E^{\text{old}}(\mathbf{x'})$, which is calculated in the same simulation as $E^{\text{old}}(\mathbf{x_0})$.}  The difficulty with using Eqn.~\ref{eq:deltaFint1} for the brute force optimization method is that a different simulation must be completed for every possible $\mathbf{x'}$, in order to calculate the Green's function $G(\mathbf{x_0},\mathbf{x'})$. 

This is where symmetry enters.  The Green's function for the Maxwell operator (as well as any linear differential operator) is symmetric: $G(\mathbf{x_0},\mathbf{x'}) = G(\mathbf{x'},\mathbf{x_0})$.  Substituting the new form of the Green's function into Eqn.~\ref{eq:deltaFint1} and re-arranging:
\begin{subequations}
\label{eq:deltaFint2}
\begin{align}
\delta F & = \operatorname{Re} \left[\overline{E^{\text{old}}(\mathbf{x_0})} G(\mathbf{x'},\mathbf{x_0})p^{\text{ind}}(\mathbf{x'}) \right] \\
& = \operatorname{Re} \left[G(\mathbf{x'},\mathbf{x_0}) \overline{E^{\text{old}}(\mathbf{x_0})} p^{\text{ind}}(\mathbf{x'}) \right]
\end{align}
\end{subequations}
Notice now that the first two terms in Eqn.~\ref{eq:deltaFint2}(b), $G(\mathbf{x'},\mathbf{x_0})\overline{E^{\text{old}}(\mathbf{x_0}})$, are \emph{exactly the fields of a dipole driven with amplitude} $\overline{ E^{\text{old}}(\mathbf{x_0}) }$.  Importantly, this dipole is driven at $\mathbf{x_0}$, radiating (or providing a local near field) to $\mathbf{x'}$.  If we define the \emph{adjoint} electric field
\begin{align}
\label{eq:adjointScalar}
\Eadj(\mathbf{x'}) = G(\mathbf{x'},\mathbf{x_0}) \overline{E^{\text{old}}(\mathbf{x_0})}
\end{align}
then we now have the simple formula
\begin{align}
\label{eq:deltaFadjoint}
\delta F = \Eadj (\mathbf{x'}) p^{\text{ind}}(\mathbf{x'})
\end{align}

\begin{figure}[]
\centering
\includegraphics[width=\textwidth]{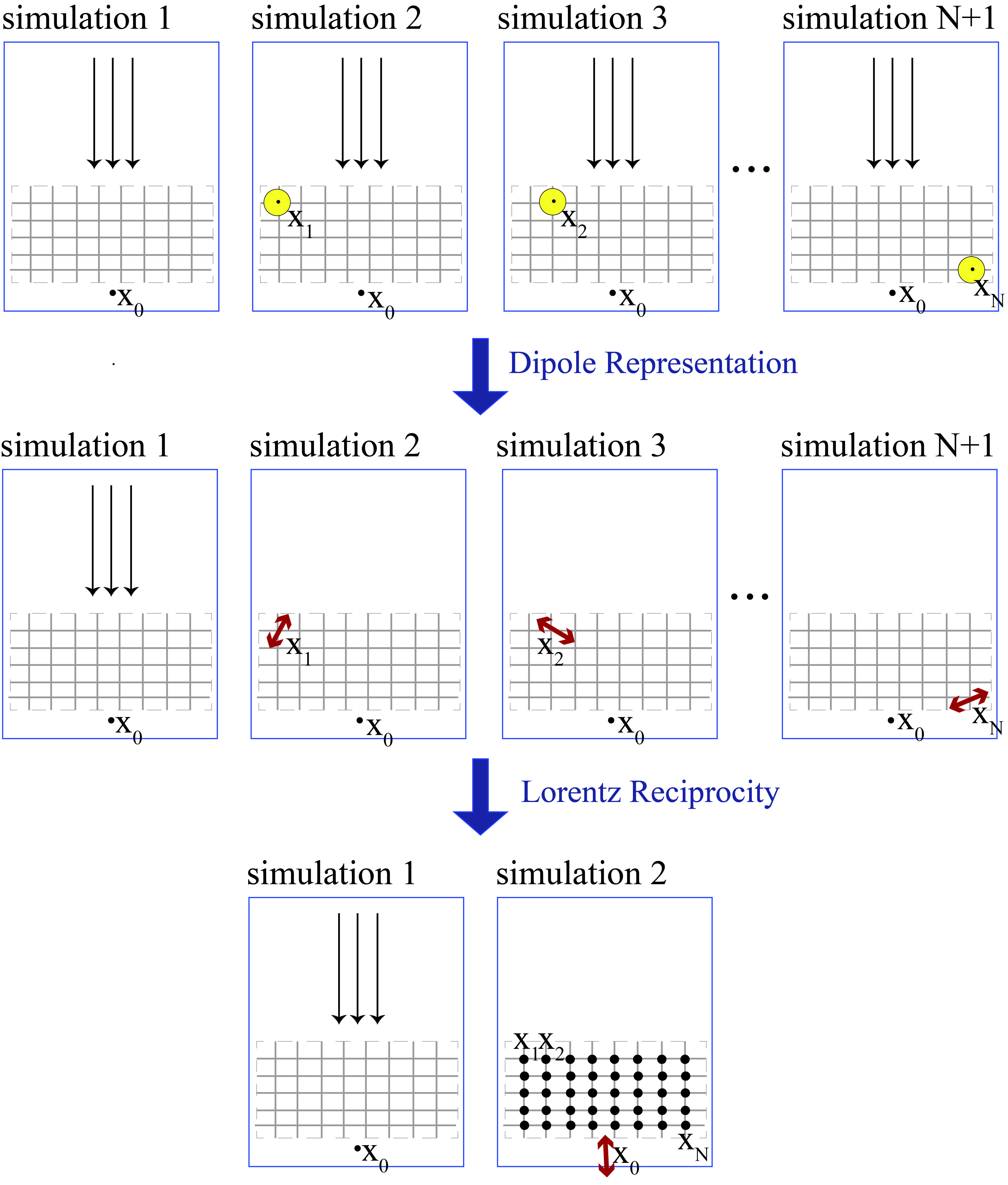}
\caption{Reduction from the $N+1$ simulations per iteration of the brute force method to the $2$ simulations per iteration of the shape calculus approach.  As long as the inclusions are small, they can be represented by their electric dipole moments.  Symmetry then enables the dipole source and measurement points to be switched, reducing the number of simulations to two, regardless of the size $N$ of the parameter space. } 
\label{fig:SimpExRed}
\end{figure}

Eqn.~\ref{eq:deltaFadjoint} is the equation we have been driving for.  In Fig.~\ref{fig:OptSimpleExample}, consider starting with a geometry of vacuum only.  First, simulate the electric field of that structure.  This gives the potential induced polarization ($p^{\text{ind}}$) for all possible inclusion locations $\mathbf{x'}$, from Eqn.~\ref{eq:IndDipScalar}.  Moreover, $E^{\text{old}}(\mathbf{x_0})$ is now known.  Therefore, in a second simulation, drive an electric dipole at $\mathbf{x_0}$ with amplitude $\overline{E^{\text{old}}(\mathbf{x_0})}$, as dictated by the equation for the adjoint field (Eqn.~\ref{eq:adjointScalar}).  The single dipole simulation provides $\Eadj(\mathbf{x'})$ for all $\mathbf{x'}$ as well.  Thus, $\delta F$ is now known for every possible $\mathbf{x'}$, with only two simulations.  Fig.~\ref{fig:SimpExRed} is a pictorial representation of the reduction in number of simulations.

The path to a gradient-based optimization method is clear.  Starting from an initial structure, two simulations (what we will call the ``direct'' and the ``adjoint'') are needed to understand the shape gradient.  The shape gradient is a map over the entire designable domain, dictating how promising every possible dielectric inclusion is.  Once the optimal inclusion has been chosen, the process iterates until an optimal design is reached.  

Note that in this section the derivative of the continuous design problem, rather than of the discretized problem, was calculated.  In the limit of infinite resolution the two derivatives are equivalent, but otherwise there are slight differences.  For electromagnetic inverse design problems, cf. \cite{Chen2003,Seliger2006,Jensen2011} for the discretize-then-differentiate formulation.  For a discussion of the discretize-then-differentiate versus the differentiate-then-discretize formulations within the context of aerodynamic inverse design problems, cf. \cite{Peter2010}.

The example chosen for this section was an unrealistically simple optimization problem.  While the next section will cover in depth the more general case, it is worth discussing some of the generalizations briefly here.

First, it seems fortuitous to have chosen to optimize the field intensity at only a single point, $\mathbf{x_0}$.  If the field intensity over a large region is to be optimized, won't many adjoint dipole simulations be required?  In fact, still only one adjoint simulation is needed.  Because the equation for $\delta F$, by construction, is linear in the term $G(\mathbf{x'},\mathbf{x_0})\overline{E^{\text{old}}(\mathbf{x_0})}$, each dipole should be included in the same simulation.

The persistent use of inclusions may seem irrelevant as well, as one typically is not optimizing for a cluster of cylindrical objects, but rather more general shapes.  However, the purposeful use of the inclusion captures the essence of the method, which is reducing geometric changes to their induced dipole moments and then exploiting their symmetry.  Any arbitrary boundary movement can also be approximated as a set of induced dipole moments, seamlessly extending the method of this section to more general optimizations.  

Finally, the simplification to scalar fields seems to have played a role in the derivation, as the re-ordering of fields from Eqn.~\ref{eq:deltaFint2}(a) to \ref{eq:deltaFint2}(b) would not be possible for vector fields.  Indeed, this is not a trivial step and is discussed further in the next section.

\subsection{General case}
In the previous section, the shape calculus method was distilled to its most basic form.  For any given geometry, a change in shape (or topology) introduces a polarization density where the change occurs.  Through symmetry, the Green's function connecting the dipole moments to the measurement points (i.e. locations for optimization) can be reversed, resulting in one extra simulation that captures the necessary information.  The resulting map dictates how the geometry should be changed, and the process iterates until an optimal design is reached.

The formulation is now generalized to a much larger class of three-dimensional photonic inverse design problems.  The electromagnetic fields are treated as vector fields, while tensor permittivity and permeability are allowed.  The merit function can be any function of the fields $F(\mathbf{E},\mathbf{H})$, instead of the electric field intensity at a single point.  Note that merit functions of the eigen-frequencies can also be optimized, but have substantially different derivations.  Their treatment can be found in, e.g., \cite{Cox1999,Cox2000,Kao2005}.  The time-dependent inverse design formalism is derived in Appendix~\ref{chap:TimeDep}.  The permeability $\mu$ will be considered fixed and non-designable, although the derivation proceeds along very similar lines; moreover, the derivatives could be obtained without any extra work through the duality relations \cite{Rothwell2001}.

The geometric constraints are also loosened.  In the previous example only dielectric inclusions were discussed, as their induced dipole moment was simple and well-known.  In the applied mathematics literature, finding the change in a merit function due to a new inclusion is known as taking the \emph{topological derivative} \cite{Masmoudi2005a,Cea2000}, as the new inclusion changes the topology of the structure.  An equally important geometric change arises from a \emph{shape derivative}, in which the boundaries of the current structure are moved, without a change in topology.  We will consider equally shape and topological derivatives in this section.

The general derivation proceeds from the same starting point as the simple example.  Given the merit function $F(\mathbf{E},\mathbf{H})$, the key is to understand the variation $\delta F$ arising under small changes in geometry.  The merit function may depend on the field values at many points or regions within the domain (as opposed to a single point $\mathbf{x_0}$), so it is helpful to re-write the merit function
\begin{align}
\label{eq:genMerit}
F(\mathbf{E},\mathbf{H}) = \int_{\chi} f \left(\mathbf{E(\mathbf{x})},\mathbf{H(\mathbf{x})}\right) \,\mathrm{d}^3 \mathbf{x}
\end{align}
where $\chi$ is the region over which the fields are to be optimized.

The variation in $F$ depends directly on the variations $\delta \mathbf{E}$ and $\delta \mathbf{H}$.  However, care must be taken with $\Eb$ and $\Hb$, as they are complex-valued.  One approach would be to distinguish between the variation in the real and imaginary parts of the fields, as they are independent variables.  A simpler, and equivalent, approach is to distinguish between $\mathbf{E}$ and $\overline{\mathbf{E}}$ (and equivalently for $\mathbf{H}$).  Consequently, the variation in $F$ can be written.\footnote{All gradients with respect to vector fields will be denoted $\partial / \partial \mathbf{V}$, for a vector field $\mathbf{V}$.  This could equivalently be written $\nabla_{\mathbf{V}}$, and simply means $(\partial / \partial V_x,\partial / \partial V_y, \partial / \partial V_z)$.}
\begin{align}
\label{eq:deltaFcompconj}
\delta F = \int_{\chi} \left[ \frac{ \partial f}{\partial \Eb } (\mathbf{x}) \cdot \delta\Eb (\mathbf{x}) + \frac{\partial f}{\partial \overline{\Eb}} (\mathbf{x}) \cdot \delta\overline{\Eb} (\mathbf{x}) + \frac{ \partial f}{\partial \Hb } (\mathbf{x}) \cdot \delta\Hb (\mathbf{x}) + \frac{\partial f}{\partial \overline{\Hb}} (\mathbf{x}) \cdot \delta\overline{\Hb} (\mathbf{x}) \right] \text{d}^3 \mathbf{x}
\end{align}
For a real-valued function (as $F$ must be), $\delta F / \delta \Eb$ is the complex conjugate of $\delta F / \delta \overline{\Eb}$ \cite{Remmert}.  The first term in Eqn.~\ref{eq:deltaFcompconj} is therefore the complex conjugate of the second term, and the third is the complex conjugate of the fourth.  The equation for the variation in $F$ simplifies:
\begin{align}
\label{eq:deltaFint3}
\delta F = 2 \operatorname{Re} \int_{\chi} \left[ \frac{ \partial f}{\partial \Eb }(\mathbf{x}) \cdot \delta\Eb (\mathbf{x}) + \frac{ \partial f}{\partial \Hb } (\mathbf{x}) \cdot \delta\Hb (\mathbf{x}) \right] \text{d}^3 \mathbf{x}
\end{align}
Eqn.~\ref{eq:deltaFint3} for the general inverse design problem is the analogue of Eqn.~\ref{eq:linearization}(c) for the simpler example.  

\begin{figure}[]
\centering
\includegraphics[width=5in]{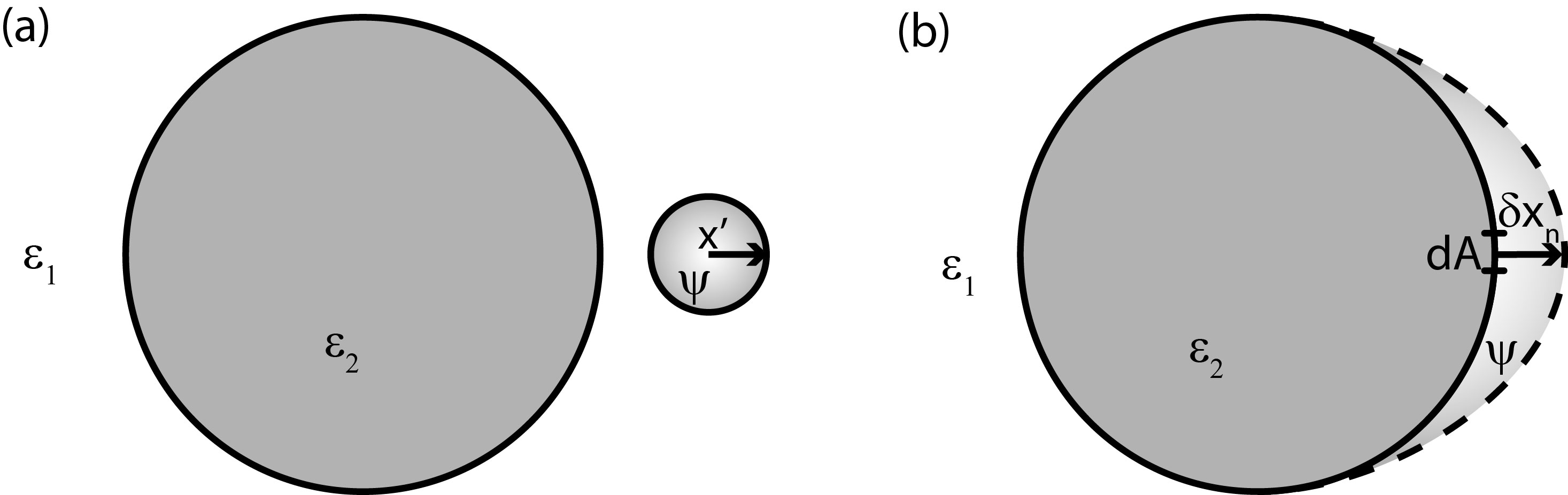}
\caption{(a) Shape and (b) topological variations, respectively.  In each case there is a region $\psi$ over which the geometry changes.  There is an induced polarization density $\mathbf{P}^{\textbf{ind}}$ on $\psi$ resulting from the change in geometry.  The ability to model $\mathbf{P^{ind}}$ and thus exploit dipole reciprocity relations is central to the shape calculus algorithm.} 
\label{fig:ShapeTopComp}
\end{figure}

The derivatives $\partial f / \partial \Eb$ and $\partial f / \partial \Hb$ are known from the merit function definition, leaving the variations $\delta \Eb$ and $\delta \Hb$ to be determined.  Instead of considering inclusions only, we now consider more general structural changes, as shown in Fig.~\ref{fig:ShapeTopComp}.  For both shape and topological derivatives, there is a confined region $\psi$ over which the permittivity $\epsr$ changes.  The effect of the change in permittivity is to induce a polarization density $\mathbf{P^{\text{ind}}}$ on $\psi$.

The variations in the electric and magnetic fields, $\delta \Eb$ and $\delta \Hb$ respectively, are the electric and magnetic fields arising from the induced polarization density.  Again, the Green's function connects the dipole moments to the fields.  We define $\overline{\overline{\mathbf{G^{\text{EP}}}}}(\mathbf{x_0},\mathbf{x'})$ to be the tensor containing the electric field at $\mathbf{x_0}$ from a unit electric dipole at $\mathbf{x'}$.  Similarly, $\overline{\overline{\mathbf{G^{\text{HP}}}}}(\mathbf{x_0},\mathbf{x'})$ is the magnetic field at $\mathbf{x_0}$ from a unit electric dipole at $\mathbf{x'}$.  Green's functions for fields from a magnetic dipole will be superscripted with an $M$ instead of a $P$, but it should be emphasized that for small changes in geometry, \emph{only the electric dipole moment} is significant enough to be retained.  Magnetic dipole moments will be needed only for symmetry reasons.

Given the Green's functions, the variations in the fields are straightforward:
\begin{align}
\label{eq:deltaEgreen}
\delta \Eb(\mathbf{x}) = \int_{\psi} \overline{\overline{\mathbf{G^{\text{EP}}}}} (\mathbf{x},\mathbf{x'}) \mathbf{P^{\text{ind}}}(\mathbf{x'}) \text{d}^3 \mathbf{x'} \\
\label{eq:deltaHgreen}
\delta \Hb(\mathbf{x}) = \int_{\psi} \overline{\overline{\mathbf{G^{\text{HP}}}}}(\mathbf{x},\mathbf{x'}) \mathbf{P^{\text{ind}}}(\mathbf{x'}) \text{d}^3 \mathbf{x'}
\end{align}
Eqns.~\ref{eq:deltaEgreen},\ref{eq:deltaHgreen} can be directly substituted into Eqn.~\ref{eq:deltaFint3}:
\begin{align}
\label{eq:deltaFgreen}
\delta F = 2 \operatorname{Re} \int_{\chi} \text{d}^3 \mathbf{x} \int_{\psi} \text{d}^3 \mathbf{x'} \left[ \frac{ \partial f}{\partial E_i} (\mathbf{x}) G_{ij}^{\text{EP}}(\mathbf{x},\mathbf{x'}) P_{j}^{\text{ind}}(\mathbf{x'})  + \frac{ \partial f}{\partial H_i } (\mathbf{x}) G_{ij}^{\text{HP}}(\mathbf{x},\mathbf{x'}) P_{j}^{\text{ind}}(\mathbf{x'}) \right] 
\end{align}
where the fields have temporarily been written in Einstein notation for easier manipulation, and summation over repeated indices is assumed.  Eqn.~\ref{eq:deltaFgreen} is the generalization of Eqn.~\ref{eq:deltaFint1}.

Again, symmetry is the key step to finding a simple equation for the variation $\delta F$.  The symmetry relations are derived in Appendix~\ref{chap:SymmMaxwell}, but they can be stated here:
\begin{subequations}
\label{eq:GreenSymm}
\begin{align}
G_{ij}^{\text{EP}}(\mathbf{x},\mathbf{x'}) = G_{ji}^{\text{EP}}(\mathbf{x'},\mathbf{x}) \\
G_{ij}^{\text{HP}}(\mathbf{x},\mathbf{x'}) = -G_{ji}^{\text{EM}}(\mathbf{x'},\mathbf{x})
\end{align}
\end{subequations}
The first relation states that the source and observation point of an electric dipole can be exchanged, as long as the polarization components are exchanged as well.  The second relation is slightly more complex, stating that the $i$-component of the \emph{magnetic} field at $\mathbf{x}$, from the $j$-component of an \emph{electric} dipole at $\mathbf{x'}$, is equal to the negative of the $j$-component of the \emph{electric} field at $\mathbf{x'}$ from a \emph{magnetic} dipole at $\mathbf{x}$.  Eqns.~\ref{eq:GreenSymm} are reciprocity relations, and it should be noted that they are consequences of the Maxwell operator being a linear operator, \emph{not} time-reversal invariance.  In fact, time-reversal invariance is not required, as Eqns.~\ref{eq:GreenSymm} apply even for lossy materials.

With the new forms for the Green's functions, as well as an interchange of the integration limits, the variation in $F$ becomes
\begin{align}
\delta F = 2 \operatorname{Re} \int_{\psi} \text{d}^3 \mathbf{x'} \text{ } P_{j}^{\text{ind}}(\mathbf{x'}) \int_{\chi} \text{d}^3 \mathbf{x} \left[ G_{ji}^{\text{EP}}(\mathbf{x'},\mathbf{x}) \frac{ \partial f}{\partial E_i} (\mathbf{x}) - G_{ji}^{\text{EM}}(\mathbf{x'},\mathbf{x}) \frac{ \partial f}{\partial H_i } (\mathbf{x}) \right] 
\end{align}
Newly re-written, the meaning of the terms in the brackets is clear.  The first term represents the electric field at $\mathbf{x'}$ due to an electric dipole at $\mathbf{x}$, with amplitude $\partial f / \partial E_i (\mathbf{x})$.  The second term represents the electric field at $\mathbf{x'}$, from a magnetic dipole at $\mathbf{x}$ and with amplitude $-(1/\mu_0)\partial f / \partial H_i (\mathbf{x})$.\footnote{The extra factor of $1/\mu_0$ comes from the normalization of the magnetic Green's function, see App.~\ref{chap:SymmMaxwell}.}  The entire integration over $\chi$ is then the field at a given $\mathbf{x'}$, from all of the electric and magnetic dipoles throughout $\chi$.  We again call this the adjoint field, denoted $\mathbf{E^A}(\mathbf{x'})$.  With this substitution, and returning to vector notation, the equation for the variation in $F$ simplifies:
\begin{align}
\label{eq:deltaFvector}
\delta F = 2 \operatorname{Re} \int_{\psi} \mathbf{P^{ind}}(\mathbf{x'}) \cdot \mathbf{E^A}(\mathbf{x'}) \text{ d}^3 \mathbf{x'}
\end{align}
Eqn.~\ref{eq:deltaFvector}, the analogue of Eqn.~\ref{eq:deltaFadjoint} of the simpler case, indicates that again only two simulations are required to understand the variation in $F$.  For all possible $\mathbf{x'}$ in the domain, the induced polarization $\mathbf{P}(\mathbf{x'})$ is known from a single simulation, as it is a function of the steady-state electric field at $\mathbf{x'}$.  Given the electric and magnetic fields over all space, the sources for the adjoint simulation ($\partial f / \partial \mathbf{E}$, $- (1/\mu_0)\partial f / \partial \mathbf{H}$) are also known.  The adjoint simulation then gives $\mathbf{E^A}(\mathbf{x'})$, and $\delta F$ is understood for any possible geometric deformation.

Eqn.~\ref{eq:deltaFvector} is not the ideal form for use with an inverse design algorithm.  For a given deformation $\psi$, it does tell you how much the merit function will change.  It does not, however, indicate which deformation maximally changes the merit function (as a gradient usually does).  It is therefore helpful to further manipulate Eqn.~\ref{eq:deltaFvector}, for easier use in inverse design.  Deformations split nicely into two categories: topological deformations (i.e. new inclusions) and shape deformations (i.e. boundary movements).  We consider the two types of deformations separately.

\subsection{Topological Derivative}
The creation of a dielectric inclusion (of arbitrary permittivity) is considered a change in topology.  For an inclusion of permittivity $\overline{\overline{\epsilon}}_2$ embedded in a region with permittivity $\overline{\overline{\epsilon}}_1$, the induced polarization is:
\begin{align}
\mathbf{P^{ind}}(\mathbf{x}) = \overline{\overline{\alpha}} \mathbf{E}(\mathbf{x})
\end{align}
where $\overline{\overline{\alpha}}$ is the shape-dependent polarizability and is related to the Claussiu-Mossotti factor \cite{Jackson1999}.  For a spherical inclusion of isotropic permittivity, for example, one would have
\begin{align}
\alpha_{sphere} = 3\epsilon_0 \left( \frac{ \epsilon_2 - \epsilon_1 } { 2 \epsilon_2 + \epsilon_1 } \right) 
\end{align}
Eqn.~\ref{eq:deltaFvector} can then be written 
\begin{align}
\delta F = 2 \operatorname{Re} \int_{inclusion} \overline{\overline{\alpha}} \mathbf{E}(\mathbf{x}) \cdot \mathbf{E^A}(\mathbf{x}) \text{ d}^3 \mathbf{x}
\end{align} 
The center of mass of the inclusion is taken to be $\mathbf{x'}$.  The derivative is defined in the limit as the size of the inclusion goes to zero, so the fields in the integrand can be approximated by their values at the center, leaving:
\begin{align}
\label{eq:TopDeriv}
\delta F = 2 V  \operatorname{Re} \left[  \overline{\overline{\alpha}} \mathbf{E}(\mathbf{x'}) \cdot \mathbf{E^A}(\mathbf{x'}) \right]
\end{align}
where $V$ is the volume of the inclusion.  Eqn.~\ref{eq:TopDeriv} yields a clear inverse design algorithm.  Simulate $\mathbf{E}$ and $\mathbf{E^{A}}$, then evaluate Eqn.~\ref{eq:TopDeriv} over the entire designable domain.  Create a new inclusion at the point for which the change in $F$ is maximum, and iterate.

\subsection{Shape Derivative}
Extending Eqn.~\ref{eq:deltaFvector} for a boundary deformation is trickier than the topological deformation.  For a rigorous derivation of boundary deformations using the calculus of variations, see \cite{Santosa1996}.  Instead, we take a less formal but more intuitive approach.

In Eqn.~\ref{eq:deltaFvector}, $\chi$ is the region over which a polarization is induced.  In Fig.~\ref{fig:ShapeTopComp}, for example, an initial boundary (solid line) and updated boundary (dashed line) are drawn, with the region between the two shaded in.  The shaded region is $\chi$, for a given deformation.  The figure should be considered a two-dimensional cross-section from a three-dimensional shape and deformation.  Integrating over $\chi$ requires integrating along the boundary of the initial shape, and over the variation in the normal direction.  We will call the differential element along the surface to be $\mathrm{d}A$, and the differential element in the normal direction to be $\mathrm{d}x_n$.  Eqn.~\ref{eq:deltaFvector} can be re-written:
\begin{align}
\delta F = 2 \operatorname{Re} \int \text{d}A \int \text{d}x_n  \left[ \mathbf{P^{ind}}(\mathbf{x'}) \cdot \mathbf{E^A}(\mathbf{x'}) \right]
\end{align}
In the limit as the size of the deformation goes to zero, the integrand can be evaluated exactly at the surface, and we can make the substitution
\begin{align}
\int \text{d}x_n \rightarrow \delta x_n(\mathbf{x'})
\end{align}
where $\delta x_n$ is the size of the deformation, in the normal direction, at each point along the boundary.  The equation for the variation in $F$ becomes
\begin{align}
\label{eq:deltaFint4}
\delta F = 2 \operatorname{Re} \int \delta x_n(\mathbf{x'}) \mathbf{P^{ind}}(\mathbf{x'}) \cdot \mathbf{E^A}(\mathbf{x'}) \text{d}A 
\end{align}
A subtle problem with Eqn.~\ref{eq:deltaFint4}, which arises only for shape derivatives, is that the electric fields at the boundary are \emph{not continuous}.  The boundary conditions from Maxwell's equations specify that the tangential component of $\mathbf{E}$ be continuous, and that the normal component of $\mathbf{D}$ be continuous.  In order to turn Eqn.~\ref{eq:deltaFint4} into a meaningful equation, it must be expressed in terms of only continuous fields.

The first step is to recognize that the polarization should be 
written:
\begin{align}
\mathbf{P^{ind}}(\mathbf{x'}) = (\epsilon_2 - \epsilon_1) \mathbf{E^{new}}(\mathbf{x'})
\end{align}
where $\mathbf{E^{new}}$ is the steady-state electric field of the deformed shape, not the original shape, and the outward normal directs from the material with permittivity $\epsilon_2$ to the material with permittivity $\epsilon_1$.  In this section the permittivity is assumed to be a scalar; for the tensor case, see \cite{Kottke2008}.  Because small deformations are assumed, we can relate $\mathbf{E^{new}}$ to $\mathbf{E^{old}}$.  However, it is NOT true that $\mathbf{E^{new}} \simeq \mathbf{E^{old}}$, because the components of $\mathbf{E}$ that are not continuous can change drastically even for small changes in the boundary.\footnote{Think of a boundary at which the electric field is perpendicular to the boundary.  The electric field on, e.g., the inside of the boundary will be much larger than the field on the outside.  If the boundary is deformed outward, a point which was outside the boundary will now be inside.  The field, which was large, will now be drastically reduced by the continuity condition, far more than linearization permits.}  Instead, one has to write:
\begin{subequations}
\begin{align}
\mathbf{E^{new}} & = \mathbf{E^{old}_{\parallel}} + \delta \mathbf{E_{\parallel}} + \frac{\mathbf{D^{old}_{\perp}} + \delta \mathbf{D_{\perp}}}{\epsilon_2} \\
& \simeq \mathbf{E^{old}_{\parallel}} + \frac{\mathbf{D^{old}_{\perp}} }{\epsilon_2}
\end{align}
\end{subequations}
where the \emph{continuous fields}, rather than the electric field components, are assumed to undergo small changes.  The adjoint field can be similarly split, without approximation:
\begin{align}
\mathbf{E^{A}} = \mathbf{E^{A}_{\parallel}} + \frac{\mathbf{D^{A}_{\perp}} }{\epsilon_1}
\end{align}
With the new expression for $\mathbf{P^{ind}}$ and $\mathbf{E^{A}}$, in terms of only continuous fields, the variation in $F$ becomes
\begin{align}
\label{eq:ShapeDeriv}
\delta F = 2 \operatorname{Re} \int \delta x_n(\mathbf{x'}) 
\left[ (\epsilon_2 - \epsilon_1) \mathbf{E_{\parallel}}(\mathbf{x'}) \cdot \mathbf{E^{A}_{\parallel}}(\mathbf{x'}) + \left( \frac{1}{\epsilon_1}-\frac{1}{\epsilon_2} \right) \mathbf{D_{\perp}}(\mathbf{x'}) \cdot \mathbf{D^{A}_{\perp}}(\mathbf{x'}) \right] \text{d}A 
\end{align}
where the ``old'' superscript has been dropped.  For a derivation of Eqn.~\ref{eq:ShapeDeriv} in the context of perturbation theory for roughness losses, see \cite{Johnson2002,Johnson2005}.  Note the similarity of Eqn.~\ref{eq:ShapeDeriv} to the variational equation of ordinary calculus
\begin{align}
\delta F = \sum_{i} \frac{\partial F}{\partial x_i} \delta x_i
\end{align}
for a finite number of variables $x_i$.  A typical gradient-based optimization algorithm would operate by setting 
\begin{align}
\delta x_i = \frac{\partial F}{\partial x_i}
\end{align}
ensuring that to first order the merit function increases every iteration.  Exactly the same reasoning leads in our case to the choice (up to a normalization constant) of
\begin{align}
\delta x_n(\mathbf{x'})  = 
2\operatorname{Re} \left[ (\epsilon_2 - \epsilon_1) \mathbf{E_{\parallel}}(\mathbf{x'}) \cdot \mathbf{E^{A}_{\parallel}}(\mathbf{x'}) + \left( \frac{1}{\epsilon_1}-\frac{1}{\epsilon_2} \right) \mathbf{D_{\perp}}(\mathbf{x'}) \cdot \mathbf{D^{A}_{\perp}}(\mathbf{x'}) \right]
\end{align}
for all $\mathbf{x'}$ on the boundary surface.  Thus, the right-hand side can be considered the infinite-dimensional shape derivative, and we have a procedure for updating a boundary with only two simulations.

\section{Information Loss?}
\label{sec:InfoLoss}
It would seem that a sleight of hand has occurred.  It has been claimed that two simulations are equivalent, to first order, the brute force $N+1$ simulations, where $N$ is arbitrarily large.  But how is it possible that so few simulations could could capture that much information?

\begin{figure}
\centering
\includegraphics[width=6in]{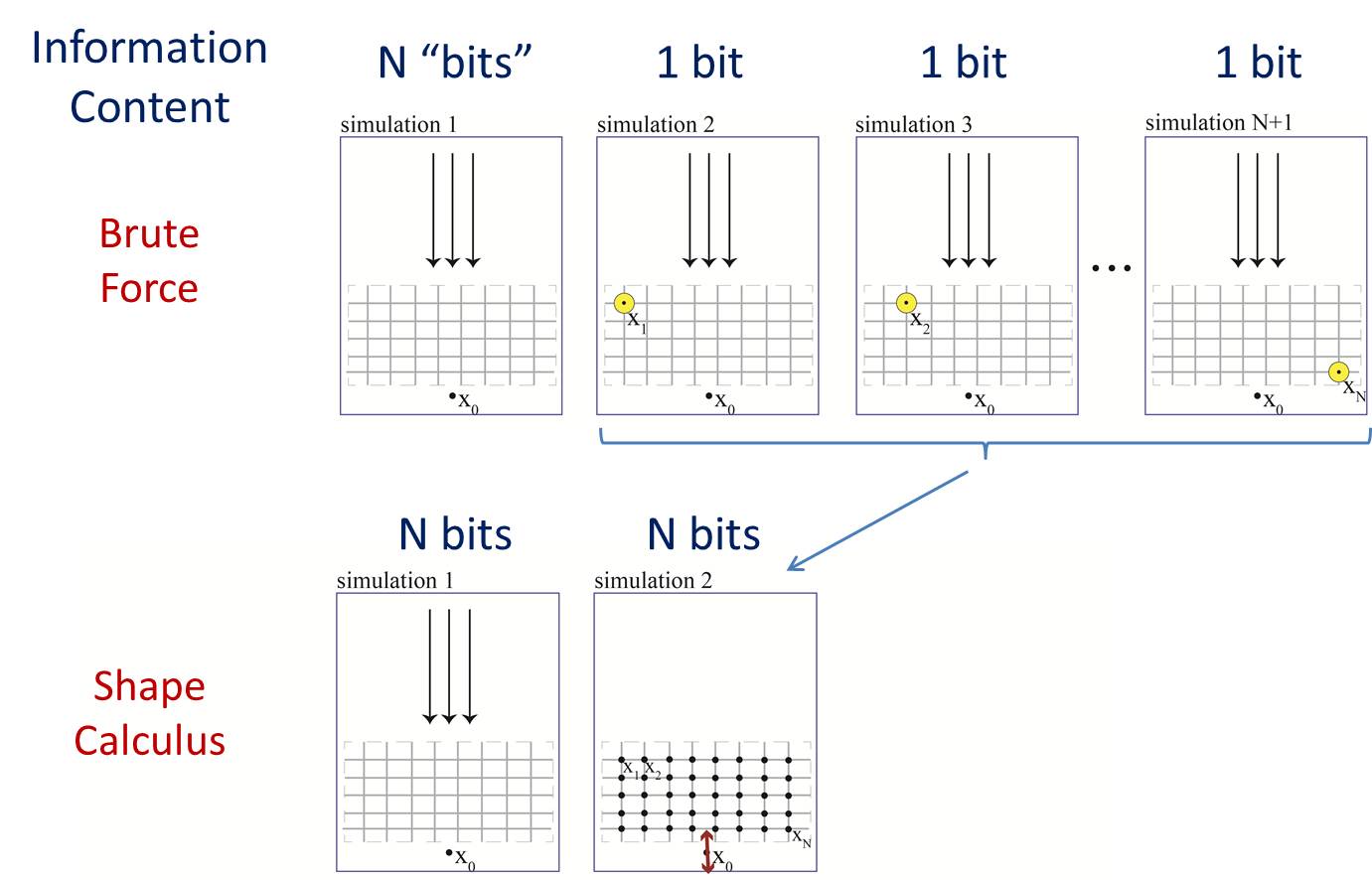}
\caption{Information content of brute force and shape calculus approaches to optimization.  The test simulations (i.e. simulations $2$ through $N$) of the brute force method are particularly wasteful.  Even though the entire domain must be simulated, ultimately only the field at $\mathbf{x_0}$ matters.  The shape calculus method exploits symmetry to avoid computation of irrelevant data.  With only two simulations, it gleans the same amount of information as the $N+1$ simulations of the brute force method, to first order.} 
\label{fig:InfoContent}
\end{figure}

To understand the answer, we re-examine the simple example of section~\ref{sec:ShapeCalc}.  Figure~\ref{fig:InfoContent} compares a single iteration of the brute force method and the shape calculus method side-by-side.   In each method, the first simulation is of the initial structure.  This provides a lot of information: not only is the field at $\mathbf{x_0}$ relevant, but the field at each of the $N$ possible inclusion points is relevant, as their dipole moments would be proportional to the fields there.  $N+1$ ``bits'' of information have been learned, where ``bit'' is not used in a rigorous sense.

Then consider one of the $N$ dielectric simulations in the brute force method.  For these simulations, the fields at $\mathbf{x_0}$ are relevant.  However, the fields at all of the other $\mathbf{x'}$ \emph{do not matter}; they provide no useful information.  Because the fields must be simulated over the whole region, there was a substantial computational cost to this irrelevant information.  The entire simulation results in only $1$ extra bit of information, the field at $\mathbf{x'}$.  The total information content of the $N+1$ brute force simulations is only $2N$, whereas approximately $N^2$ data points had to be simulated.

In contrast, the adjoint simulation does not waste information.  In the second simulation, the fields at every of the $N$ points within the designable region provide useful information.  For each point, the fields are the relevant dipole fields that would have been radiated from a local dipole to the measurement point.  With the shape calculus approach, the $2N$ bits of information require simulations of only about $2N$ data points, representing the most efficient way of acquiring the information.

\section{Level Set Geometric Representation}
\label{sec:LevelSet}
For arbitrarily-shaped boundaries, level set methods \cite{Osher1988} provide a natural, flexible representation.  The basic framework will be discussed here; for a detailed presentation see \cite{Osher2003,Sethian1999}.

For two-phase material structures, the permittivity and permeability are known everywhere once the boundary is specified.  The boundary, therefore, contains all the necessary information.  However, implementing computational schemes in which the boundary is composed of little particles moving independently is a cumbersome process.  Functions like merging or pulling apart objects are exceptionally difficult.  Instead, embedding the boundary into a function defined across the entire domain turns out to be a far easier and more natural representation.  The boundary is defined as the zero level set of a \emph{level set function} $\phi$, such that $\phi = 0$ at points on the boundary.  Away from the boundary, it is numerically convenient to set $\phi$ equal to a signed distance function; that is, at every point $\phi$ equals the distance to the closest boundary point.  Moreover, points inside the boundary are taken to be the negative of the distance, enabling one to keep track of which material is where.

\begin{figure}
\centering
\includegraphics[width=4.5in]{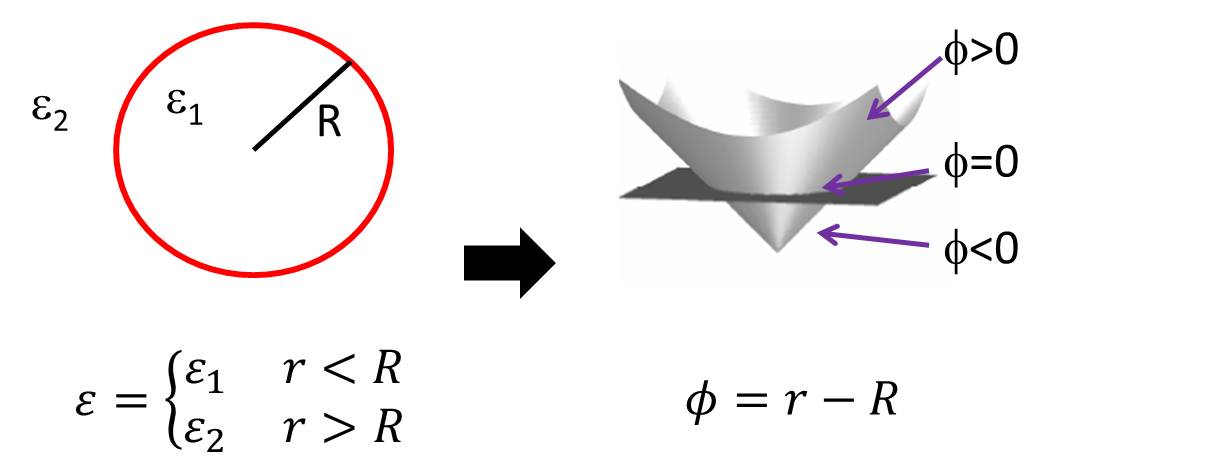}
\caption{Illustration of signed distance function for a simple circular boundary.  The boundary distinguishes between the volume with $\epsilon=\epsilon_1$ and the volume with $\epsilon=\epsilon_2$.  The signed distance function $\phi = r-R$ equals $0$ at the boundary.  The magnitude of $\phi(\mathbf{x})$ is the distance from $\mathbf{x}$ to the closest point on the boundary, and the sign of $\phi$ distinguishes the interior and exterior of the boundary.} 
\label{fig:CirclePhi}
\end{figure}

As an example, consider a boundary consisting of a circle of radius $R$ centered at $(0,0)$, as shown in Fig.~\ref{fig:CirclePhi}.  The level set function describing the circle would be:
\begin{align}
\phi^{circle} = \sqrt{x^2 + y^2} - R
\end{align}
On the boundary, $\phi=0$.  At a point interior to the circle, at say a radius $R_1 < R$, $\phi = -|R_1-R|$.  At a point exterior to the circle, at $R_2 > R$, $\phi = |R_2-R|$.  Thus $\phi$ is a proper signed distance function over all space.

The utility of such a function is that it enables easy tracking and merging.  Instead of trying to maintain a set of boundary particles through many deformations, one simply has to continually update the level set function, then find the $\phi=0$ level set.  To update the level set function, start with the equation
\begin{align}
\label{eq:PhiEq0}
\phi(\mathbf{x}(t),t) = 0
\end{align}
where $\mathbf{x}(t)$ is the set of points along the boundary. The geometry can be deformed continuously through the artificial time parameter $t$, and time $t=0$ represents the initial geometry.  Taking the total time derivative of Eqn.~\ref{eq:PhiEq0} results in the equation
\begin{align}
\frac{\partial \phi}{\partial t} + \frac{d \mathbf{x}}{dt} \cdot \nabla \phi = 0
\end{align} 
The term $d\mathbf{x} / dt$ is referred to as the \emph{velocity}, representing the speed at which a point along the boundary is moving, and in which direction.  Only movement normal to the boundary is relevant; movement along the boundary is simply a re-parametrization.  Reducing the velocity to a scalar field $V_n$ in the normal direction:
\begin{align}
\label{eq:HamJacEqn}
\frac{\partial \phi}{\partial t} + V_n \left|\nabla \phi\right| = 0
\end{align}
Eqn.~\ref{eq:HamJacEqn}, a Hamilton-Jacobi partial differential equation, governs the deformation of level set functions.  An initial structure defines the level set function for $t=0$.  Starting from that initial condition, and given a velocity $V_n(\mathbf{x})$, the PDE is solved through time, until the boundary has deformed as much as desired.

The shape calculus of Chap.~\ref{chap:Algorithm} can be extended to the level set framework \cite{Osher2001,Allaire2004,Wang2004,Burger2005,Kao2005}.  The primary result of the shape calculus derivation was Eqn.~\ref{eq:ShapeDeriv}, denoting the variation in a merit function $F$ for deviations in points along the boundary $\delta x_n$.  Now, we would like to write the variation in terms of a deformation in $\phi$.  The first step is to recognize that
\begin{equation}
\delta x_n = \frac{d\mathbf{x}}{dt} \cdot \hat{\mathbf{n}} \delta t = V_n \delta t
\end{equation}
where $\hat{\mathbf{n}}$ is the normal vector to the boundary.  Thus one can re-write Eqn.~\ref{eq:ShapeDeriv} as
\begin{align}
\label{eq:ShapeDerivPhi}
\delta F = 2\delta t \operatorname{Re} \int V_n(\mathbf{x'}) 
\left[ (\epsilon_2 - \epsilon_1) \mathbf{E_{\parallel}}(\mathbf{x'}) \cdot \mathbf{E^{A}_{\parallel}}(\mathbf{x'}) + \left( \frac{1}{\epsilon_1}-\frac{1}{\epsilon_2} \right) \mathbf{D_{\perp}}(\mathbf{x'}) \cdot \mathbf{D^{A}_{\perp}}(\mathbf{x'}) \right] \text{d}A 
\end{align}
Within the level set framework, the optimization would proceed as follows.  After completing the two required simulations, one would set the velocity $V_n$ equal to the term in brackets.  One would then solve the Hamilton-Jacobi equation, Eqn.~\ref{eq:HamJacEqn}, given $V_n$ and over some artificial time $\delta t$.  This would give an updated shape and update $\phi$, and the process iterates.

In addition to providing a flexible framework for dealing with arbitrary shapes, level set methods also enable simpler derivations of the shape derivative for many highly constrained shapes.  Directly using Eqn.~\ref{eq:ShapeDeriv} is difficult, as the normal vector is needed to construct $\delta x_n$.  Instead, we can indirectly use level set calculus.  Instead of Eqn.~\ref{eq:ShapeDerivPhi}, one could also write the shape derivative directly in terms of the change in $\phi$ by making the replacement
\begin{equation}
V_n \delta t = -\frac{\delta\phi}{|\nabla \phi|}
\end{equation}
where $\delta \phi$ is the variation in the signed distance function from one geometry to the next.  For a geometry consisting of a finite collection of objects, the variation in $\phi$ would decompose into the sums over the variations in $\phi_i$, where $\phi_i$ is the level set function corresponding to the $i^{\textrm{th}}$ individual shape.  If the level set function $\phi$ is known in terms of the shape parameters, the variation in $\phi_i$ can then be written
\begin{equation}
\delta \phi_i = \frac{\partial \phi_i}{\partial \mathbf{p}_i} \cdot \delta \mathbf{p}_i
\end{equation}
where $\mathbf{p}_i$ represents the shape parameters for each distinct shape.  Finally, Eqn.~\ref{eq:ShapeDerivPhi} can be re-written as
\begin{align}
\label{eq:ShapeDerivPhi2}
\delta F = -2 \operatorname{Re} \displaystyle\sum_i \int \delta \mathbf{p}_i \cdot \frac{\partial \phi_i}{\partial \mathbf{p}_i} \frac{1}{|\nabla \phi_i|} 
\left[ (\epsilon_2 - \epsilon_1) \mathbf{E_{\parallel}}(\mathbf{x'}) \cdot \mathbf{E^{A}_{\parallel}}(\mathbf{x'}) + \left( \frac{1}{\epsilon_1}-\frac{1}{\epsilon_2} \right) \mathbf{D_{\perp}}(\mathbf{x'}) \cdot \mathbf{D^{A}_{\perp}}(\mathbf{x'}) \right] \text{d}A 
\end{align}
If $\phi_i$ is known, Eqn.~\ref{eq:ShapeDerivPhi2} provides a simple prescription for optimizing the parameters $\mathbf{p}_i$.

Consider, for example, a geometry that is constrained to allow only a set of N two-dimensional ellipses $\{e_1,e_2,...,e_N\}$, where the $i^{\textrm{th}}$ ellipse has center point $(x_i,y_i)$ and radii $(r_1,r_2)$.  We can bundle the shape parameters into the four-vector $\mathbf{p}_i$ for each ellipse.  The geometry then has $4N$ parameters for which a shape derivative is needed.  Eqn.~\ref{eq:ShapeDerivPhi2} suggests that as long as a level set representation can be found, the shape derivatives will naturally follow.  One possible level set function would be
\begin{equation}
\label{eq:EllipsePhi}
\phi_i = \left( \frac{x-x_0^{(i)}}{r_1^{(i)}} \right)^2 + \left( \frac{y-y_0^{(i)}}{r_2^{(i)}} \right)^2 - 1
\end{equation}
Although Eqn.~\ref{eq:EllipsePhi} is not a signed distance function, it is a level set function, as $\phi_i=0$ on the boundary, $\phi_i>0$ outside the boundary, and $\phi_i<0$ inside the boundary.  The derivative $\partial \phi_i / \partial \mathbf{p}_i$ is straightforward, with individual components
\begin{subequations}
\begin{align}
\frac{\partial \phi_i}{\partial x_0^{(i)}} & = -\frac{2}{r_1^{(i)}} \left( \frac{x-x_0^{(i)}}{r_1^{(i)}} \right) \\
\frac{\partial \phi_i}{\partial y_0^{(i)}} & = -\frac{2}{r_2^{(i)}} \left( \frac{y-y_0^{(i)}}{r_2^{(i)}} \right) \\
\frac{\partial \phi_i}{\partial r_1^{(i)}} & = -\frac{2}{r_1^{(i)}} \left( \frac{x-x_0^{(i)}}{r_1^{(i)}} \right)^2 \\
\frac{\partial \phi_i}{\partial r_2^{(i)}} & = -\frac{2}{r_2^{(i)}} \left( \frac{y-y_0^{(i)}}{r_2^{(i)}} \right)^2
\end{align}
\end{subequations}
The magnitude of the gradient $|\nabla \phi|$ is also straightforward, directly yielding the shape derivatives through Eqn.~\ref{eq:ShapeDerivPhi2}.

Thus level set methods are a natural setting for shape calculus.  For some applications, they provide a flexible and efficient shape representation for complex geometries.  In others, they simplify the mathematical framework for rapid calculation of the shape derivatives.  They will be used in both forms in the upcoming chapters.

\section{Computational Implementation}
\label{sec:CompImpOpt}
\begin{figure}
\centering
\includegraphics[width=3.5in]{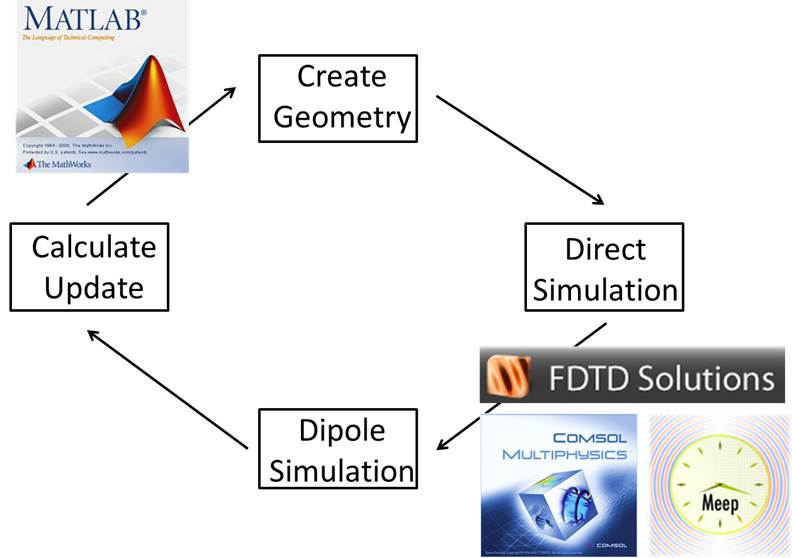}
\caption{Schematic implementation of shape calculus optimization algorithm.  The full-wave simulations are completed in solvers such as Lumerical FDTD Solutions \cite{Lumerical2012}, Comsol Multiphysics \cite{Comsol2012}, and Meep \cite{Oskooi2010}.  The geometric representation and shape/topological derivatives are computed in Matlab \cite{Matlab2010}.} 
\label{fig:AlgSchematic}
\end{figure}
Fig.~\ref{fig:AlgSchematic} schematically illustrates the computational implementation of the shape calculus algorithm presented in this chapter.  The over-arching inverse design is performed in Matlab \cite{Matlab2010}, which couples to full-wave electromagnetics solvers such as Lumerical FDTD Solutions \cite{Lumerical2012}, Comsol Multiphysics \cite{Comsol2012}, and Meep \cite{Oskooi2010}.  The full algorithm in pseudocode is presented in Alg.~\ref{alg:CompImp}.  The geometry is represented in a data structure in Matlab, with a solver-specific code that ports the geometry into the electromagnetic solver.  The solver performs the two simulations, returning all of the field data.  The shape derivatives and updates are performed in Matlab, at which point the algorithm iterates.

There are many possible geometric representations.  There are level set functions, as presented in Sec.~\ref{sec:LevelSet}, for a general class of smooth shapes.  One could also imagine a set of parameterized shapes, including ellipses, rectangles, and other simple structures that could be used to build up more complex geometries.  There are also splines, gridded boundaries, and many other types of geometries.  Once one has built up a set of functions for initialization, calculation of shape derivatives, and updates for a large class of shapes, the inverse design approach presented here is very generic, applicable to any optimization problem with Maxwell's equations as constraints.  Ostensibly, future commercial electromagnetics solvers will have an optimization functionality in the same way they currently have a simulation functionality.  Shape calculus will almost certainly play a major role.

\begin{algorithm}
\caption{Inverse Design Algorithm.}
\label{alg:CompImp}
\begin{algorithmic}
\State Initialize geometry
\While{$stopping$ $condition$ $not$ $met$} 
	\State Direct simulation: compute $\mathbf{E},\mathbf{H}$
	\State Remove all sources
	\State Add electric and magnetic dipole sources such that
	\State\hspace{\algorithmicindent} $\mathbf{P}(\mathbf{x}) = \frac{\partial f}{\partial \mathbf{E}} \qquad \forall \mathbf{x} \in \chi$ 
	\State\hspace{\algorithmicindent} $\mathbf{M}(\mathbf{x}) = -\frac{1}{\mu_0}\frac{\partial f}{\partial \mathbf{H}} \qquad \forall \mathbf{x} \in \chi$
	\State Dipole simulation: compute adjoint fields $\mathbf{E^A},\mathbf{H^A}$
	\State Calculate shape/topological derivatives according to Eqns.~\ref{eq:ShapeDeriv},\ref{eq:TopDeriv}
	\State Update geometry
\EndWhile
\end{algorithmic}
\end{algorithm} 

%% file: Cloaking.tex
\chapter{Optical Cloak Optimization}
\label{chap:OptCloak}
\epigraph{With four parameters I can fit an elephant, and with five I can make him wiggle his trunk.}{John von Neumann}

\noindent
Optical cloaking has received considerable interest in the past five years.  Clever material design for optical cloaking has been desired since the seminal papers of Pendry \cite{Pendry2006} and Leonhardt \cite{Leonhardt2006b}.  Conceptually, optical cloaking occurs when incident light rays are re-directed around an object, such that to an observer the rays appear unperturbed.  This effect has been demonstrated experimentally at various frequencies and with varying degrees of generality (i.e. angular bandwidth, frequency bandwidth, etc.) many times in the subsequent years \cite{Schurig2006a,Cai2007,Liu2009a,Valentine2009a,Chen2011}.

\section{Transformation Optics}
Transformation optics is the key idea driving optical cloaking schemes \cite{Rahm2008b,Leonhardt2009,Kildishev2008}.  The basic idea is as follows.  Consider a Euclidean coordinate system $x^{\alpha}$ through which light rays pass unperturbed.  Now imagine transforming the coordinate system by a transformation matrix $T_{\alpha}^{\alpha'}$.  The new coordinate system $x^{\alpha'}$ is related to the original coordinate system by the relation
\begin{equation}
x^{\alpha'} = T_{\alpha}^{\alpha'} x^{\alpha}
\end{equation}
Maxwell's equations are form-invariant under coordinate transformations, such that in the new coordinate system Maxwell's equations still hold.  Moreover, the coordinate transformation is equivalent to transformations of the permittivity $\epsilon^{ij}$ and the permeability $\mu^{ij}$:
\begin{subequations}
\label{eq:MatTrans}
\begin{align}
\epsilon^{i'j'} & = \left[ \det \left( T_i^{i'} \right) \right]^{-1} T_i^{i'}T_j^{j'} \epsilon^{ij} \\
\mu^{i'j'} & = \left[ \det \left( T_i^{i'} \right) \right]^{-1} T_i^{i'}T_j^{j'} \mu^{ij}
\end{align}
\end{subequations}
Thus, if one can find a coordinate transformation that redirects every light ray as desired, then there will be a transformation of the material parameters given by Eqn.~\ref{eq:MatTrans} that produces the same effect.

The inherent simplicity of transformation optics can also provide significant practical difficulties.  First, for a given transformation both $\epsilon$ \emph{and} $\mu$ must be transformed - one cannot choose, for example, to use non-magnetic materials (with the limited exception of quasi-conformal mappings \cite{Li2008}).  Worse, the permittivity and permeability are generally fully nine-component matrices, with the local coordinate systems for diagonalization differing throughout the space.  The values are also exotic, often less than zero or tending to infinity.  Fundamentally, for a given coordinate transformation there is only a single transformation of $\epsilon$ and $\mu$, and there is no straightforward way to impose simple material constraints.

In this chapter we apply inverse design techniques to optical cloaking design.  Instead of taking a coordinate transformation-based design approach, we use the shape calculus of Chap.~\ref{chap:Algorithm} to design a material with cloaking functionality.  We show that inverse design enables simple reduced-material design for reduced functionality, with the exact tradeoffs determined by the designer.  Real materials, such as imperfect conductors, can also be seamlessly incorporated.

\begin{figure}
\centering
\includegraphics[width=3in]{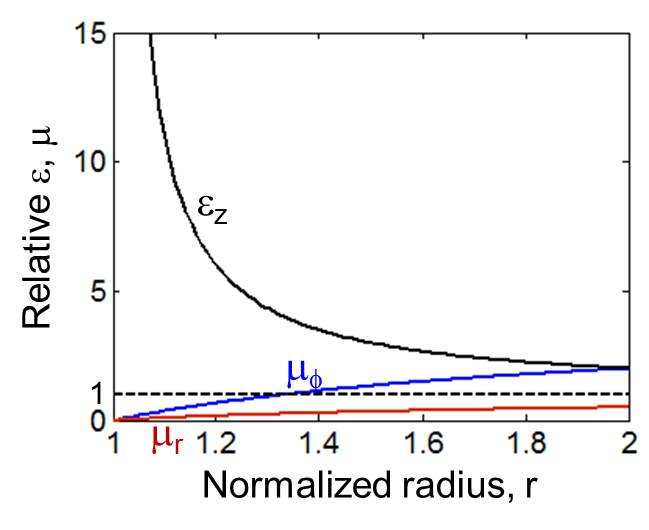}
\caption{Values of the permittivity $\epsilon$ and permeability $\mu$ for a cylindrical cloak of $TE$-polarized waves, from Eqn.~\ref{eq:EpsMuTOEqn}, designed in \cite{Cummer2006a}.  Note that $\epsilon_z$ tends to infinity approaching the inner radius, the material is magnetic and anisotropic in its permeability, and the permeability values are less than $1$ over a significant radius.} 
\label{fig:EpsMuTO}
\end{figure}

The first cloak to be optimized has translational symmetry in one dimension (say, $z$).  To compare and contrast the advantages of the inverse design technique, comparisons with the work of Cummer et. al.  
\cite{Cummer2006a} will be drawn.  In \cite{Cummer2006a}, full-wave simulations of a two-dimensional cylindrical cloak were performed.  Cloaking was achieved by a coordinate transformation that stretched a single point out to some radius $R_1$, resulting in a transformation of the material parameters for $TE$-polarized waves of
\begin{subequations}
\label{eq:EpsMuTOEqn}
\begin{align}
\epsilon_z & = \left( \frac{R_2}{R_2-R_1} \right)^2 \frac{r-R_1}{r} \\
\mu_r & = \frac{r-R_1}{r} \\
\mu_{\phi} & = \frac{r}{r-R_1}
\end{align}
\end{subequations}
where $R_1$ and $R_2$ are the inner and outer radii of the cloak, respectively.  Fig.~\ref{fig:EpsMuTO} plots the material parameters in the cloaking region for $R_2 = 2R_1$.  Note that the material parameters are exotic: $\epsilon_z$ tends to infinity approaching the inner radius.  The material is magnetic; moreover, it is anisotropic in a radial basis.  And finally, the permeability values are less than $1$ over a sizeable distance, significantly increasing the difficulty of fabricating the cloak.

The advantages of the transformation-optics design lie in its analytical nature; once the coordinate transformation was known, finding the values of $\epsilon$ and $\mu$ was simply a matter of multiplying matrices.  Also, the design works for all incidence angles and frequencies, although in practice having permeability values less than one generally restricts the frequency bandwidth.  The disadvantage, as mentioned previously, is the exotic nature of the materials required through the coordinate transformations.

\section{Shape Calculus}
\begin{figure}
\centering
\includegraphics[width=3.5in]{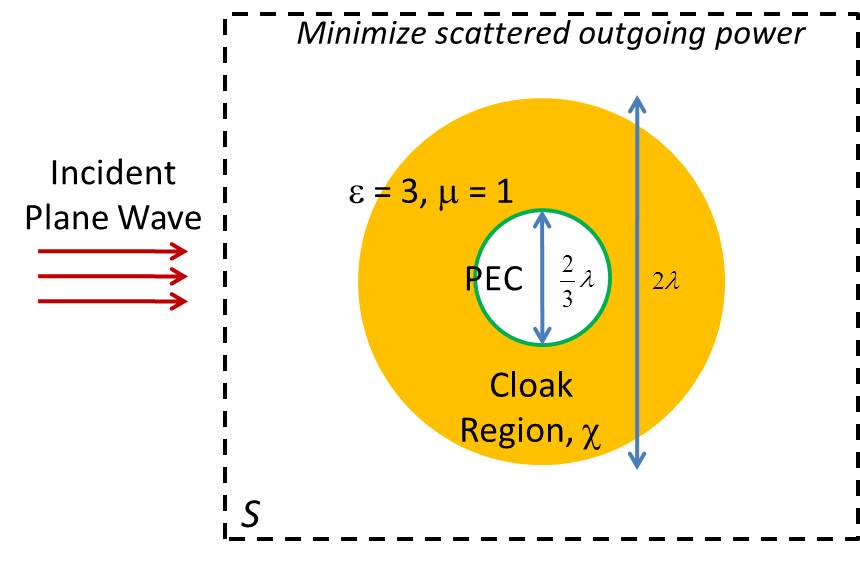}
\caption{Setup for cloaking optimization for a single frequency and single incidence angle.  The arbitrary material to cloak sits within the perfectly conducting ring with diameter $(2/3)\lambda$.  A $TM$-polarized (i.e. scaler $H$-field) is incident from the left.  Varying degrees of symmetry can be imposed on the cloaking region to increase the angular acceptance of the design.  The merit function to be minimized is the power scattered through a bounding region $S$.} 
\label{fig:2dCloakSetup}
\end{figure}
Fig.~\ref{fig:2dCloakSetup} shows the inverse design approach to achieving optical cloaking.  With the wavelength as a free parameter (the Maxwell equations are scale-invariant), an object with diameter of $(2/3)\lambda$ is inside a perfectly conducting ring, surrounded by a dielectric cloaking region.  The incident plane wave is $TM$-polarized (i.e. scalar $H$-field).  The merit function is the negative of the scattering width $\sigma$, which is taken to be the power in the scattered fields divided by the power incident on the diameter of the perfect conductor:
\begin{equation}
\label{eq:FCloak}
F(\mathbf{E},\mathbf{H}) = -\frac{\frac{1}{2} \int_S \operatorname{Re} \left( \mathbf{E}_s \times \overline{\mathbf{H}}_s \right) \cdot \hat{\mathbf{n}} \mathrm{d}s}{P_{inc}}
\end{equation}
where $S$ is the bounding region in Fig.~\ref{fig:2dCloakSetup}, $\hat{\mathbf{n}}$ is the outward surface normal, and the $s$ subscript denotes a scattered field.  The negative is taken for consistency with Chap.~\ref{chap:Algorithm}, where the merit functions were to be maximized, not minimized (the two formulations are equivalent up to a sign).  Comparing with the notation of Eqn.~\ref{eq:genMerit}, for this application the cloaking region is $\chi$ and the function $f$ is the integrand of Eqn.~\ref{eq:FCloak}, including the factors $1/2$ and $1/P_{inc}$.  The variation $\delta f$ is
\begin{equation}
\delta f = -\frac{1}{2P_{inc}}\operatorname{Re} \left(\delta \mathbf{E}_s \times \overline{\mathbf{H}}_{s} \cdot \hat{\mathbf{n}}  + \mathbf{E}_s \times \delta \overline{\mathbf{H}}_s \cdot \hat{\mathbf{n}} \right)
\end{equation}
which can be re-arranged as
\begin{equation}
\delta f = -\frac{1}{2P_{inc}}\operatorname{Re} \left(\delta \mathbf{E}_s \cdot \overline{\mathbf{H}}_{s} \times \hat{\mathbf{n}}  + \delta \mathbf{H}_s \cdot \hat{\mathbf{n}} \times \overline{\mathbf{E}}_s \right)
\end{equation}
through vector identities and taking the complex conjugate within the $\operatorname{Re}$ operator for the $\delta \mathbf{H}_s$ term.  The optimization proceeds as generally described in Sec.~\ref{sec:CompImpOpt}, with the sources for the adjoint simulation set as
\begin{subequations}
\label{eq:PMCloak}
\begin{align}
\mathbf{P} &= \frac{\partial f}{\partial \mathbf{E}} = \frac{\hat{\mathbf{n}} \times \overline{\mathbf{H}}_s}{2P_{inc}} \\
\mathbf{M} &= -\frac{1}{\mu_0} \frac{\partial f}{\partial \mathbf{H}} = - \frac{\hat{\mathbf{n}} \times \overline{\mathbf{E}}_s }{2\mu_0 P_{inc}}
\end{align}
\end{subequations}
throughout the surface $S$ through which the scattered power is measured.  The total field $\mathbf{E}$ is the sum of the incident field $\mathbf{E}_{inc}$ and the scattered field $\mathbf{E}_s$.  Because the incident field is constant throughout the optimization, it was justified in Eqn.~\ref{eq:PMCloak} to set $\partial f / \partial \mathbf{E} = \partial f / \partial \mathbf{E}_s$, and likewise for $\mathbf{H}$.

The optimization starts with $\epsilon=3$ and $\mu=1$.  Throughout the optimization, the material is constrained to have $\mu=1$, isotropic $\epsilon$, and $\epsilon>1$.  The permittivity can take on any value between $1$ and $12$; this is the only example of this dissertation for which two-phase materials are not used.  As such, the shape derivative formula is actually simpler.  Now, for small perturbations, the electric fields \emph{will} approximately be continuous.  Because there is no sharp boundary, there is no need to separate the fields into their normal and tangential components.  Consequently, the induced polarization from a small perturbation can be modeled as $\mathbf{P}^{ind} \approx \delta \epsilon \mathbf{E}$.  In this case, instead of Eqn.~\ref{eq:ShapeDeriv}, the shape derivative is given by
\begin{equation}
\delta F = 2\operatorname{Re} \int_{\chi} \delta \epsilon\left(\mathbf{x'}\right) \mathbf{E}(\mathbf{x'}) \cdot \mathbf{E^A}(\mathbf{x'}) \mathrm{d}V
\end{equation}
and the choice of update consequently given by $\delta\epsilon(\mathbf{x'}) = c\operatorname{Re} \left[\mathbf{E}(\mathbf{x'})\cdot \mathbf{E^A}(\mathbf{x'}) \right]$, with $c$ normalized to ensure a small step size.

\begin{figure}
\centering
\includegraphics[width=4in]{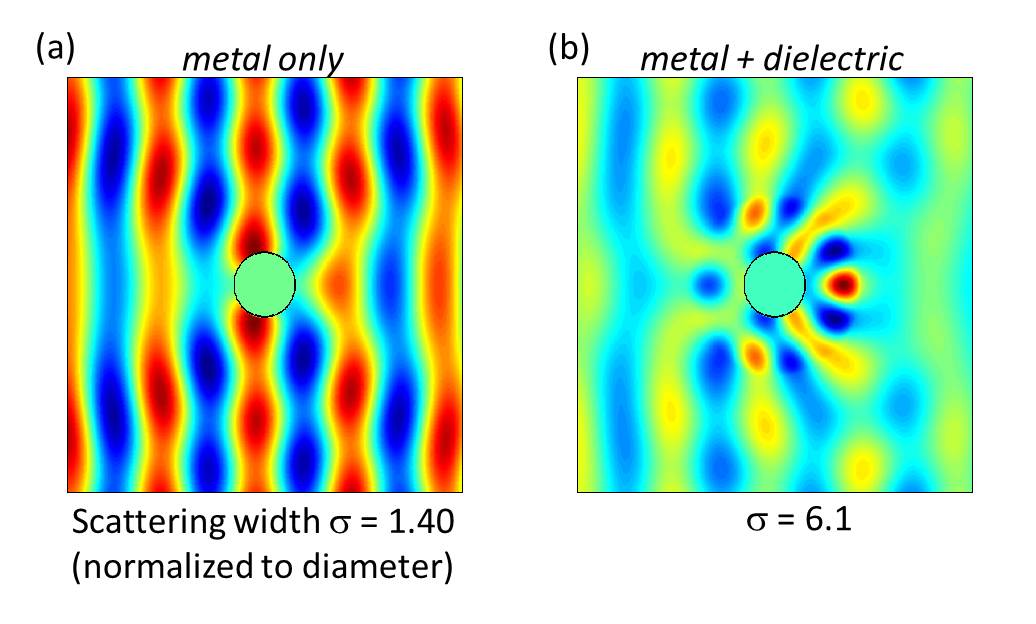}
\caption{Magnetic field for cloaking setup, with (a) metal only and (b) metal plus surrounding dielectric, for $\epsilon=3$ in the cloaking region.  Adding the constant dielectric severely penalizes the scattering width, increasing it to more six times the power incident upon the metallic cylinder.} 
\label{fig:InitialFields}
\end{figure}

The magnetic fields of the uncloaked object are shown in Fig.~\ref{fig:InitialFields}.  For the perfect conductor alone, the phase fronts are clearly distorted and the scattering width is slightly larger than one.  Upon adding the dielectric material to be optimized, the scattering worsens and there is now significant scattering throughout the volume.  For this optimization, all simulations were completed in Lumerical's FDTD Solutions \cite{Lumerical2012}.

\begin{figure}
\centering
\includegraphics[width=5in]{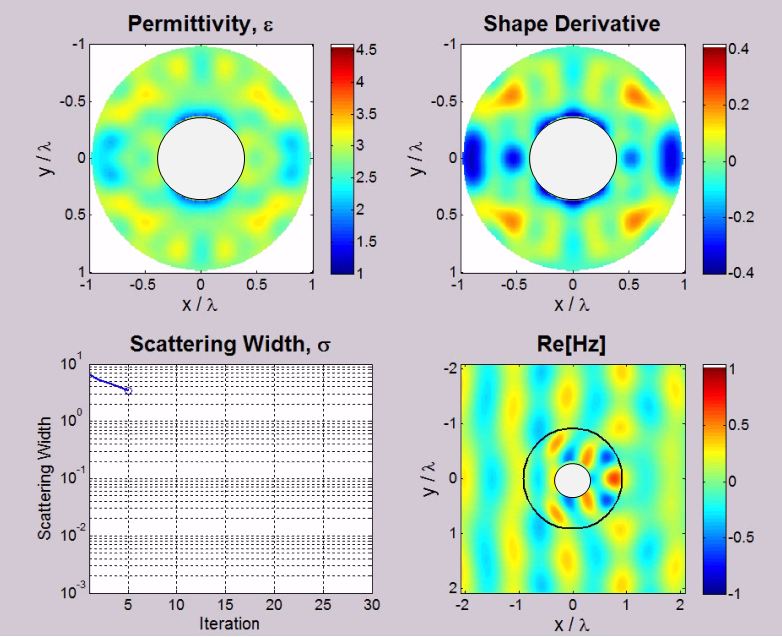}
\caption{Fifth iteration of the optimization.  The scattering width has decreased, the shape derivative dictates how to change $\epsilon$, and the fields appear closer to the desired plane wave.} 
\label{fig:2dIteration5}
\end{figure}

\section{Optimization Results}
Fig.~\ref{fig:2dIteration5} illustrates the inverse design procedure as applied to the optical cloak design.  At the fifth iteration, the permittivity has started to deviate from the constant value of three everywhere.  The shape derivative dictates how it should change from iteration to iteration, and the scattering width is decreasing rapidly.  The magnetic field looks more like a plane wave, compared to the starting field of Fig.~\ref{fig:InitialFields}(b).  

\begin{figure}
\centering
\includegraphics[width=6.5in]{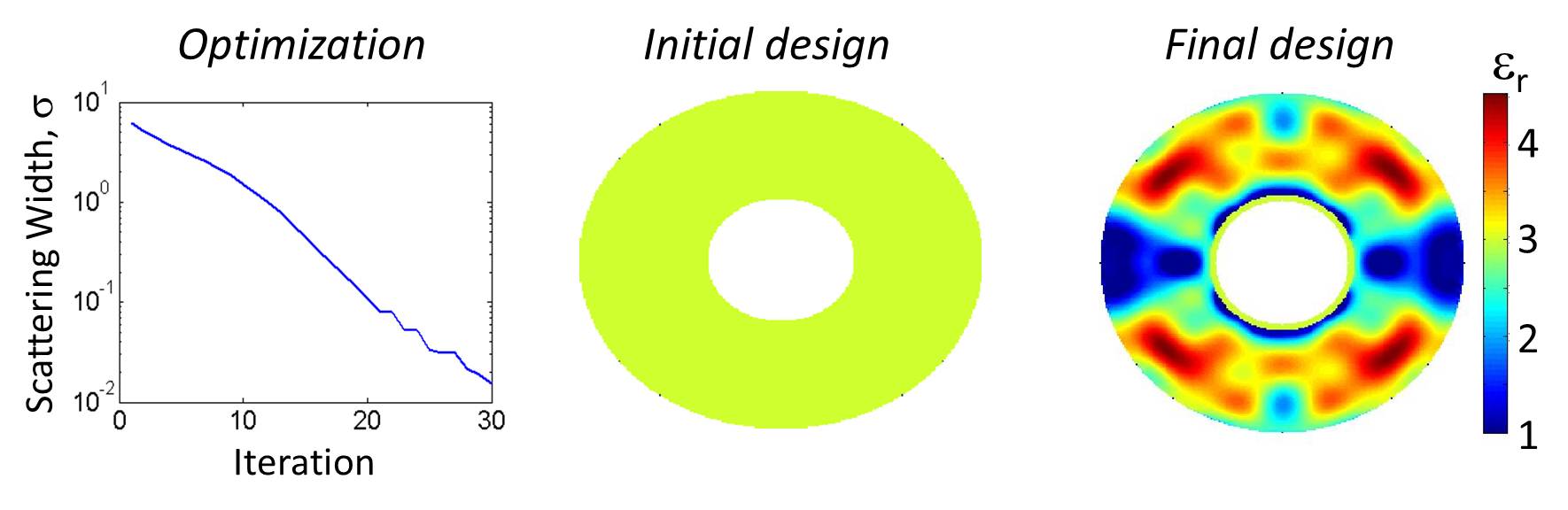}
\caption{Results of two-dimensional optimization at a single frequency and incidence angle.  The initial design transforms from a constant permittivity to a complex structure, re-directing the waves around the internal perfect conductor.  By the $30^{th}$ iteration, the scattering width has decreased by three orders of magnitude.  The contrast of the final design remains relatively small, reducing potential experimental complexity.} 
\label{fig:2dOptimization}
\end{figure}

Fig.~\ref{fig:2dOptimization} shows the results of the optimization from start to finish.  Within thirty iterations, the scattering width has reached a value of about $10^{-2}$, almost three orders of magnitude smaller than the initial value.  The final design is a complex pattern of varying permittivity, practically impossible to intuit and in such a large parameter space that it would be difficult to optimize through a stochastic approach.  The relatively low contrast of the permittivity has been maintained.

\begin{figure}
\centering
\includegraphics[width=6.5in]{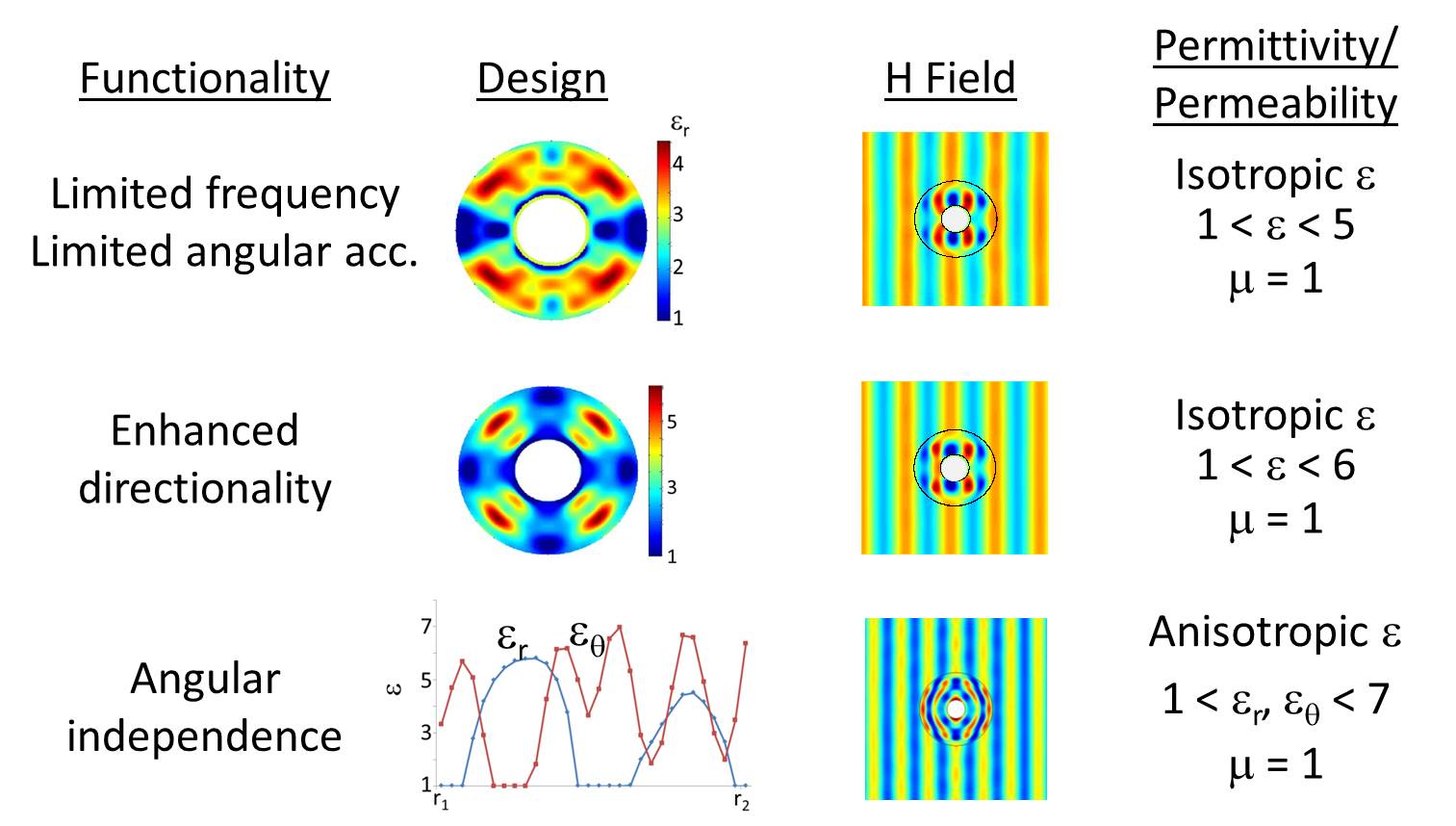}
\caption{Portfolio of designs and tradeoffs for varying degrees of cloaking functionality.  For the simplest functionality of operation over small frequency and angular bandwidths, the permittivity can remain isotropic and with maximum value less than five.  For enhanced directionaliry, imposed through symmetry constraints, the same general material parameters suffice, albeit with slightly stronger contrast.  To achieve full angular independence through rotational symmetry, the permittivity restriction must be loosened, allowing for an anisotropic $\epsilon$ (still with $\mu=1$).} 
\label{fig:CloakTradeoffs}
\end{figure}

While the optimization of Fig.~\ref{fig:2dOptimization} was interesting, the design reduces scattering for only a limited angular bandwidth.  However, the strength of the inverse design method is its flexibility - the designer can choose the tradeoffs desired.  Fig.~\ref{fig:CloakTradeoffs} shows optimization results for a variety of desired functionalities, and the resulting permittivities.  In each case, the permeability was fixed at $\mu=1$.  To achieve enhanced directionality, in this case four-fold symmetry, a slightly higher contrast permittivity was needed.  For perfect rotational symmetry, optimizing with isotropic $\epsilon$ was not sufficient.  However, performing the optimization with an anistropic $\epsilon$ achieved a sufficiently small scattering width to be considered a cloak, this time with complete angular independence.  The anisotropic permittivity simulations were performed in Comsol Multiphysics \cite{Comsol2012}, with the radial and angular components of the permittivity converted to Cartesian coordinates by the transformation
\begin{equation}
\begin{pmatrix}
\epsilon_{xx} & \epsilon_{xy} \\
\epsilon_{yx} & \epsilon_{yy} 
\end{pmatrix} = 
\begin{pmatrix}
\cos\theta & -\sin\theta \\
\sin\theta & \cos\theta
\end{pmatrix}
\begin{pmatrix}
\epsilon_{rr} & 0 \\
0 & \epsilon_{\theta\theta}
\end{pmatrix}
\begin{pmatrix}
\cos\theta & \sin\theta \\
-\sin\theta & \cos\theta
\end{pmatrix}
\end{equation}
which can be found by ensuring the both $\mathbf{D}$ and $\mathbf{E}$ transform properly under the coordinate transformation.

\begin{figure}
\centering
\includegraphics[width=5.5in]{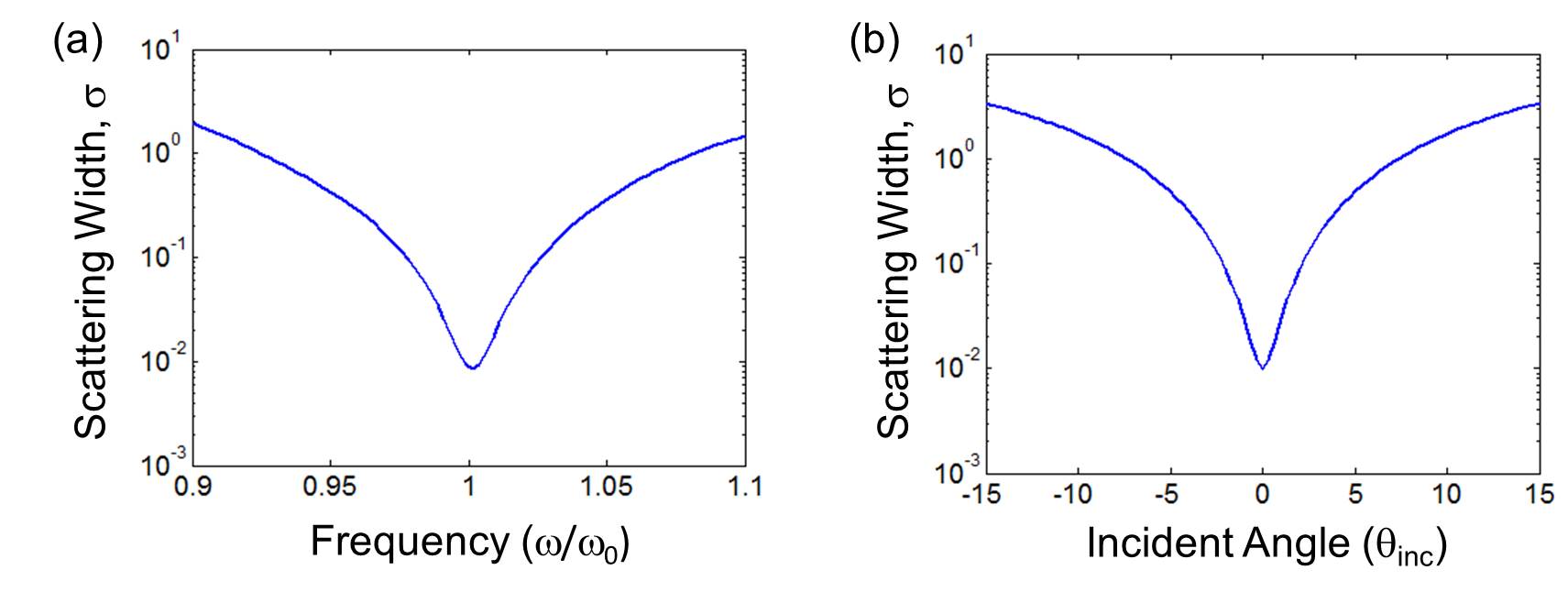}
\caption{(a) Frequency dependence and (b) angular dependence for cloaking design with four-fold symmetry (second example in Fig.~\ref{fig:CloakTradeoffs}).  The object has a scattering width $\sigma<1$ for $\Delta \omega / \omega_0 \approx 1/6$, or for a variation in incidence angle of less than $8^\circ$.} 
\label{fig:FreqAngleDep}
\end{figure}

Fig.~\ref{fig:FreqAngleDep} shows the scattering width as a function of the frequency (for $\theta_{inc}=0$) and of the incidence angle (for $\omega = \omega_0$) for the design with four-fold symmetry.  For a scattering width $\sigma < 0.1$, the relative bandwidth $\Delta \omega / \omega_0$ is approximately $1/20$.  The relative bandwidth over which $\sigma<1$ is $\Delta\omega / \omega_0 \approx 1/6$.  For the angular response, $\sigma<0.1$ is achieved for $|\theta_{inc}|<2^\circ$, while $\sigma<1$ is achieved for $|\theta_{inc}|<8^\circ$.

\begin{figure}
\centering
\includegraphics[width=5in]{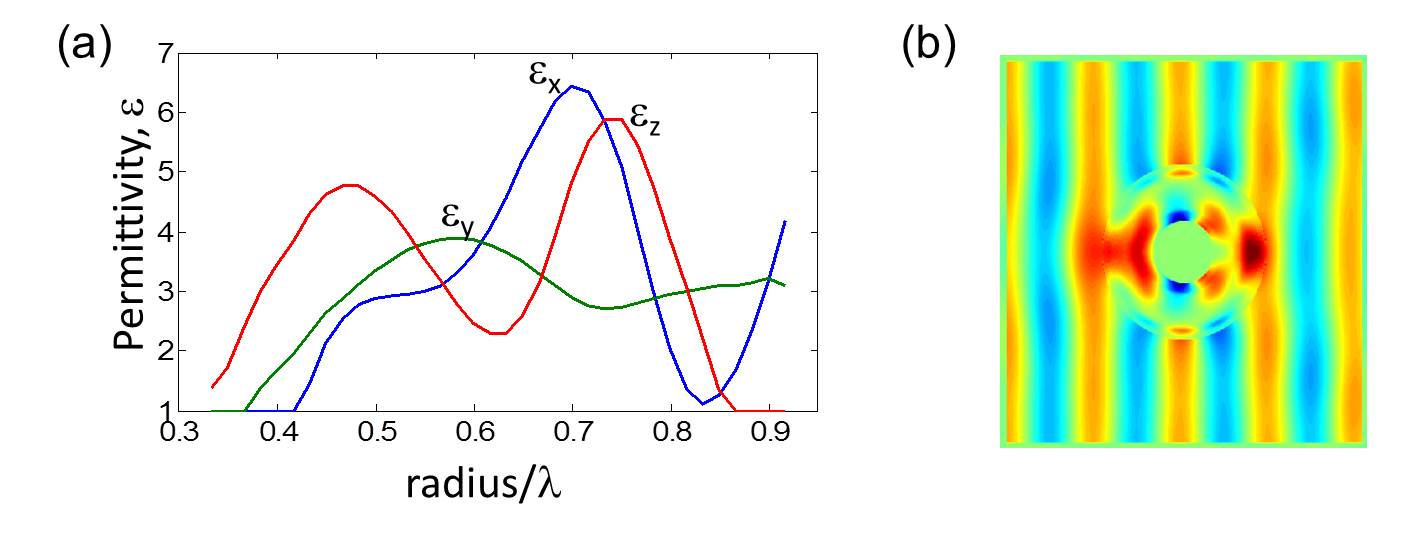}
\caption{(a) Optimal design and (b) two-dimensional cross section of electric field for a three-dimensional cloak optimization.  The structure is simulated with concentric spheres with varying values of $\epsilon_x$, $\epsilon_y$, and $\epsilon_z$.} 
\label{fig:3dCloakOptimization}
\end{figure}

Similar to the third example of Fig.~\ref{fig:CloakTradeoffs}, an anistropic permittivity can also be designed for three-dimensional structures.  The inverse design method scales seamlessly to three dimensions, with the increased computational cost coming only as a result of the larger simulations.  A three-dimensional optimization was completed with the same setup as the two-dimensional optimization, but extended to a sphere instead of a cylinder.  All simulations were performed in Lumerical's FDTD Solutions \cite{Lumerical2012}.  Consequently, the permittivity could not be optimized in the $(r,\theta,\phi)$ basis; instead, the $(x,y,z)$ basis had to be used, and the cloak therefore does not have rotational symmetry.  Nevertheless, for a small angular bandwidth the cloak does achieve a small scattering width, as seen by the field profile in Fig.~\ref{fig:3dCloakOptimization}(b).

\begin{figure}
\centering
\includegraphics[width=6in]{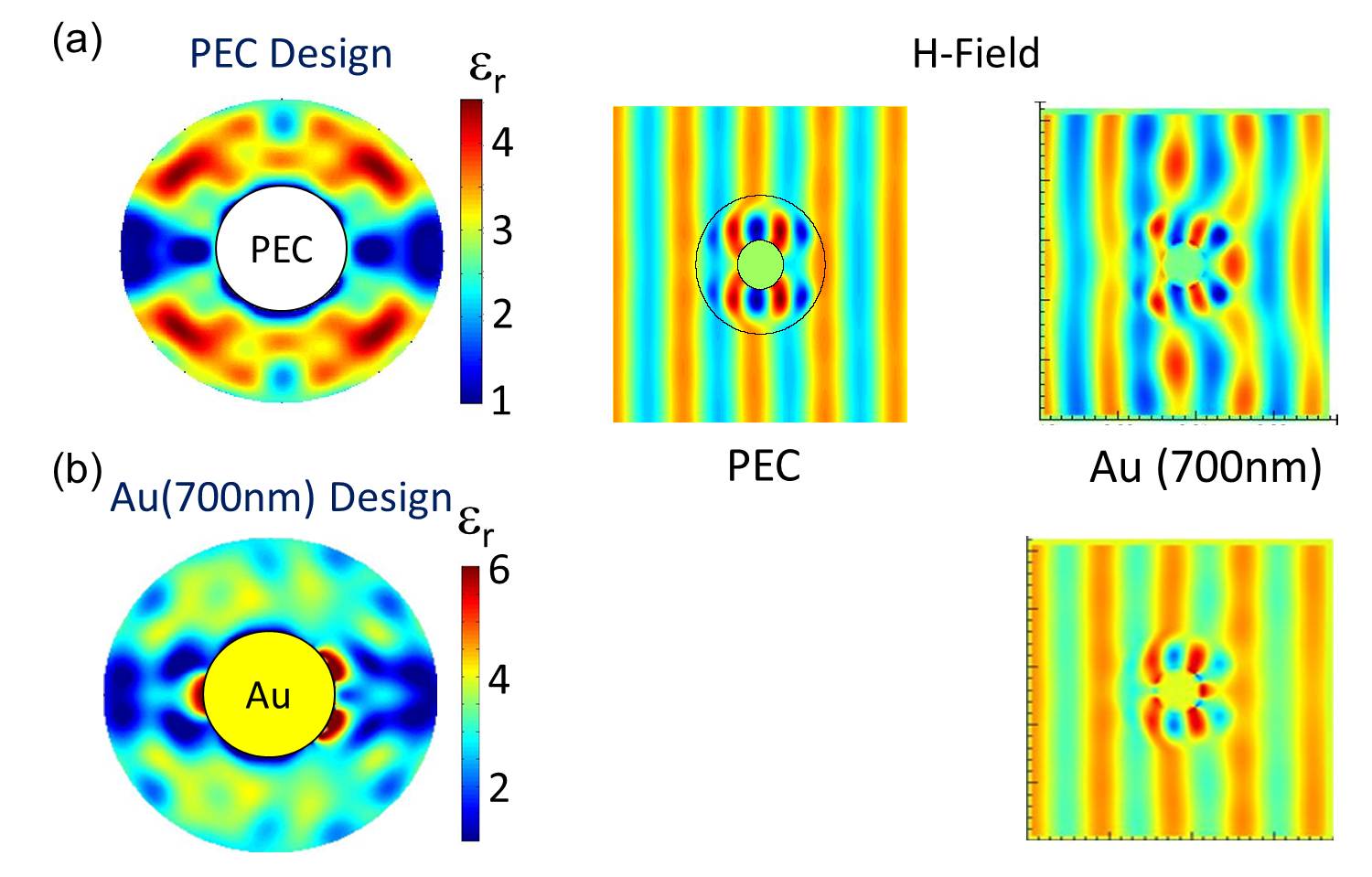}
\caption{The inverse design method can incorporate real material parameters directly into the optimization.  The performance of the design optimized with a PEC interior boundary, shown in (a), degrades significantly if a real material such as gold is used instead.  (b) By including the gold directly into the optimization, a new design can be found that minimizes the effect of the gold.  For this example, the permittivity of $Au$ at $700nm$ wavelength was used.} 
\label{fig:CloakAu}
\end{figure}

An important property of the inverse design method is its intelligence with regard to the Maxwell equations.  Because it depends on full-wave simulations each iteration, it can directly take into account real material properties.  For example, the coordinate transformation method of transformation optics often works for the case of a perfect electrical conductor (PEC), subject to the boundary condition $\hat{\mathbf{n}}\times \mathbf{E} = 0$.  However, especially at optical frequencies, the assumption of perfect conductors is unrealistic.  At $1.8eV$ (${\sim}700nm$), for example, gold has a permittivity $\epsilon_{Au}(700nm) = -14.4+0.9i$ \cite{Palik1998}, very different from a perfect conductor.  Fig.~\ref{fig:CloakAu}(a) shows the optimal design for a cloak assuming a PEC interior boundary.  When the boundary is a perfect conductor, the magnetic field passes essentially unperturbed.  However, if the real Au parameters are used with the design, the phase fronts are distorted and the cloaking behavior is severely diminished.  With the inverse design approach, however, this can be easily remedied: simply re-run the optimization, with the real Au parameters included.  Fig.~\ref{fig:CloakAu}(b) shows the resulting design from such an optimization.  Notice the sharp distinction with the design assuming a PEC.  In the optimization with real gold, there is a build-up of dielectric directly in front of the gold, ostensibly to re-direct waves around it \emph{without} strongly interacting with the gold itself.  In spite of the lossiness of the gold, the optimal design is still able to achieve a significantly reduced scattering width and recover the cloaking functionality.

This chapter demonstrated the flexibility of inverse design through shape calculus.  The example of optical cloaking proved fertile ground for exploration of designs with varying degrees of functionality.  Structures achieving cloaking over only a limited angular and frequency bandwidths could be achieved with remarkably simple materials; for more complete functionality, more complex materials could also be optimized.  The method scales seamlessly to three dimensions, and inherently incorporates the real material parameters that would be needed experimentally or commercially.  Even for a single application, the possible merit functions are often manifold, requiring an optimization method flexible and extensible enough to accommodate every possible circumstance.  Inverse design through shape calculus is such a method.

%% file: SolarOpt.tex
\chapter{Surface Textures for Sub-Wavelength Solar Cells}
\label{chap:SolarOpt}
\epigraph{God made the bulk; surfaces were invented by the devil.}{Wolfgang Pauli}

\noindent
Light trapping is a concept of fundamental importance to solar cells.  Light trapping by surface texturing can provide significant absorption enhancement; equally as important is the significant voltage boost that is also provided (cf. Sec.~\ref{sec:RandText}).  In the ray optics regime, there is a well-known $4n^2$ absorption enhancement limit for a randomly textured surface that cannot be surpassed \cite{Yablonovitch1982a}.  The limit does not apply in the wave optics regime, however, and it is an open question as to what enhancement is possible.  There have been suggestions that plasmonic effects could enable significant enhancement \cite{Atwater2010,Ferry2010a}, but specific structures with significant enhancements have not yet been proposed.  Large enhancements have been theoretically demonstrated for periodic structures in the sub-wavelength regime \cite{Yu2010,Green2011c}, but the proposed designs utilize specialized modal enhancement that occurs only for low-index materials.  For the high-index semiconductors relevant for solar cells, a different structure will be needed.

Actually beating the $4n^2$ limit in the sub-wavelength regime may not be necessary.  Realistic solar materials are either such poor absorbers that they cannot feasibly reach the sub-wavelength regime, or they are sufficiently strong absorbers that even for, e.g., $2n^2$ enhancement they can reach thicknesses ${\ll}100nm$.  The optical absorption depth in bulk crystalline Silicon, for example, is ${\sim} 1000\mu m$, requiring a broadband enhancement factor of greater than $1000$ to even reach sub-micron thicknesses.  Conversely, the optical absorption depth in GaAs is approximately $1\mu m$.  Even just $4n^2$ enhancement would imply thicknesses of $20nm$, beyond which smaller thickness may not be commercially relevant.  The situation with organic materials is similar, as their large absorption cross-sections also also require less than $4n^2$ enhancement to reach nanometer thicknesses.  The goal, therefore, is not necessarily to beat the $4n^2$ limit, but rather to understand how to achieve large enhancements in thin-film, high-index structures pertinent to future photovoltaic technology.


\section{Problem Formulation}
There are two factors determining the absorptivity of photovoltaic cell: the material absorption coefficient, and the geometry coupling incident plane waves to internal modes.  To tease out the effects of the geometry, we consider here the enhancement factor $EF$ of a cell, which can be written
\begin{equation}
\label{eq:EnhFact}
EF = \frac{\frac{1}{\omega_2-\omega_1}\int_{\omega_1}^{\omega_2}a_{avg}(\omega)\,\mathrm{d}\omega}{\alpha L}
\end{equation}
where $(\omega_1,\omega_2)$ is the frequency range considered, $\alpha L$ is the absorption depth, and $a(\omega)$ is the absorptivity of the cell (i.e. probability of absorbing an incident photon).

The enhancement factor must apply for all possible incidence angles (there is significant haze in the sky), meaning the absorptivity $a_{avg}(\omega)$ must be angle-averaged.  However, completing three-dimensional absorption simulations over many angles requires extraordinary computational power.  Instead, we chose the following merit function
\begin{equation}
\label{eq:SolarMeritInit}
F =  \underset{(\omega_1,\omega_2)}{\text{min}} \frac{a_0(\omega)}{\alpha L}
\end{equation}
which is the minimum enhancement over the frequency range $(\omega_1,\omega_2)$, where $a_0$ is the absorptivity at normal incidence.  The idea is that if the minimum enhancement factor is increased over a large bandwidth, the device operation will not depend on highly resonant behavior.  The device, although optimized for normal incidence, will then likely perform well even at skew angles.  A numerical example bearing this out is given later in the chapter.

\begin{figure}
\centering
\includegraphics[width=3.5in]{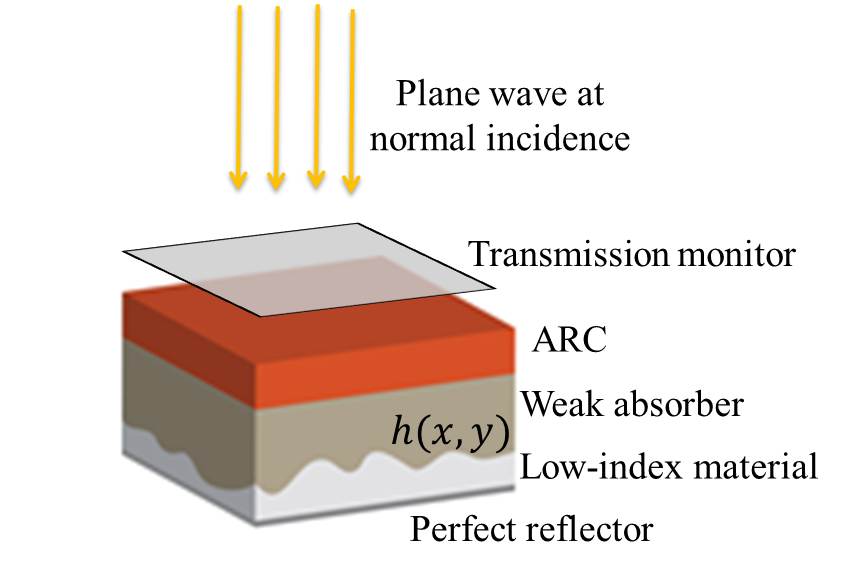}
\caption{Inverse design setup for thin-film solar cell.  An initial anti-reflection coating reduces the index mismatch at the front surface.  The rear surface of the weak absorber is textured, above a low-index material.  A perfect reflector is assumed to ensure maximal absorption enhancement.  A simple computational method of calculating the absorption is to calculate the transmission through the transmission monitor, shown above the anti-reflection coating.} 
\label{fig:SolarSetup}
\end{figure}

Fig.~\ref{fig:SolarSetup} shows the setup for the inverse design of a thin-film absorber.  A plane wave is incident upon the device, which consists of a textured thin-film surrounded by an anti-reflection coating and a low-index material (for fabrication purposes).  A perfect reflector ensures photons are not lost through the rear.  The textured surface is given by $h(x,y)$, and is the geometry to be optimized.  Instead of directly measuring absorption, which would require recording three-dimensional field data, the transmission $T(\omega)$ through a front monitor is measured, and because there is no transmission through the rear surface, $a(\omega) = T(\omega)$.  A weak absorber is chosen for the thin-film in an attempt to understand the limits of absorption enhancement; the basic understanding gleaned from the optimization should also be of help for more strongly absorbing, material-specific systems.

The transmission through the front monitor is the power flowing through the monitor divided by the total incident power.  Given the equivalence between absorptivity and transmission, the merit function can be written
\begin{equation}
\label{eq:SolarMerit}
F = \underset{(\omega_1,\omega_2)}{\text{min}} \left[ -\frac{1}{2P_{inc}} \frac{\operatorname{Re}\int_S \mathbf{E}\times \overline{\mathbf{H}} \cdot \hat{\mathbf{z}} \,\mathrm{d}A}{\alpha L} \right]
\end{equation}
where $\hat{\mathbf{z}}$ is the surface normal to the transmission monitor $S$ and is opposite the direction of the plane wave.  $P_{inc}$ is the incident power.  As derived in Chap.~\ref{chap:Algorithm}, the function $f(\mathbf{x})$ is the integrand of Eqn.~\ref{eq:SolarMerit}.  Similar to Eqn.~\ref{eq:PMCloak}, the electric and magnetic dipoles are set by the derivatives $\partial f/\partial \mathbf{E}$ and $-(1/\mu_0)\partial f/\partial \mathbf{H}$, respectively, resulting in:
\begin{subequations}
\label{eq:PMSolar}
\begin{align}
\mathbf{P} &= \frac{\hat{\mathbf{z}} \times \overline{\mathbf{H}}}{2\alpha LP_{inc}} \\
\mathbf{M} &= -\frac{\hat{\mathbf{z}} \times \overline{\mathbf{E}} }{2\alpha L \mu_0 P_{inc}}
\end{align}
\end{subequations}
The dipoles are located throughout the plane defined by the transmission monitor.

The surface $h(x,y)$ is the design variable for this structure, with $x$ and $y$ the in-plane axes.  To ensure smoothness, the surface constructed from a Fourier basis of sines and cosines,
\begin{align}
h(x,y) = h_0 + & \sum_{i=1}^N\sum_{j=1}^N \left\{ c_{ij1} \sin(\frac{2\pi i x}{L_x})\sin(\frac{2\pi j y}{L_y}) + c_{ij2}\sin(\frac{2\pi i x}{L_x})\cos(\frac{2\pi j y}{L_y}) \right. \nonumber \\
& \qquad \qquad \left. + c_{ij3}\cos(\frac{2\pi i x}{L_x})\sin(\frac{2\pi j y}{L_y}) + c_{ij4}\cos(\frac{2\pi i x}{L_x})\cos(\frac{2\pi j y}{L_y}) \right\}
\end{align}
written as a discrete sum because the surface is assumed to have periodicity $L_x$ in the $x$-direction and $L_y$ in the $y$-direction.  The coefficients $c_{ij}$ (in addition to the constant thickness $h_0$) determine the surface profile.  

The topological derivative, corresponding to generation of new holes, is not required for this design, because a single interface between the absorber and rear dielectric is required.  Therefore only the shape derivative must be calculated.  Eqn.~\ref{eq:ShapeDeriv} is a simple starting point.  For simplicity it will be re-written here
\begin{align}
\delta F = \int \delta x_n(\mathbf{x'}) 
\Lambda(\mathbf{x'}) \,\mathrm{d}A 
\end{align}
where $\Lambda(\mathbf{x'})$ is two times the real part of the term in square brackets in Eqn.~\ref{eq:ShapeDeriv}, containing the direct and adjoint fields.  Instead of parameterizing the integral by the surface element $\mathrm{d}A$ and the variation normal to the surface $\delta x_n$, it is simpler to integrate over the change in height $\delta h$ over the $x$-$y$ plane
\begin{align}
\label{eq:ShapeDerivShort}
\delta F = \int \delta h(x,y) 
\Lambda(x,y) \,\mathrm{d}x\mathrm{d}y
\end{align}
Simplifying the height function into a sum over generalized basis functions:
\begin{equation}
h(x,y) = \sum_i c_i f_i(x,y)
\end{equation}
the variation in height can be replaced by the variation in coefficients through
\begin{align}
\label{eq:deltaHSolar}
\delta h(x,y) = \sum_i \delta c_i f_i(x,y)
\end{align}
Inserting Eqn.~\ref{eq:deltaHSolar} into Eqn.~\ref{eq:ShapeDerivShort} finally yields the shape derive of $F$ with respect to the height coefficients $c_i$:
\begin{align}
\delta F = \sum_i \delta c_i \int f_i(x,y) \Lambda(x,y)\,\mathrm{d}x\mathrm{d}y
\end{align}
Each iteration, the height coefficients are updated by the equation (properly normalized)
\begin{align}
\label{eq:CoeffShapeDeriv}
\delta c_i = \int f_i(x,y) \Lambda(x,y)\,\mathrm{d}x\mathrm{d}y
\end{align}

\section{Surface Texture Optimization}
\begin{figure}
\centering
\includegraphics[width=3in]{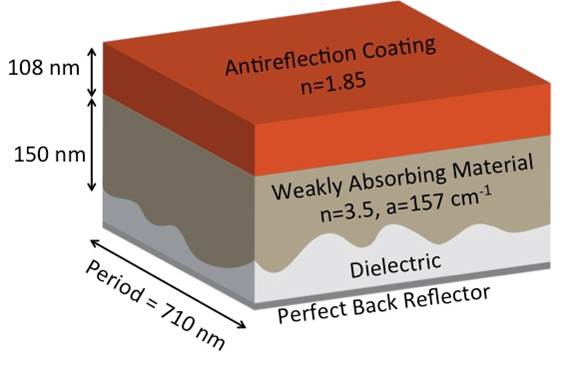}
\caption{Dimensions and materials for surface texture optimization.  A generic weak absorber requires significant enhancement to even approach full absorption.  The average thickness $h_0$ is taken to be $150nm$.  The rear dielectric has $n=1$ for simplicity, and the period of $710nm$ in the $x$ and $y$-directions} 
\label{fig:SolarOptSetup}
\end{figure}
Fig.~\ref{fig:SolarOptSetup} shows the specific geometry optimized.  A quarter-wave anti-reflection coating (with index $n_{ARC}\approx \sqrt{n_{air}n_{semi}}$) reduces reflection due simply to the impedance mismatch of air and the absorber.  Nominally\footnote{Nominal values because scale-invariance permits re-scaling.} the energy range is taken to be $1.45$-$1.65eV$, very close to the band-edge of GaAs.  More importantly, the energy bandwidth $\Delta \omega / \omega_0 \approx 1/8$, a sizeable bandwidth.  The corresponding wavelength range is $750$-$850nm$.  The sub-wavelength absorber is taken be $150nm$ thick, with the absorption coefficient chosen such that $\alpha L \approx 0.02$. The rear surface is textured to avoid influencing the effect of the anti-reflection coating on strongly absorbed photons.\footnote{i.e. photons that require less than a single pass through the cell to be absorbed, which may be outside the frequency range of the optimization.}  The rear dielectric is chosen to have an index $n=1$, for simplicity and to understand optimal texturing effects.

\begin{figure}
\centering
\includegraphics[width=6.5in]{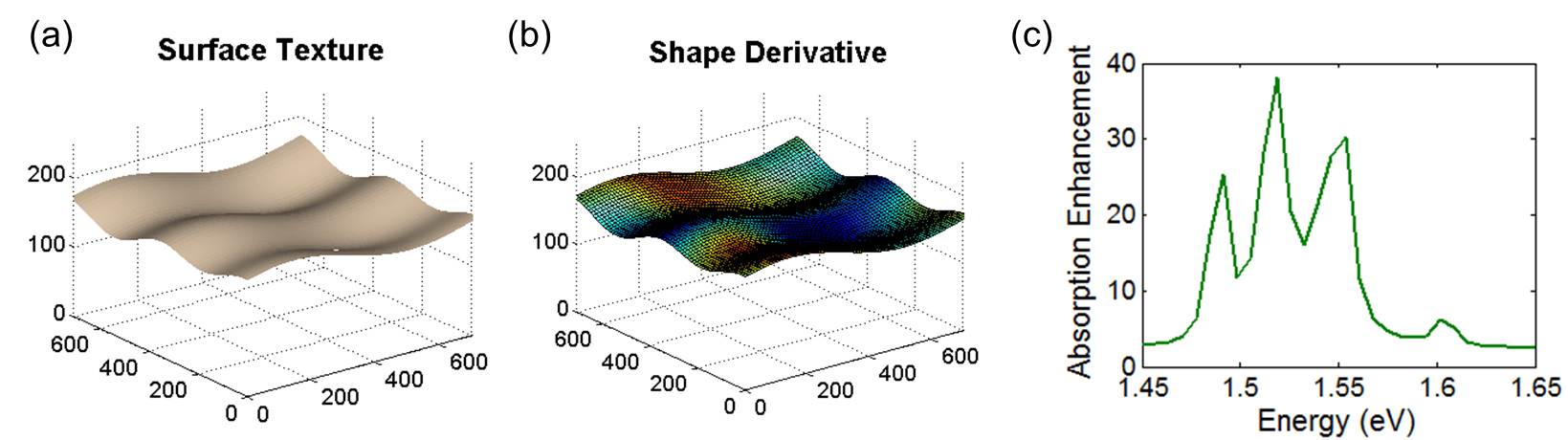}
\caption{(a) Initial surface for optimization, with small random roughness to break the symmetry of the device.  (b)  Shape derivative of the initial shape.  Red areas indicate that the height should increase, whereas blue indicates the the height should decrease.  (c)  Absorption enhancement as a function of energy.  Aside from the resonant peaks, the background enhancement is roughly $2$, consistent with the anti-reflection coating and perfect rear mirror.}
\label{fig:InitSurface}
\end{figure}
The initial surface for the optimization is shown in Fig.~\ref{fig:InitSurface}(a).  The exact structure is not important; for roughly random initial structures, the final result tends to be very similar.  A small perturbation (on the order of $5nm$ in the Fourier coefficients) is used to break the symmetry and enable the optimization to proceed.  Given the direct and adjoint simulations for the given structure,\footnote{All three-dimensional simulations were completed in Lumerical \cite{Lumerical2012}, on an 80-core cluster.} the shape derivative $\Lambda(x,y)$, from Eqn.~\ref{eq:ShapeDerivShort}, is shown in Fig.~\ref{fig:InitSurface}(b).  Red (blue) areas indicate that the height should increase (decrease).  This derivative is projected onto the Fourier basis by Eqn.~\ref{eq:CoeffShapeDeriv}.

Fig.~\ref{fig:InitSurface}(c) shows the absorption enhancement, $a(\omega)/\alpha L$, for the initial structure as a function of energy.  Resonant peaks indicating resonant coupling are seen, but the background enhancement is roughly two, as expected from the double-pass permitted by the rear mirror.

\begin{figure}
\centering
\includegraphics[width=5in]{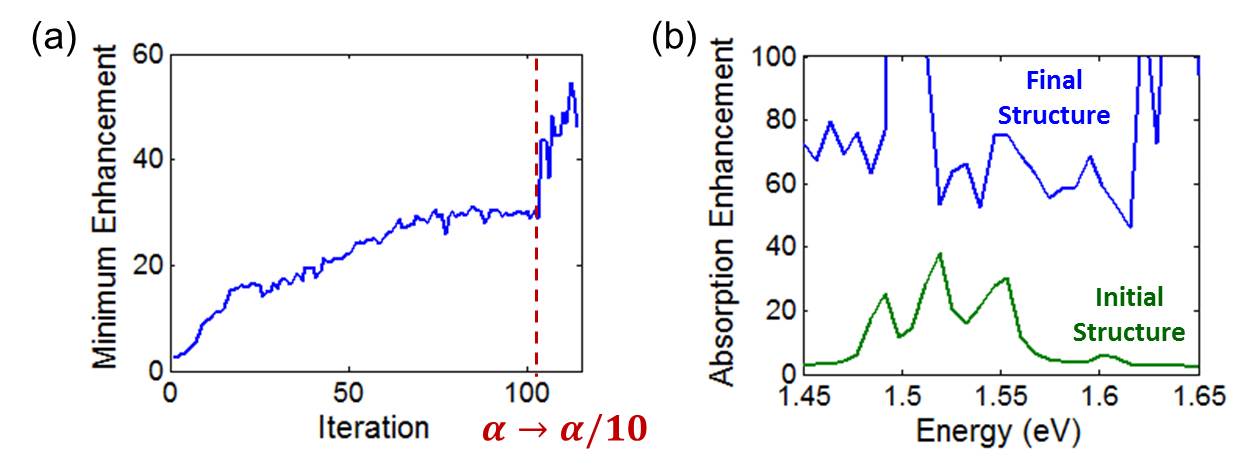}
\caption{(a) Minimum absorption enhancement as a function of iteration.  By the $104^{\mathrm{th}}$ iteration, the enhancement had improved sufficiently such that it was no longer a weak absorber.  The absorption coefficient was reduced by a factor of ten, such that the structure was again a weak absorber.  (b) Comparison of initial and final structure enhancement factors as a function of energy.  Whereas there is a large bandwidth over which the initial structure has an enhancement of roughly $2$, the final structure has an enhancement greater than $50$ over the entire bandwidth.}
\label{fig:EnhancementIter}
\end{figure}
The optimization proceeds by attempting to increase the worst-performing frequency at each iteration.  The direct simulation injects a plane wave across all frequencies within the relevant range, and the enhancement is measured as a function of frequency.  The frequency at which the minimum enhancement occurs is selected for the adjoint simulation, and sets the frequencies of the dipoles given by Eqn.~\ref{eq:PMSolar}.  Note that this simplification of the ``minimax'' problem formulation likely works well only because of the broad bandwidth and many resonances of the problem; for more general approaches to the minimax problem cf. \cite{Polak1997,Clarke1990,Murray1980}.

The evolution of the merit function, the minimum enhancement over the $1.45$-$1.65eV$ range, is shown in Fig.~\ref{fig:EnhancementIter}(a).  The minimum enhancement increases substantially.  At the $104^{th}$ iteration, once the minimum merit function stabilized, the absorption coefficient was reduced by a factor of ten.  There was such improvement that the structure was no longer in the weakly absorbing regime; reducing the absorption coefficient by ten restored the weakly absorbing nature and consequently increased the enhancement factors.  By the end of the optimization, the \emph{minimum} enhancement factor was $54.5$.    

The absorption enhancement as a function of energy is shown in Fig.~\ref{fig:EnhancementIter}(b).  The enhancement for the initial structure is shown in green, while the enhancement for the final structure is shown in blue.  Note that because of the choice of the $min$ merit function, the enhancement across the entire energy range has improved, such that even at non-resonant energies the enhancement is greater than $50$.  The average enhancement of the final structure is $78$.

\begin{figure}
\centering
\includegraphics[width=5in]{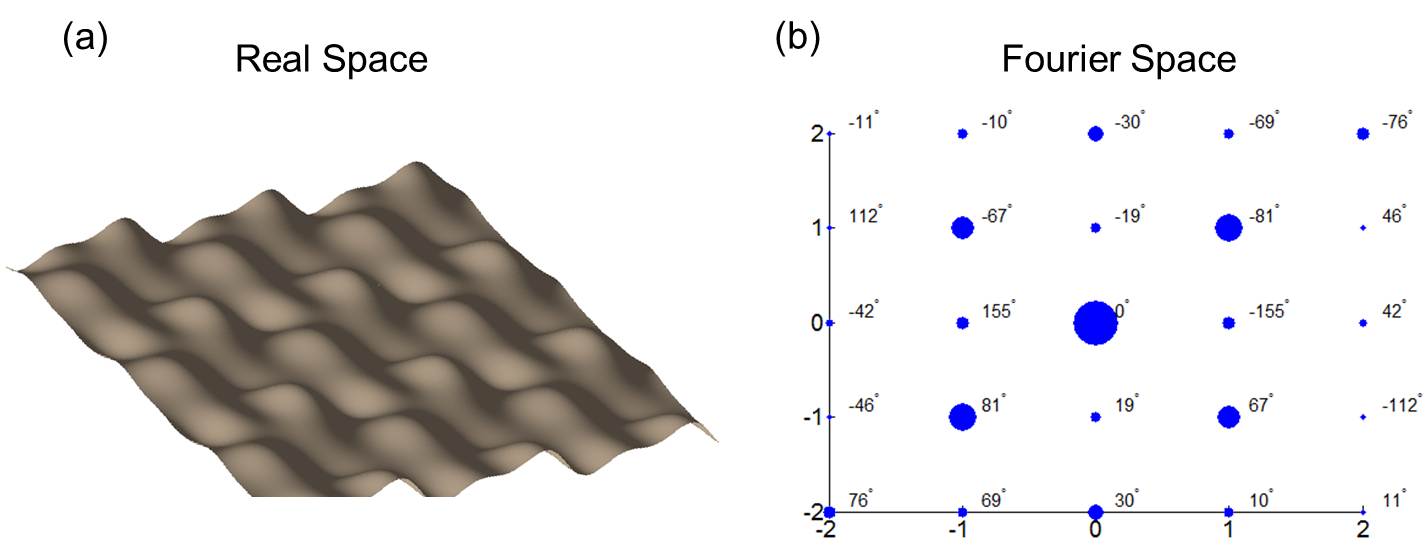}
\caption{(a) Real space and (b) Fourier space representations of the optimal surface.  The size of the dots in the Fourier space representation corresponds to the magnitude, while the phase is written adjacent.}
\label{fig:RealKSpace}
\end{figure}
The spatial variation of the surface is shown in Fig.~\ref{fig:RealKSpace}(a).  The peak-to-valley distance of the surface is ${\approx}160nm$, centered around the $150nm$ thickness.  The coefficients $c_{ij}$ of the optimal structure are shown in Fig.~\ref{fig:RealKSpace}(b).  They have been converted to an exponential basis, with basis functions of the form $\exp\left[2\pi i (mx/L_x + ny/L_y)\right]$.  The size of the circles in the figure is proportional to the magnitude of the coefficients, while the phase is displayed in the adjacent text.  Note that there have been no symmetry conditions imposed, but because the surface height is real-valued the relation $\gamma_{m,n} = \overline{\gamma}_{-m,-n}$ must hold, where $\gamma$ is the coefficient in the exponential basis.

An average enhancement of $78$, as mentioned previously and seen in Fig.~\ref{fig:EnhancementIter}(b), is far greater than the $4n^2$ limit of ray optics.  However, it is only for normal incidence and a single polarization.  As mentioned in the previous section, the optimization was performed at normal incidence and one polarization to reduce the computational complexity.  By virtue of the non-resonant improvement over a large frequency range, it is expected that even at skew angles or different polarizations the behavior of the structure should not be that different. 

\begin{figure}
\centering
\includegraphics[width=5in]{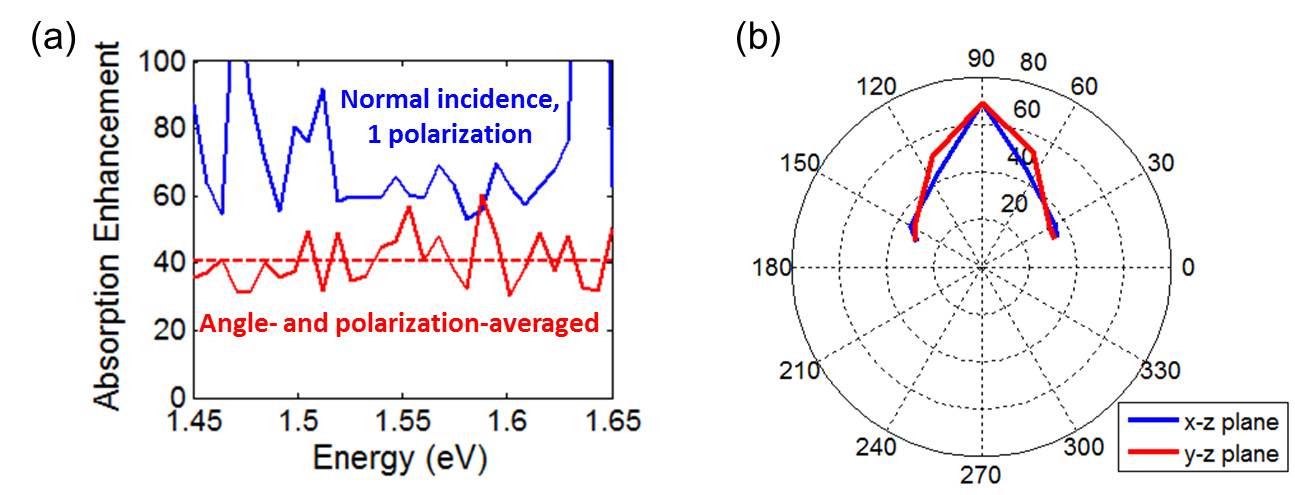}
\caption{(a) Absorption enhancement as a function of energy for the optimized structure at normal incidence and for a single polarization (blue) and averaged over angles and polarization (red).}
\label{fig:EnhEnergyAngle}
\end{figure}
Fig.~\ref{fig:EnhEnergyAngle}(a) shows the two quantities on the same plot: the enhancement versus energy for the optimization incidence angle and polarization, and the angle- and polarization-averaged enhancement.  The energy- and polarization-averaged enhancement, as a function of incident angle, is displayed in Fig.~\ref{fig:EnhEnergyAngle}(b).  One can see that although there is a penalty at off-angles, the enhancement factor remains above $20$ for any orientation, and the overall angle- and polarization-averaged enhancement is $40$.  The angle-averaging is performed by partitioning the hemisphere into equal solid angles, and averaging according to $\int EF \cos\theta \,\mathrm{d}\Omega$, assuming the Lambertian $\cos\theta$ weighting factor.

The angle- and polarization-averaging results justify the approach of optimizing at normal incidence.  Performing the optimization over the full angular and polarization space would have increased the computational time of the optimization by a factor of about $25$, increasing the time to optimize from $3$ days to $75$, which is clearly unfeasible.

\begin{figure}
\centering
\includegraphics[width=3in]{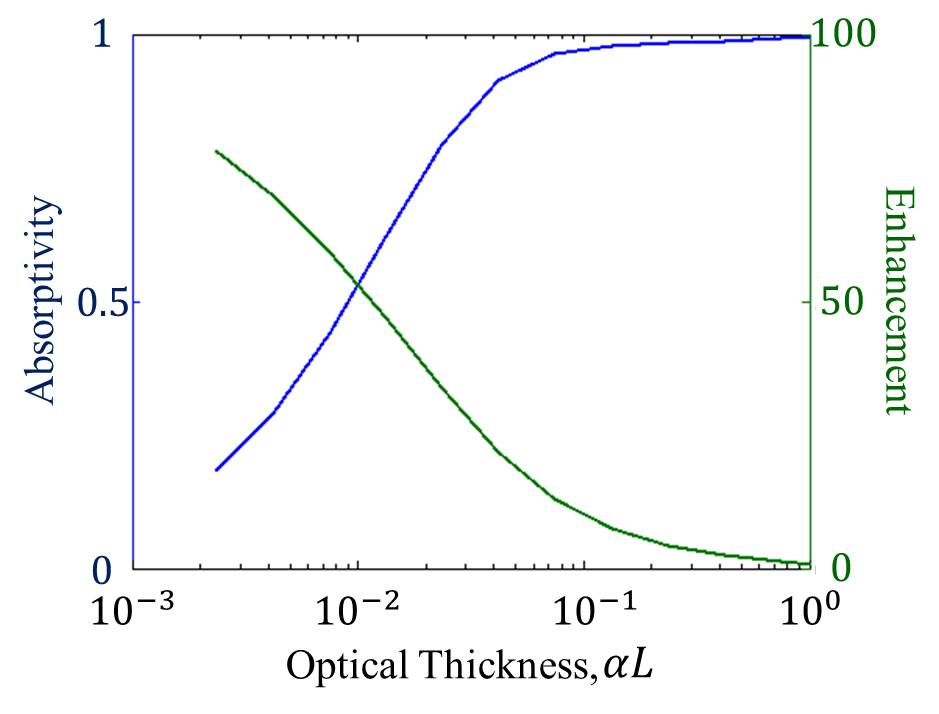}
\caption{Frequency-averaged absorptivity at normal incidence as a function of varying the optical thickness, $\alpha L$.  The structure performs well in the strongly-absorbing regime, surpassing $90\%$ absorption even at $\alpha L = 0.1$.}
\label{fig:StronglyAbsorbed}
\end{figure}

An important question to investigate is how the structure performs in the strongly absorbing regime.  While the weakly absorbing regime illuminates the underlying physics, it is not relevant for commercial technology, as realistic solar cells must absorb close to $100\%$ of the available incident spectrum.  Fig.~\ref{fig:StronglyAbsorbed} shows the frequency-averaged, normal-incidence enhancement factor and total absorption as a function of the optical thickness, $\alpha L$.  Because the structure was designed for the specific thickness $L_{avg}=150nm$, the optical thicknesses are simulated through varying the absorption coefficient $\alpha$.  Note that the structure performs very well as the optical thickness increases, asymptotically approaching $100\%$ absorption as the optical thickness increases.  Even for $\alpha L = 0.1$, the absorption surpasses $90\%$ on average across the entire bandwidth.  As an example of the relevance of this calculation, this suggests a $1.5\mu m$ thick, high-efficiency GaAs solar cell could be reduced to $150nm$ thick with little degradation of efficiency, assuming the surface can be manufactured.  Such an improvement would greatly reduce the associated material costs and enable cheaper electricity going forward.

\begin{figure}
\centering
\includegraphics[width=5.5in]{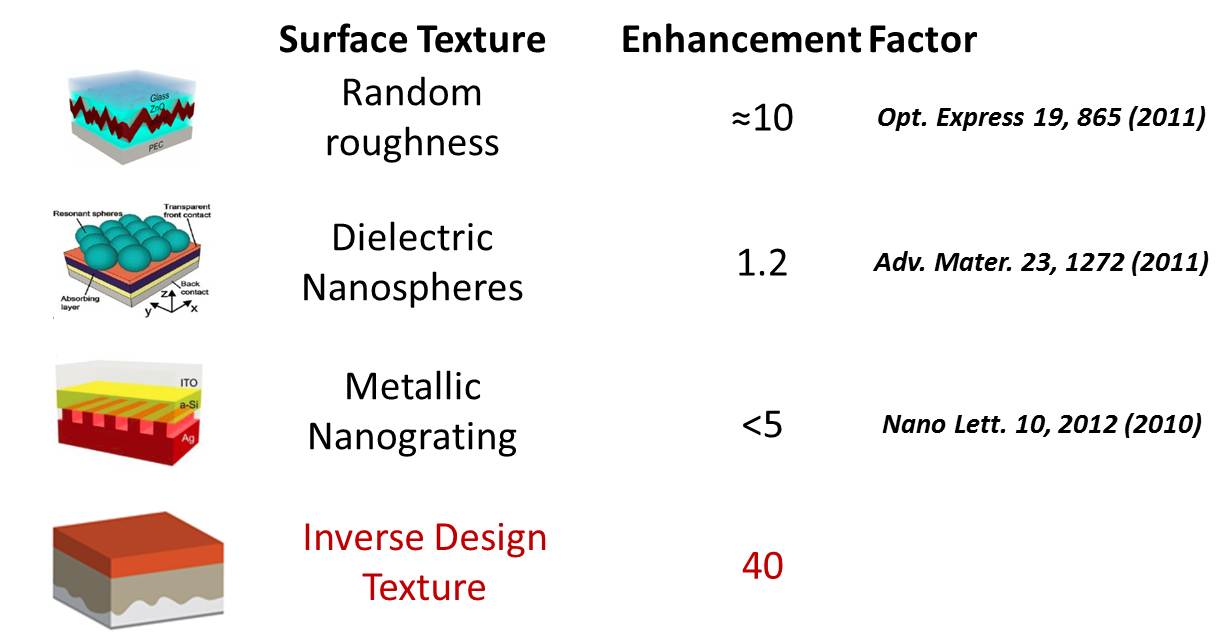}
\caption{Comparison of published enhancement methods and the inverse design texture presented in this work.  The non-intuitive, shape calculus-determined design presented here shows far superior performance than the heuristic structures designed elsewhere.}
\label{fig:DesignComps}
\end{figure}
An average enhancement factor of $40$ is close to the ray optics limit of $4n^2\approx 50$.  Although it does not surpass the ray optics limit, as discussed in the introduction to this chapter, surpassing the ray optics limit is not necessarily important for relevant solar cell technologies.  Indeed, a factor of $40$ enhancement would enable a currently $1\mu m$ thick cell to be scaled down to $25nm$, far enough to greatly reduce costs.

Fig.~\ref{fig:DesignComps} compares the surface texture optimized here with published results from the literature.  Although there have been only a few papers proposing methods for enhancing absorption with sub-wavelength, high-index materials, the improvement provided by inverse design is decisive.  The next highest enhancement factor, calculated for a randomly rough surface, is only about $10$ \cite{Fahr2011}.  Even though \cite{Fahr2011} approaches the Lambertian limit, it occurs in a highly-absorbing structure, leaving open the question of how a randomly roughened structure would perform in the weakly absorbing limit.  The two other proposals, dielectric nanospheres \cite{Grandidier2011} and nanometallic gratings \cite{Wang2010}, have even smaller enhancement factors and re-illustrate the point that symmetric or intuitive structures will tend to have far sub-optimal performance.

Stuart and Hall \cite{Stuart1997} calculated the thermodynamic limit to light trapping in thin planar structures.  They use the real modal structure of a planar thin-film, while assuming perfect coupling into and out of the modes.  For a thickness of $150nm$ and the relevant refractive indices, their calculated enhancement factor limit is $31$ \cite{Kosten2012}.  The fact that the inverse design structure as presented here surpasses this limit signifies that not only is the texture acting as a coupler, but it is also important shaping the bandstructure and modal properties directly, increasing the density of states and reducing the group velocity as needed.

%% file: MaxwellSymm.tex
\chapter{Symmetries of the Maxwell Equations}
\label{chap:SymmMaxwell}

\begin{figure}[]
\centering
\includegraphics[width=3.5in]{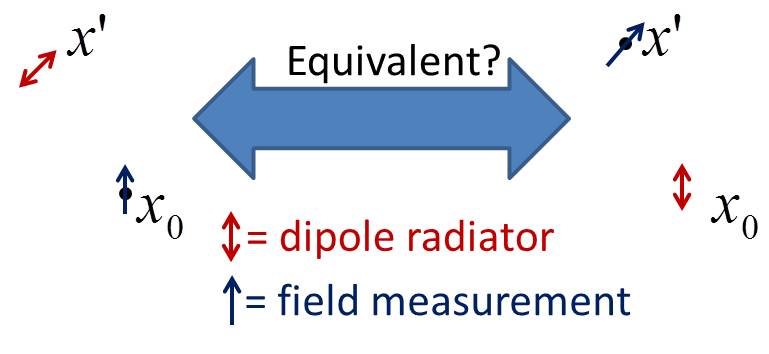}
\caption{The focus of Appendix~\ref{chap:SymmMaxwell}: can the source and measurement points be exchanged for a dipole (electric or magnetic) at $\mathbf{x}$, radiating to $\mathbf{x'}$ ($\Eb$ or $\Hb$)? } 
\label{fig:DipoleEquivalence}
\end{figure}

\noindent
The shape calculus of Part II of this thesis is predicated on understanding how the solutions to Maxwell's equations change under small perturbations to the geometry.  Being able to capture that information with only two simulations is possible because of two simplifications. First, the response to the perturbation can be understood through only the lowest-order multipole moment, that of the electric dipole.  Second, a reciprocity relation is needed, as an answer to the following question:
\begin{quote}
Can the fields at a point $x'$, from a dipole at $x_0$, be recovered by placing a dipole at $x'$ and measuring at $x_0$ instead?  
\end{quote}
This question is represented pictorially in Fig.~\ref{fig:DipoleEquivalence}.

The equations answering the previous question are known as the Rayleigh-Carson reciprocity relations.  Although they have been derived elsewhere \cite{Rayleigh1945,Carson1924,Tai1992,Rothwell2001}, we will provide an alternate derivation here, as they are crucial components of the inverse design procedure.

The two divergence relations in Maxwell's equations can be produced from the two curl equations (for $\omega \neq 0$), and initial conditions can be specified in place of the divergence equations \cite{Rothwell2001}.  We focus then on the curl equations.  It is simpler to work with one second-order differential equation than two first-order differential equations, for both the electric and magnetic fields:
\begin{subequations}
\label{eq:2ndOrderDiff}
\begin{align}
\left[ \nabla \times \mur^{-1} \nabla \times - \omega^2 \epsr \right] \mathbf{E} & = \omega^2 \mathbf{P} - j\omega \mu_0 \nabla \times \mur^{-1} \mathbf{M} \\ 
\left[ \nabla \times \epsr^{-1} \nabla \times - \omega^2 \mur \right] \mathbf{H} & = j\omega \nabla \times \epsr^{-1} \mathbf{P} + \mu_0 \omega^2 \mathbf{M}
\end{align}
\end{subequations}
Each equation is of the form $\overline{\overline{\mathbf{A}}}\mathbf{x}=\mathbf{b}$, where $\overline{\overline{\mathbf{A}}}$ is a linear differential operator, $\mathbf{x}$ is the desired field, and $\mathbf{b}$ is the set of specified sources.  Motivated to understand symmetries with respect to dipole moments, the sources have been explicitly written in terms of a polarization density $\mathbf{P}$ and a magnetization density $\mathbf{M}$, using the relations $\mathbf{D} = \epsr \mathbf{E} + \mathbf{P}$ and $\mathbf{B} = \mur \mathbf{H} + \mu_0 \mathbf{M}$.

It will be useful to take one more step, tranforming the differential equations \ref{eq:2ndOrderDiff} into their \emph{weak forms}.  Commonly used in finite element methods, weak forms have less strict differentiability requirements and more naturally employ symmetry.  If we consider a domain $D$ over which equations \ref{eq:2ndOrderDiff} are to be solved, the weak forms can be derived as follows: take the dot product of the equation with a test function $V$, and integrate over $D$.  This results in the equations:
\begin{subequations}
\label{eq:MaxwellWeak}
\begin{align}
\int_D \left[  \mur^{-1} \nabla \times \mathbf{E} \cdot \nabla \times \mathbf{V} - \omega^2 \epsr \mathbf{E} \cdot \mathbf{V} \right] \text{d}^3 \mathbf{x} & = \int_D \left[ \omega^2 \mathbf{P} \cdot \mathbf{V} - j\omega\mu_0 \nabla \times \mur^{-1} \mathbf{M} \cdot \mathbf{V} \right] \text{d}^3 \mathbf{x} \\ 
\int_D \left[ \epsr^{-1} \nabla \times \mathbf{H} \cdot \nabla \times \mathbf{V} - \omega^2 \mur \mathbf{H} \cdot \mathbf{V} \right] \text{d}^3 \mathbf{x} & = \int_D \left[ j\omega \nabla \times \epsr^{-1} \mathbf{P} \cdot \mathbf{V} + \mu_0 \omega^2 \mathbf{M} \cdot \mathbf{V} \right] \text{d}^3 \mathbf{x}
\end{align}
\end{subequations}
for all $\mathbf{V} \in H_0(\text{curl}; D)$, where $H_0(\text{curl}; D)$ is the space of allowable electric and magnetic fields.\footnote{The requirements on $\Eb$ and $\Hb$ are that they be Lebesgue-integrable; consequently, $H_0(\text{curl};D)$ is the space of vector fields that are Lebesgue-integrable and for which the curl of the field is Lebesgue-integrable.}  Integration by parts enabled the first term on the left-hand side of each equation to be written as a single curl operator in each of the fields.  The boundary conditions $\mathbf{n} \times \mathbf{E}=0$ and $\mathbf{n} \times \mathbf{H}=0$ on the boundary of $D$ have been assumed, but this is merely for clarity and intuition, and more general boundary conditions do not impair any of the upcoming steps.  It should be clear that Eqns.~\ref{eq:MaxwellWeak} are equivalent to \ref{eq:2ndOrderDiff}, as \ref{eq:MaxwellWeak} must be valid for all possible $\mathbf{V}$.  One could therefore choose $\mathbf{V}$ to approach in the limit a delta function at any point in space, recovering Eqns.~\ref{eq:2ndOrderDiff} explicitly.

There are in general four cases to consider: the electric and magnetic fields from an electric dipole, and the electric and magnetic fields from a magnetic dipole.  It turns out that the magnetic fields from an electric dipole and the electric fields from a magnetic dipole are related, reducing the number of unique cases to three.

\subsection{Case I: The electric field from an electric dipole}
For this case, we will only need Eqn.~\ref{eq:MaxwellWeak}(a).  Two separate situations are considered.  In the first, a dipole with amplitude $1 \text{ C}\cdot\text{m}$ and polarization $\mathbf{\hat{\epsilon_1}}$ is placed at the point $\mathbf{x_1}$.  In the second, a dipole with amplitude $1 \text{ C}\cdot\text{m}$ and polarization $\mathbf{\hat{\epsilon_2}}$ is placed at the point $\mathbf{x_2}$:
\begin{align}
\mathbf{P_1} = \mathbf{\hat{\epsilon_1}} (1 \text{ C$\cdot$m})\delta ^3(\mathbf{x}-\mathbf{x_1}) \\
\mathbf{P_2} = \mathbf{\hat{\epsilon_2}} (1 \text{ C$\cdot$m})\delta ^3(\mathbf{x}-\mathbf{x_2})
\end{align}
The fields resulting from $\mathbf{P_1}$ and $\mathbf{P_2}$ are $\mathbf{E_1}$ and $\mathbf{E_2}$, respectively.  The background $\epsr$ and $\mur$ are kept fixed.  We re-write Eqn.~\ref{eq:MaxwellWeak}(a) for each of the two scenarios:
\begin{align}
\label{eq:w11}
\int_D \left[  \mur^{-1} \nabla \times \mathbf{E_1} \cdot \nabla \times \mathbf{V} - \omega^2 \epsr \mathbf{E_1} \cdot \mathbf{V} \right]\text{d}^3 \mathbf{x} & = \omega^2 \mathbf{\hat{\epsilon_1}} \cdot \mathbf{V}(\mathbf{x_1}) \\ 
\label{eq:w12}
\int_D \left[  \mur^{-1} \nabla \times \mathbf{E_2} \cdot \nabla \times \mathbf{V} - \omega^2 \epsr \mathbf{E_2} \cdot \mathbf{V} \right] \text{d}^3 \mathbf{x} & = \omega^2 \mathbf{\hat{\epsilon_2}} \cdot \mathbf{V}(\mathbf{x_2})
\end{align}
We now employ the power of the weak forms.  Eqns.~\ref{eq:w11},\ref{eq:w12} must be true for all possible $\mathbf{V}$; we can therefore choose a particular $\mathbf{V}$ as needed.  In Eqn.~\ref{eq:w11}, we take $\mathbf{V} = \mathbf{E_2}$, and in Eqn.~\ref{eq:w12}, we take $\mathbf{V} = \mathbf{E_1}$.  The previous equations now read:
\begin{align}
\label{eq:w13}
\int_D \left[  \mur^{-1} \nabla \times \mathbf{E_1} \cdot \nabla \times \mathbf{E_2} - \omega^2 \epsr \mathbf{E_1} \cdot \mathbf{E_2} \right] \text{d}^3 \mathbf{x} & = \omega^2 \mathbf{\hat{\epsilon_1}} \cdot \mathbf{E_2}(\mathbf{x_1}) \\ 
\label{eq:w14}
\int_D \left[  \mur^{-1} \nabla \times \mathbf{E_2} \cdot \nabla \times \mathbf{E_1} - \omega^2 \epsr \mathbf{E_2} \cdot \mathbf{E_1} \right] \text{d}^3 \mathbf{x} & = \omega^2 \mathbf{\hat{\epsilon_2}} \cdot \mathbf{E_1}(\mathbf{x_2})
\end{align}
We want to take advantage of the fact that the left-hand sides of Eqns.~\ref{eq:w13},\ref{eq:w14} are identical; however, for the most general case they actually are not.  The tensors $\epsr$ and $\mur$ prevent such equivalence (because, for example, $(\epsr \mathbf{E_1})\cdot \mathbf{E_2} \neq \mathbf{E_1} \cdot (\epsr \mathbf{E_2})$).  In what scenario do we have such equivalence?  In Einstein notation, the previous relation becomes $\epsilon_{ij} E^{(1)}_j E^{(2)}_i \overset{?}{=} E^{(1)}_j \epsilon_{ji} E^{(2)}_i$.  The equality holds only if $\epsilon_{ij} = \epsilon_{ji}$, or in other words $\epsr = \epsr^\top$.  Additionally, we must also have $\mur = \mur^\top$.  A scalar permittivity and permeability will always work, but even anisotropic, non-diagonal permittivities and permeabilities are permissible as long as they are symmetric.  Hereafter, symmetric permittivities and permeabilities will be assumed.\footnote{Even symmetric permittivities and permeabilities are not actually necessary, for the purpose of adjoint-based optimization.  For example, if the permittivity for the second dipole, $\mathbf{P_2}$, had been given by $\epsr^\top$ instead of $\epsr$, the left-hand sides of Eqns.~\ref{eq:w13},\ref{eq:w14} would be equivalent without any assumptions on $\epsr$.  Likewise for $\mur$.  For non-symmetric permittivities and permeabilities, the adjoint simulation would simply require transposed material properties relative to the direct simulation.}

As discussed in Sec.~\ref{sec:ShapeCalc}, the electric field from a unit dipole is the Green's function.  $\overline{\overline{\mathbf{G^{\text{EP}}}}}(\mathbf{x_0},\mathbf{x'})$ is the electric field at $\mathbf{x_0}$ from an unit electric dipole at $\mathbf{x'}$.  Given a symmetric permittivity and permeability, the left-hand sides of Eqns.~\ref{eq:w13},\ref{eq:w14} are identical, and the right-hand sides must also be equal: $ \mathbf{\hat{\epsilon_1}} \cdot \mathbf{E_2}(\mathbf{x_1})  = \mathbf{\hat{\epsilon_2}} \cdot \mathbf{E_1}(\mathbf{x_2}) $.  In Green function notation, this relation is:
\begin{equation}
\mathbf{G}_{ij}^{\text{EP}}(\mathbf{x_2},\mathbf{x_1}) = \mathbf{G}_{ji}^{\text{EP}}(\mathbf{x_1},\mathbf{x_2})
\end{equation}  
This is one of the crucial symmetry relations needed for inverse design.  In words, it says
\begin{quote}
The $i$-component of the electric field at $\mathbf{x_2}$, from a $j$-oriented electric dipole at $\mathbf{x_1}$, exactly equals the $j$-component of the electric field at $\mathbf{x_1}$ from an $i$-oriented electric dipole at $\mathbf{x_2}$.
\end{quote}
The relation confirms the equivalence of the exchanged dipole and measurement points in Fig.~\ref{fig:DipoleEquivalence}, for the case of electric fields from electric dipoles.

\subsection{Case II: The magnetic field from a magnetic dipole}
In this case only Eqn.~\ref{eq:MaxwellWeak}(b) is needed.  We again consider two situations; this time, two magnetic dipoles:
\begin{align}
\label{eq:w21}
\mathbf{M_1} = \mathbf{\hat{\epsilon_1}} \frac{1}{\mu_0} (1 \text{ Wb$\cdot$m})\delta ^3(\mathbf{x}-\mathbf{x_1}) \\
\label{eq:w22}
\mathbf{M_2} = \mathbf{\hat{\epsilon_2}} \frac{1}{\mu_0} (1 \text{ Wb$\cdot$m})\delta ^3(\mathbf{x}-\mathbf{x_2})
\end{align}
We now employ $\mathbf{V}=\mathbf{H_2}$ in Eqn.~\ref{eq:MaxwellWeak}(b) for the first scenario, and $\mathbf{V}=\mathbf{H_1}$ in Eqn.~\ref{eq:MaxwellWeak}(b) for the second scenario:
\begin{align} 
\int_D \left[ \epsr^{-1} \nabla \times \mathbf{H_1} \cdot \nabla \times \mathbf{H_2} - \omega^2 \mur \mathbf{H_1} \cdot \mathbf{H_2} \right] \text{d}^3 \mathbf{x} & =  \omega^2 \mathbf{\hat{\epsilon_1}} \cdot \mathbf{H_2}(\mathbf{x_1}) \\
\int_D \left[ \epsr^{-1} \nabla \times \mathbf{H_2} \cdot \nabla \times \mathbf{H_1} - \omega^2 \mur \mathbf{H_2} \cdot \mathbf{H_1} \right] \text{d}^3 \mathbf{x} & =  \omega^2 \mathbf{\hat{\epsilon_2}} \cdot \mathbf{H_1}(\mathbf{x_2})
\end{align}
Assuming symmetric permittivity and permeability tensors, we arrive at the reciprocity relation for the magnetic field from a magnetic dipole: $\mathbf{\hat{\epsilon_1}} \cdot \mathbf{H_2}(\mathbf{x_1}) = \mathbf{\hat{\epsilon_2}} \cdot \mathbf{H_1}(\mathbf{x_2})$.  In Green's function notation:
\begin{equation}
\mathbf{G}_{ij}^{\text{HM}}(\mathbf{x_2},\mathbf{x_1}) = \mathbf{G}_{ji}^{\text{HM}}(\mathbf{x_1},\mathbf{x_2})
\end{equation}
This is very similar to the relation for the electric field from an electric dipole, and can be written in words:
\begin{quote}
The $i$-component of the magnetic field at $\mathbf{x_2}$, from a $j$-oriented magnetic dipole at $\mathbf{x_1}$, exactly equals the $j$-component of the magnetic field at $\mathbf{x_1}$ from an $i$-oriented magnetic dipole at $\mathbf{x_2}$.
\end{quote}

\subsection{Case III: The electric field from a magnetic dipole (and vice versa)}
For the third case we want to understand the behavior of the magnetic field radiated from an electric dipole.  It is \emph{not true} that one can simply switch the location of the dipole and the measurement point.  To see this, we again consider two scenarios, but this time we place an electric dipole at $x_1$ and a magnetic dipole at $x_2$:
\begin{align}
\mathbf{P_1} & = \mathbf{\hat{\epsilon_1}} (1 \text{ C$\cdot$m})\delta ^3(\mathbf{x}-\mathbf{x_1}) \\
\mathbf{M_2} & = \mathbf{\hat{\epsilon_2}} \frac{1}{\mu_0} (1 \text{ Wb$\cdot$m})\delta ^3(\mathbf{x}-\mathbf{x_2})
\end{align}
We use Eqn.~\ref{eq:MaxwellWeak}(a) for both scenarios.  Care must be taken with the magnetic dipole term in Eqn. (8.4).  Through integration by parts, one can make the substitution $\nabla \times \mur^{-1} \mathbf{M} \cdot \mathbf{V} = \mur^{-1} \mathbf{M} \cdot \nabla \times \mathbf{V}$, where it is implicitly assumed there are no magnetic dipoles on the outer boundary of the domain.  The two equations are then,
\begin{align}
\label{eq:w31}
\int_D \left[  \mur^{-1} \nabla \times \mathbf{E_1} \cdot \nabla \times \mathbf{V} - \omega^2 \epsr \mathbf{E_1} \cdot \mathbf{V} \right] \text{d}^3 \mathbf{x} & = \omega^2 \mathbf{\hat{\epsilon_1}} \cdot \mathbf{V}(\mathbf{x_1}) \\ 
\label{eq:w32}
\int_D \left[  \mur^{-1} \nabla \times \mathbf{E_2} \cdot \nabla \times \mathbf{V} - \omega^2 \epsr \mathbf{E_2} \cdot \mathbf{V} \right] \text{d}^3 \mathbf{x} & = -j  \omega \mur^{-1} \mathbf{\hat{\epsilon_2}} \cdot \nabla \times \mathbf{V}(\mathbf{x_2})
\end{align}
Now one can make the substitutions $\mathbf{V} = \mathbf{E_2}$ in Eqn.~\ref{eq:w31} and $\mathbf{V} = \mathbf{E_1}$ in Eqn.~\ref{eq:w32}.  Eqn.~\ref{eq:w32} is further simplified through the relation $\nabla \times \mathbf{E_1} = -j\omega \mur \mathbf{H_1}$.  The right-hand side then reads $-\omega^2 \mur^{-1} \mathbf{\hat{\epsilon_2}} \cdot \mur \mathbf{H_1}$.  In proper matrix notation, the previous expression would be written $-\omega^2 \left(\mur^{-1} \mathbf{\hat{\epsilon_2}}\right)^\top \mur \mathbf{H_1} = -\omega^2 \mathbf{\hat{\epsilon_2}}^\top (\mur^{-1})^\top \mur \mathbf{H_1} = -\omega^2 \mathbf{\hat{\epsilon_2}} \cdot \mathbf{H_1}$, where the assumption of symmetric $\mur$ also implied $\mur^{-1}$ was symmetric.  This results in the final reciprocity relation:
$ \mathbf{\hat{\epsilon_1}} \cdot \mathbf{E_2}(\mathbf{x_1})  = - \mathbf{\hat{\epsilon_2}} \cdot \mathbf{H_1}(\mathbf{x_2})$.  In terms of the Green's functions:
\begin{equation}
\mathbf{G}_{ij}^{\text{EM}}(\mathbf{x_2},\mathbf{x_1}) = -\mathbf{G}_{ji}^{\text{HP}}(\mathbf{x_1},\mathbf{x_2})
\end{equation}
The reciprocity relation for the magnetic field from an electric dipole can be stated
\begin{quote}
The $i$-component of the magnetic field at $\mathbf{x_2}$, from a $j$-oriented electric dipole at $\mathbf{x_1}$, exactly equals the negative of the $j$-component of the electric field at $\mathbf{x_1}$ from an $i$-oriented magnetic dipole at $\mathbf{x_2}$.
\end{quote}

%% file: TimeDependentShapeCalculus.tex
\chapter{Time-Dependent Merit Functions}
\label{chap:TimeDep}

The formulation of Chap.~\ref{chap:Algorithm} focused on optimal design for steady-state, continuous-wave signals.  Here, we generalize to time-dependent signals.  For derivations of the discretized time-dependent inverse problem, cf. \cite{Chung2000,Kang2002}.

Analogous to Eqn.~\ref{eq:genMerit} of Chap.~\ref{chap:Algorithm}, a time-dependent merit function can be written as an arbitrary function over the fields not only over a spatial domain $\chi$, but also over a time domain $T$:
\begin{equation}
\label{eq:tdMerit}
F(\mathbf{E},\mathbf{H}) = \int_T \int_{\chi} f \left(\mathbf{E}(\mathbf{x},t),\mathbf{H}(\mathbf{x},t)\right) \,\mathrm{d}^3 \mathbf{x}\,\mathrm{d}t
\end{equation}
For example, if one wanted to maximize absorption with the first $100ns$, one could set $f$ equal to the time-dependent absorption over the time-domain $T=[0,100ns]$.  

The derivation of the shape derivative proceeds along the same lines as in Sec.~\ref{sec:ShapeCalc}.  First, the variation of $F$ is specified with respect to the variation in the fields:\footnote{for an underlying variation in the geometry.}
\begin{equation}
\label{eq:deltaFTD}
\delta F = \int_T \int_{\chi} \left[ \frac{\partial f}{\partial \mathbf{E}} \cdot \delta \mathbf{E}(\mathbf{x},t) + \frac{\partial f}{\partial \mathbf{H}} \cdot \delta \mathbf{H}(\mathbf{x},t) \right] \,\mathrm{d}^3 \mathbf{x}\,\mathrm{d}t
\end{equation}
where the fields are now real-valued, and there is not an imaginary component to consider.  The variation in the fields is related to the induced polarization from a change in geometry through the Green's function, which is now time-dependent:
\begin{subequations}
\label{eq:deltaEHgreenTD}
\begin{align}
\delta \Eb(\mathbf{x}) = \int_{\mathcal{T}}\int_{\psi} \overline{\overline{\mathbf{G^{\text{EP}}}}}(\mathbf{x},t,\mathbf{x'},t') \mathbf{P^{\text{ind}}}(\mathbf{x'},t') \text{d}^3 \mathbf{x'} \,\mathrm{d}t' \\
\delta \Hb(\mathbf{x}) = \int_{\mathcal{T}}\int_{\psi} \overline{\overline{\mathbf{G^{\text{HP}}}}}(\mathbf{x},t,\mathbf{x'},t') \mathbf{P^{\text{ind}}}(\mathbf{x'},t') \text{d}^3 \mathbf{x'} \,\mathrm{d}t'
\end{align}
\end{subequations}
where $\psi$ is again the region over which the permittivity changes.  $\mathcal{T}$ is the region over which a polarization is induced, and causality requires that it contain only times before $t$, such that one generally has $\mathcal{T}=(-\infty,t)$.  Eqn.~\ref{eq:deltaEHgreenTD} can be inserted into Eqn.~\ref{eq:deltaFTD}.  The crucial simplification arises from the symmetry of the Green's functions, which for time-dependent fields becomes
\begin{figure}[]
\centering
\includegraphics[width=6in]{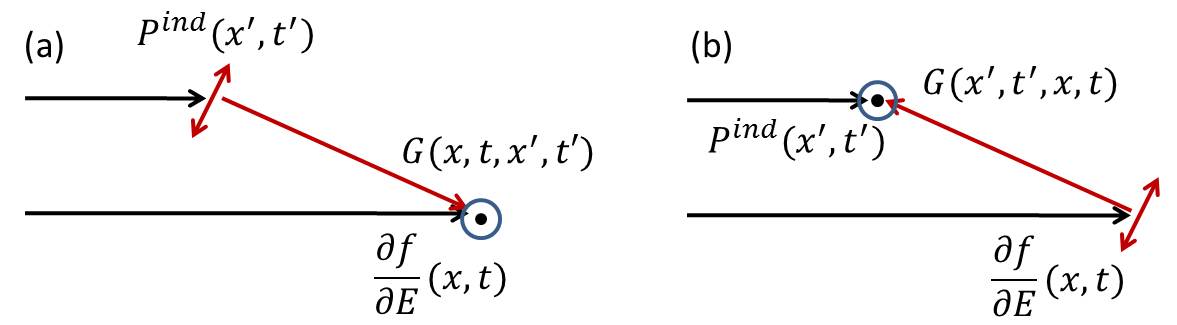}
\caption{Symmetry implications for time-dependent merit functions.  Instead of multiplying the fields at a current time $t$, the fields are multiplied at a past time $t'$, requiring the adjoint simulation to be run backwards in time.} 
\label{fig:ReversalFig}
\end{figure}
\begin{subequations}
\label{eq:GreenSymmTD}
\begin{align}
G_{ij}^{\text{EP}}(\mathbf{x},t,\mathbf{x'},t') = G_{ji}^{\text{EP}}(\mathbf{x'},t',\mathbf{x},t) \\
G_{ij}^{\text{HP}}(\mathbf{x},t,\mathbf{x'},t') = -G_{ji}^{\text{EM}}(\mathbf{x'},t',\mathbf{x},t)
\end{align}
\end{subequations}
Eqns.~\ref{eq:GreenSymmTD} have the same physical picture as in Chap.~\ref{chap:Algorithm}, except that the symmetrized version of the Green's function has a source radiating at time $t$ to a measurement at $t'$, where $t'<t$.  This can only be true if Maxwell's equations \emph{run backwards in time}, from the present time $t$ to some previous time $t'$.  This result is actually quite intuitive.  Consider the case of scalar fields for simplicity.  If there is a unit electric dipole source at $(\mathbf{x'},t')$ that creates an electric field $G$ at $(\mathbf{x},t)$, then one could imagine winding the clock backwards from $t$ to $t'$, with a unit electric dipole source at $(\mathbf{x},t)$ thus causing the same electric field $G$ at $(\mathbf{x'},t')$.  It was emphasized in Sec.~\ref{sec:ShapeCalc} that the Green's function symmetry does not require time-reversal invariance; nevertheless, the physical intuition provided by time-reversal aids in understanding the mechanism.

Insertion of Eqn.~\ref{eq:deltaEHgreenTD} into Eqn.~\ref{eq:deltaFTD} along with taking the symmetric counterpart of the Green's functions results in the equation
\begin{align}
\delta F = \int_{\mathcal{T}} \mathrm{d}t' \int_{\psi} \text{d}^3 \mathbf{x'} \text{ } P_{j}^{\text{ind}}(\mathbf{x'},t') \int_T \mathrm{d}t \int_{\chi} \text{d}^3 \mathbf{x} \left[ G_{ji}^{\text{EP}}(\mathbf{x'},t',\mathbf{x},t) \frac{ \partial f}{\partial E_i} (\mathbf{x},t) - G_{ji}^{\text{EM}}(\mathbf{x'},t',\mathbf{x},t) \frac{ \partial f}{\partial H_i } (\mathbf{x},t) \right] 
\end{align}
The term in brackets represents the fields at $(\mathbf{x'},t')$ from sources $\left(\mathbf{P},\mathbf{M}\right) = \left( \partial f / \partial \mathbf{E}, -\partial f / \partial \mathbf{H} \right)$ at $(\mathbf{x},t)$.  Integrating over all possible $\mathbf{x}$ and $t$ (i.e. over $\chi$ and $\mathcal{T}$) results in the adjoint field $\mathbf{E^A}(\mathbf{x'},t')$.  Such a replacement yields the final, time-dependent derivative:
\begin{align}
\label{eq:DerivTD}
\delta F = \int_{\mathcal{T}} \int_{\psi}  \text{ } \mathbf{P^{\text{ind}}}(\mathbf{x'},t') \cdot \mathbf{E^A}(\mathbf{x'},t')\text{d}^3 \mathbf{x'} \mathrm{d}t'
\end{align}
Eqn.~\ref{eq:DerivTD} is the shape/topological derivative for time-dependent merit functions.  It is the logical extension of Eqn.~\ref{eq:deltaFvector}, with the caveat that the sources for the adjoint simulation are driven by the time-dependent electric and magnetic fields, and the adjoint simulation must be run \emph{backwards in time}.